\documentclass{llncs}
\usepackage{graphicx} 
\usepackage{amsmath} 
\usepackage{amssymb}  
\usepackage{hyperref}
\usepackage{bigstrut}
\usepackage{tikz}
\usepackage{algorithm2e}
\usepackage{booktabs,caption}
\usepackage[flushleft]{threeparttable}
\usepackage{array}
\usepackage{multirow}
\usepackage{caption}
\usepackage{subcaption}
\usepackage{fancyhdr}
\newcommand\copyrighttext{%
  \footnotesize \emph{Submitted to Springer as final chapter in a monograph entitled ``Social Cybersecurity''}}
\newcommand\copyrightnotice{%
\begin{tikzpicture}[remember picture,overlay]
\node[anchor=south,yshift=20pt] at (current page.south) {\mbox{\parbox{\dimexpr\textwidth-\fboxsep-\fboxrule\relax}{\copyrighttext}}};
\end{tikzpicture}%
}
\begin{document}

\title{Social Cybersecurity Chapter 13: Casestudy with COVID-19 Pandemic }

\titlerunning{COVID-19 Case Study}  
%
\author{David M. Beskow \and Kathleen M. Carley}
\authorrunning{D. Beskow et al.} 
%
%
\institute{School of Computer Science \\ 
Carnegie Mellon University \\ 5000 Forbes Ave, Pittsburgh, PA 15213, USA \\
Email: \email{dbeskow@andrew.cmu.edu} and  \email{kathleen.carley@cs.cmu.edu}}

\maketitle              

\begin{abstract}
The purpose of this case study is to leverage the concepts and tools presented in the preceding chapters and apply them in a real world social cybersecurity context.  With the COVID-19 pandemic emerging as a defining event of the 21st Century and a magnet for disinformation maneuver, we have selected the pandemic and its related social media conversation to focus our efforts on.  This chapter therefore applies the tools of information operation maneuver, bot detection and characterization, meme detection and characterization, and information mapping to the COVID-19 related conversation on Twitter.  This chapter uses these tools to analyze a stream containing 206 million tweets from 27 million unique users from 15 March 2020 to 30 April 2020.  Our results shed light on elaborate information operations that leverage the full breadth of the BEND maneuvers and use bots for important shaping operations. 
\end{abstract}
\copyrightnotice{}
\section{Introduction}

The COVID-19 pandemic is a defining event of the modern era, and there are few events more appropriate to apply social cybersecurity tools and concepts.  At the time of this writing, the pandemic has reached almost every society in the world, with massive impact not only on the lives of those who contract it but on the social, economic, and institutional fabric of these societies.  The pandemic and differing opinions on how to react to it have created a virtual battle of ideas across social media.  Actors ranging from soccer moms to well-resourced nation states have entered the virtual marketplace of beliefs and ideas trying to sway the beliefs, actions, and decisions of both leaders and followers.   With the pandemic as the backdrop of life as we write this book, it seemed appropriate to use the social cybersecurity tools that we discussed in the previous chapters to identify and understand information operations related to COVID-19.  

There are still many questions as well as competing narratives about the origins and nature of the COVID-19 coronavirus disease.  The disease is caused by the severe acute respiratory syndrome coronavirus (SARS-CoV-2), and was first identified in Wuhan China in December 2019.  While still spreading, as of 15 May 2020 the pandemic has grown to 4,651,119 confirmed cases with 312,119 global deaths according to the John Hopkins University COVID-19 Data Map\footnote{\url{https://coronavirus.jhu.edu/map.html}}.    

The virus and resulting pandemic have radically altered the world landscape and daily life for many people.  With a large portion of the world's consumers sheltering in or near their home, the world's markets have slowed, inducing the greatest global recession since the Great Depression \cite{TheGreat93:online}.  It has led to the cancellation or deferment of most travel activities, sporting and entertainment events, religious and political gatherings.  As of 14 April 2020, schools and universities had closed or otherwise  been disrupted in 191 countries affecting 1.57 billion students (pre-primary, primary, lower-secondary, and upper secondary levels of education), or 90.1\% of the total enrolled students for these categories \cite{Schoolcl50:online}.  

The COVID-19 pandemic has induced information warfare at multiple levels, both within and between nations.  At the individual level, much of the information is in regards to safety and health during the pandemic.  Social media often doesn't have structural filters for good ideas, and at times poor ideas regarding health and safety have been promoted by segments of the crowd.  Within countries, the pandemic has created two warring factions: 1) those who think policy should prioritize the health and safety of citizens and 2) those who think policy should prioritize the national economy and maintaining the jobs and livelihoods of citizens.  These factions often seem to fall along existing political party lines, with conservatives emphasizing the economy and liberals emphasizing health and safety.  This policy oriented friction has contributed a large portion of the information conflict within nations.  Finally, the origin and nature of the COVID-19 coronavirus has aggravated existing geopolitical fault lines between nations, particularly between the United States and China, with the European Union, Russia, Brazil, and other nations participating in the conflict of narratives.  As we look at information warfare and social cybersecurity in the COVID-19 pandemic conversation, we attempt to illuminate information conflict across this spectrum. 

Each country's response has varied based on its society, forms of government, and health care system.  The COVID-19 pandemic has served as a worldwide test of the resiliency of these systems.  Nations are therefore turning to information warfare to strengthen their test results while attacking the response and results of other nations, particularly those whose form of government differs from theirs.  This is particularly true in the information conflict between China and United States.  

This chapter will showcase the use of social cybersecurity tools and theory to identify and characterize information operations in the COVID-19 related Twitter Stream.  To do this, we will start by discussing data collection and initial exploratory data analysis.  Then we will conduct bot detection and characterization, merging bot classification with other quantitative methods.  We will then look at meme classification and analysis, highlighting the use of multi-modal data for information warfare.  Finally, we will briefly illustrate the use of Sketch-IO.  Throughout this chapter will highlight the importance of BEND as a foundational construct for understanding modern information warfare.

\section{Collecting Data}

Like most event oriented social cybersecurity data collection on Twitter, our team started by establishing a Twitter Stream with select COVID-19 related terms, periodically expanding the terms as appropriate.  This gives us the main foundation of data for assessment.  Once we began analysis of the stream, we turned to the Twitter REST API to collect other data as necessary.  This additional data is often account oriented and includes \textit{timelines}, \textit{friends}, \textit{followers}, and at times account ID ``re-hydration''.  We will discuss each of these below.  

For COVID-19, our team began collecting data on 27 January 2020 with a limited list of keywords and expanded this list on 13 March 2020.  For the data that we will focus on in this chapter, the list of keywords are: ``coronaravirus'', ``coronavirus'', ``wuhan virus'', ``wuhanvirus'', ``2019nCoV'', ``NCoV'', ``NCoV2019'', ``covid-19'', ``covid19'', ``covid 19''.  The resulting stream provided the primary data that we will focus our study on.  In this chapter, we will focus on the data from 15 March 2020 to 30 April 2020.  The summary statistics of this data is provided in Table \ref{tab:stream_statistics}.  This is not the entire COVID-19 conversation.  Many tweets may not contain these key words, or may be in other languages.  Additionally, our stream is limited to 50 tweets per second, or approximately 4.3 million tweets per day.  The temporal analysis below will highlight the 4.3 million daily limit on our stream.  

As we go through our analysis, we will return at times to the Twitter REST API to collect additional data that is not found in the stream.  For example, the stream only contains an account's COVID-19 content, but not all other content and topics.  To get access to all content shared by an account, we will at times collect their account \emph{timeline} (aka account \emph{history}).  There are also times we need to identify an account's \textit{friends} and \textit{followers}, which will also require us to call the Twitter REST API.  Finally, at the end of our exploratory data analysis, we will try to find out if any accounts have been suspended by Twitter since contributing content to our stream.  We will do this by ``re-hydrating'' account IDs in order to see if the account still exists.  If the account doesn't exist, we will then test to see if the account was suspended or deleted by the user.

\begin{table}[htbp]
  \centering
  \caption{Statistics for COVID-19 Twitter Stream (15 March 2020 to 30 April 2020)}
    \begin{tabular*}{\textwidth}{l@{\extracolsep{\fill}}r}
    Statistic & Value \\ \hline
    Total Tweets &        206,330,119  \\
    Total Unique Users &          27,153,875  \\
    Original Tweets &          23,502,825  \\
    Retweet Tweets &        126,878,369  \\
    Reply Tweets &            8,705,000  \\
    Quote Tweets &          47,243,925  \\
    Hashtags &            1,854,437  \\
    Bots @ Tier1 &          11,381,861  \\
    Images &          10,704,232  \\
    URLs  &          50,088,836  \\ 
    Unique Domains &               305,597  \\
    State Sponsored Media Tweets &                   9,120  \\
    Retweet/Mentions of State Sponsored Media &               435,557  \\
    Tweets from Verified Accounts &            4,031,779  \\ \hline
    \end{tabular*}%
  \label{tab:stream_statistics}%
\end{table}%

\section{Initial Exploratory Data Analysis}

All data analysis begins with a thorough exploration of the data.  The analyst should explore all of the distributions and possible relationships in the data.  In our case, this means exploring the temporal, categorical, geospatial, and quantitative distributions of various fields in the Twitter Data.  We collected and parsed the data using the \texttt{twitter\_col}\footnote{\url{https://github.com/dmbeskow/twitter_col}} Python package that we created to assist with collecting and manipulating Twitter data.  

The primary statistics for the data are provided in Table \ref{tab:stream_statistics}.  We have 206 million tweets produced by 27 million users.  We see that the majority of the tweets are actually retweets. In fact, only 23 million tweets (11\% of the stream) is original content produced by the respective account.  It is also interesting that quotes significantly outnumber replies.  A quote is produced when a user references another tweet and comments on it, starting a new thread.  Replies, by contrast, are addressed to the original author and don't start a new thread.  We see a modest number of state sponsored tweets, with substantially more amplification of state sponsored accounts.  Finally, for reference, this primary stream is 177 gigabytes in compressed \texttt{gzip} format or 1.54 terabytes in uncompressed format.

\subsection{Temporal Analysis}

Next we analyze the temporal distribution of our stream.  This is primarily done to study the peaks and valleys to understand when the conversation surges and ebbs. This analysis will also ensure that our tweets fall within the scope of our study (from 15 March 2020 to 30 April 2020).  

\begin{figure}[htb]
\centering
  \includegraphics[width=\linewidth]{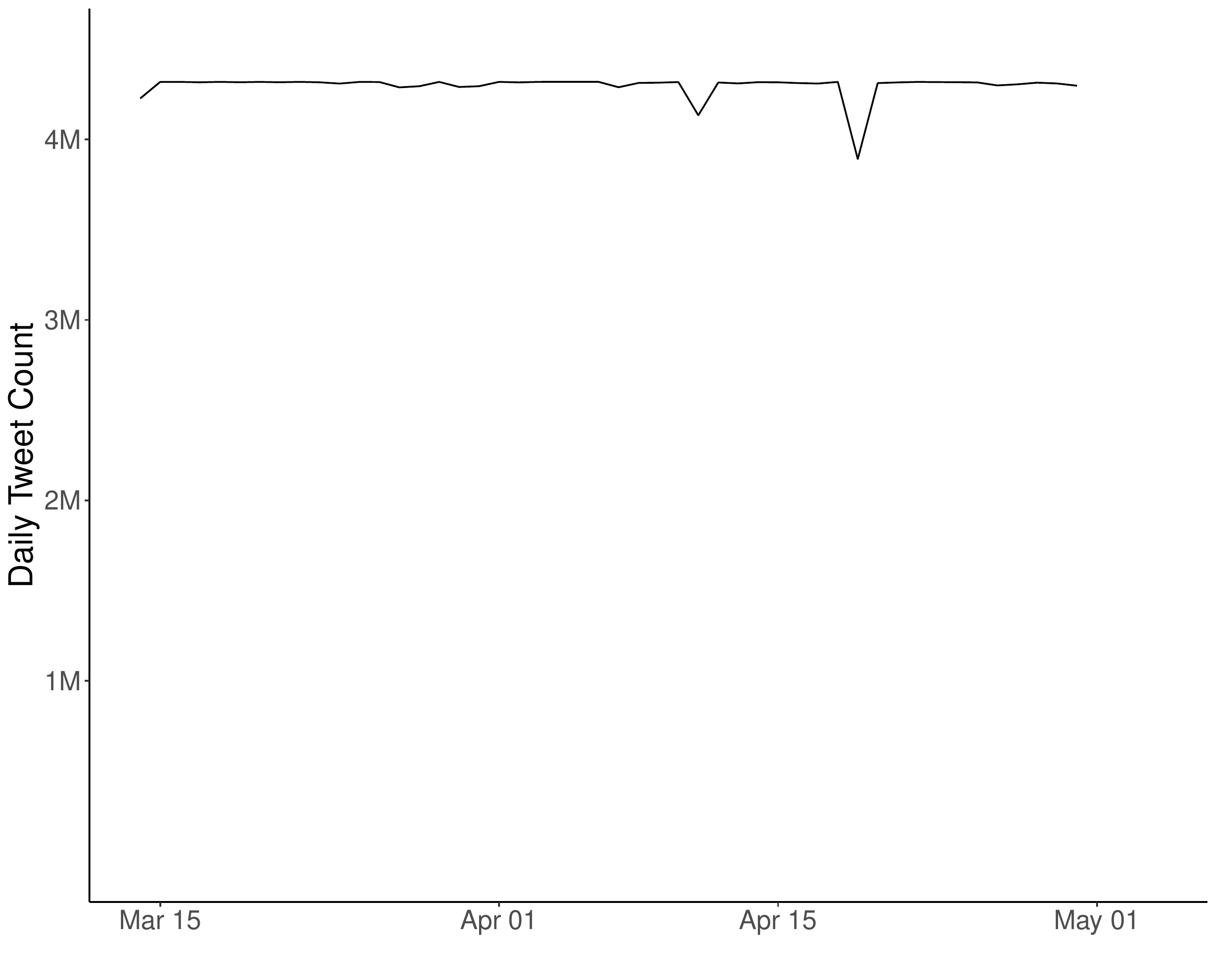}
  \caption{Daily Tweet Count in Stream}
  \label{fig:casestudy_timeline}
\end{figure}

The temporal distribution of our stream is provided in Figure \ref{fig:casestudy_timeline}.  This is not what we expect to see, and is definitely not what you would observe in smaller streams.  Given the size of our stream, we are artificially limited by Twitter rate limiting.  Twitter limits basic streaming API's to no more than 50 tweets per second, or 4.3 million tweets per day.  It is unknown what portion of the total conversation we are getting, but given that several days fall below the 4.3 million, we believe we are collecting the majority.  Twitter did open up a unlimited COVID-19 Stream in mid-May \cite{Overview64:online}, but to our knowledge this was not available during the time frame we were interested in.

\subsection{Geospatial Density}

\begin{figure}[htbp]
\centering
  \includegraphics[width=\linewidth]{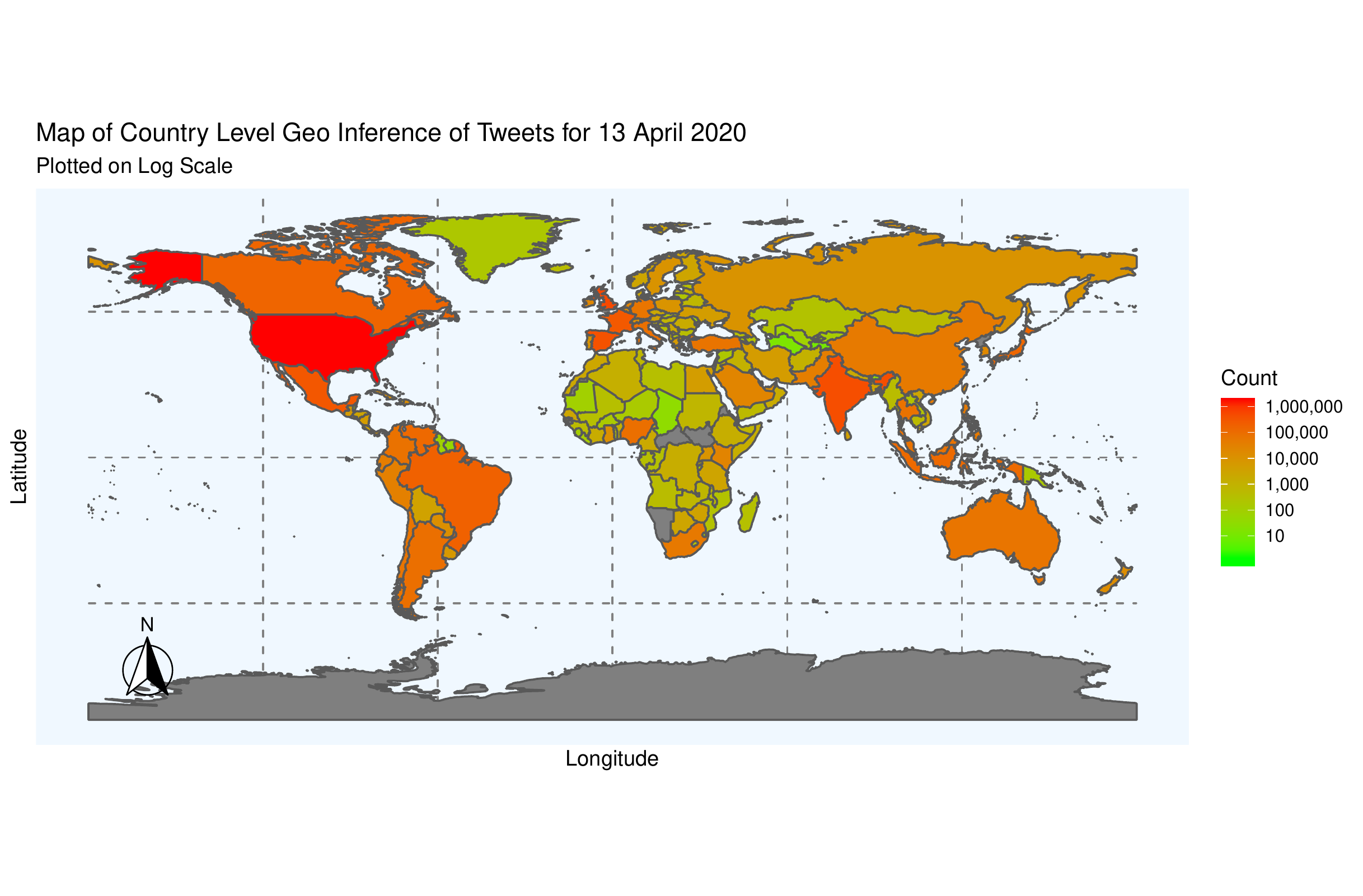}
  \caption{Geospatial Denstiy of Geo-Inferred Location for Tweets on 13 April 2020}
  \label{fig:location_map}
\end{figure}

In order to measure the geospatial density of the data, we used a country level geo-inference algorithm created by Huang and Carley \cite{huang2017predicting}.  This model infers the country level location of tweets with 92.1\% accuracy.  We ran this on data from 13 April and plotted it on a chloropleth map in Figure \ref{fig:location_map} with a logarithmic scale.  This shows that the vast majority of the data is from the United States, with notable contributions from Canada, South America, Western Europe, Nigeria, South and Southeast Asia.  This Geospatial inference and visualization provides yet another facet of our exploratory data analysis.  Given both the keyords that we used and Twitter users by country, this spatial distribution is essentially what one would expect.

\subsection{Categorical Data Analysis}

Its important to explore categorical data distributions in Twitter data.  In addition to the retweet/reply/quote/original content distribution that we explored in Table \ref{tab:stream_statistics}, we also want to look at top languages, hashtags, mentions, and URL domains.   

\begin{figure}[htbp]
\centering
  \includegraphics[width=\linewidth]{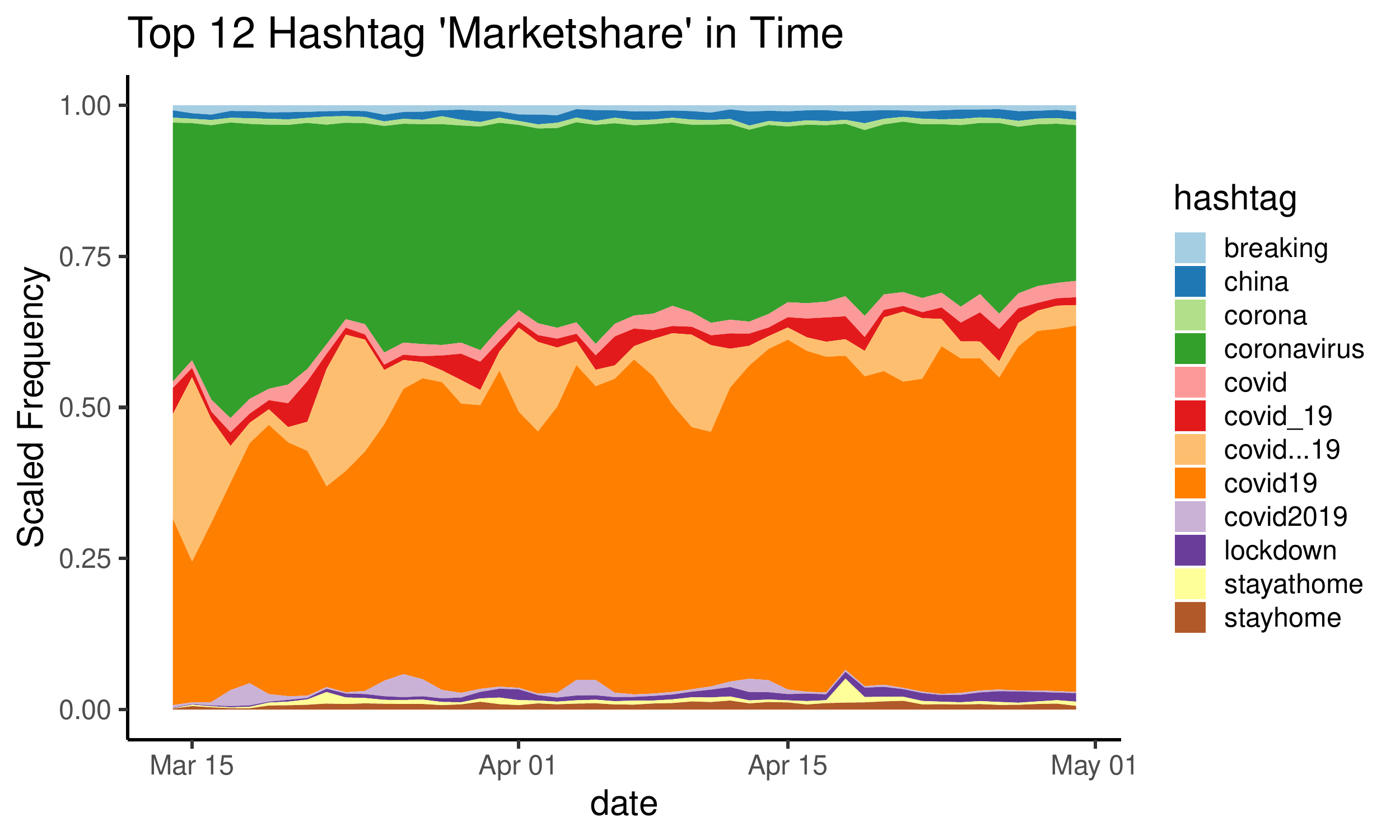}
  \caption{Prominent Hashtag ``marketshare'' over time.}
  \label{fig:hashtag_marketshare}
\end{figure}

The importance of hashtags ebbs and flows with time and events.  We've developed the visualization found in Figure \ref{fig:hashtag_marketshare} to understand this changing dynamic for the top 12 hashtags found in the stream.  We observe decreasing use of ``corona'' and increasing use of ``COVID19'' as a hashtag to identify pandemic related content and conversation.  This reflects both the naming of the virus and the increasing knowledge in the world about the virus.

The full counts for languages and domains is provided in Table \ref{tab:language_domain}. We see that the majority of the content is in English, followed by many of the prominent world languages.  The only major world language that is underrepresented is Chinese, and that's because Twitter is blocked by the ``Great Firewall of China.''  We note that much of the data from China in this data set is actually from state sponsored chinese media.  The domains include several link shorteners, multimedia companies, as well as a spattering of news media companies, including both traditional and alternative news, independent and state owned media, and both conservative and liberal leaning news sites.  

\begin{table}[htb]
  \centering
  \caption{Top Languages and Domains for the COVID-19 Twitter Stream}
    \begin{tabular*}{\textwidth}{l@{\extracolsep{\fill}}r@{\hskip 1.5in}lr}
    \multicolumn{1}{c}{Language} & \multicolumn{1}{l}{Count} & \multicolumn{1}{c}{Domain} & \multicolumn{1}{c}{Count} \\ \hline
    English &        119,706,946  & bit.ly &        1,353,911  \\
    Spanish &          37,269,276  & youtu.be &           567,527  \\
    French &            9,434,176  & dlvr.it &           523,276  \\
    Indonesian &            6,688,718  & theguardian.com &           506,881  \\
    Portuguese &            5,941,304  & nytimes.com &           455,153  \\
    Thai  &            3,917,547  & ow.ly &           310,834  \\
    Japanese &            3,434,460  & pscp.tv &           279,671  \\
    Italian &            2,281,549  & trib.al &           276,783  \\
    Hindi &            2,140,788  & ift.tt &           265,921  \\
    Turkish &            2,000,057  & reuters.com &           258,747  \\
    Catalan &            1,608,929  & cnn.com &           239,437  \\
    German &            1,204,235  & washingtonpost.com &           195,058  \\
    Tagalog &               869,062  & foxnews.com &           192,790  \\
    Arabic &               590,716  & nypost.com &           183,850  \\
    Dutch &               534,198  & buff.ly &           177,518  \\
    Tamil &               352,563  & independent.co.uk &           157,276  \\
    Russian &               349,828  & bbc.in &           149,790  \\
    Korean &               285,612  & rt.com &           144,361  \\
    Polish &               203,136  & youtube.com &           133,724  \\
    Urdu  &               195,197  & breitbart.com &           131,119  \\
    Chinese &               184,123  & sky.com &           130,732  \\
    Greek &               116,296  & latimes.com &           129,909  \\
    Romanian &               114,787  & uol.com.br &           121,347  \\
    Swedish &                 97,838  & bbc.co.uk &           118,402  \\
    Estonian &                 95,398  & rawstory.com &           115,461  \\
    Marathi &                 94,987  & paper.li &           112,391  \\
    Farsi &                 85,832  & justthenews.com &           112,243  \\
    Lithuanian &                 63,168  & hill.cm &           111,731  \\ \hline
    \end{tabular*}%
  \label{tab:language_domain}%
\end{table}%

\subsection{Visualize the Network}

By its very nature social media creates links between people, and these links combine to create various types of networks.  Most social media have a \textit{friend} or \textit{follow} functionality, which creates the most obvious social network on these platforms.  Additionally, the online conversation itself will create links between accounts.  Whenever an account \textit{retweets},\textit{ mentions}, \textit{replies}, or \textit{quotes} another account they are creating a type of conversational link.  These links create a second type of social media network...the \textit{conversation} network. 

With Twitter data, collecting friend/follower links is limited by strict rate-limiting on the part of Twitter.  Most developer accounts are only allowed to scrape 5,000 friends/followers for one account every minute.  The Twitter REST API will only return 5,000 friends/followers per request.  For example, at the time of this writing the primary Twitter account for the United States Center for Disease Control (@CDCgov), has 2.7 million followers.  To get all of the followers for @CDCGov would take approximately $\frac{2,700,000}{5,000} = 540$ minutes.  In other words, it would take 9 hours to scrape the followers for this single account.  Scraping links for all of the accounts in the COVID-19 conversation (and most conversations) is therefore not realistic.  For this reason we often visualize parts of the \textit{conversation} network, for which we already have the data and is more useful in understanding the conversation.  In our visualization of the network, we visualized the \emph{mention} network, though at other times the \emph{retweet} or \emph{reply} network may be appropriate. 

The Twitter JSON structure has all of the information we need to create \textit{mention}, \textit{retweet}, or \textit{reply} networks.  As discussed at other times in this book, we created the public facing \texttt{twitter\_col}\footnote{\url{https://github.com/dmbeskow/twitter_col}} Python package to collect and manipulate Twitter data.  The \texttt{twitter\_col} package has several functions that make it easy to parse networks from raw Twitter JSON data.  The ORA software \cite{carley2014ora,carley2020ora}, both ORA-PRO \footnote{\url{http://netanomics/products-2}} and ORA-Lite software\footnote{\url{http://www.casos.cs.cmu.edu/projects/ora/}} has functions for parsing Twitter data into networks.  These functions were used on the COVID-19 stream.  

Visualizing large networks can be difficult.  The COVID-19 mention network contains 25,673,160 nodes and 152,481,141 edges ($density = 0.000000231$).  Very few software packages can visualize a network of this size, and none of the software solutions that our team commonly uses can visualize this network.  For this reason, we chose to visualize just the core of the network.  To do this, we found the k-core of the network.  The k-core of graph $G$ is the maximal connected subgraph of $G$ where all the nodes (vertices) have a degree of at least $k$.  While experimenting with $k$ with our mention network, we found the $k=100$ was adequate.  This means we will visualize the core of the \emph{mention} network in which all nodes will have a degree of at least 100.  This core network is dense ($density = 0.000947$), which means there are more edges than we are able to visualize.  To sparsify the graph, we sampled one million edges to visualize.  This final core network contains 159,533 nodes and 1,000,000 edges with $density = 0.0000393$.

\begin{figure}[htbp]
\centering
  \includegraphics[width=\linewidth]{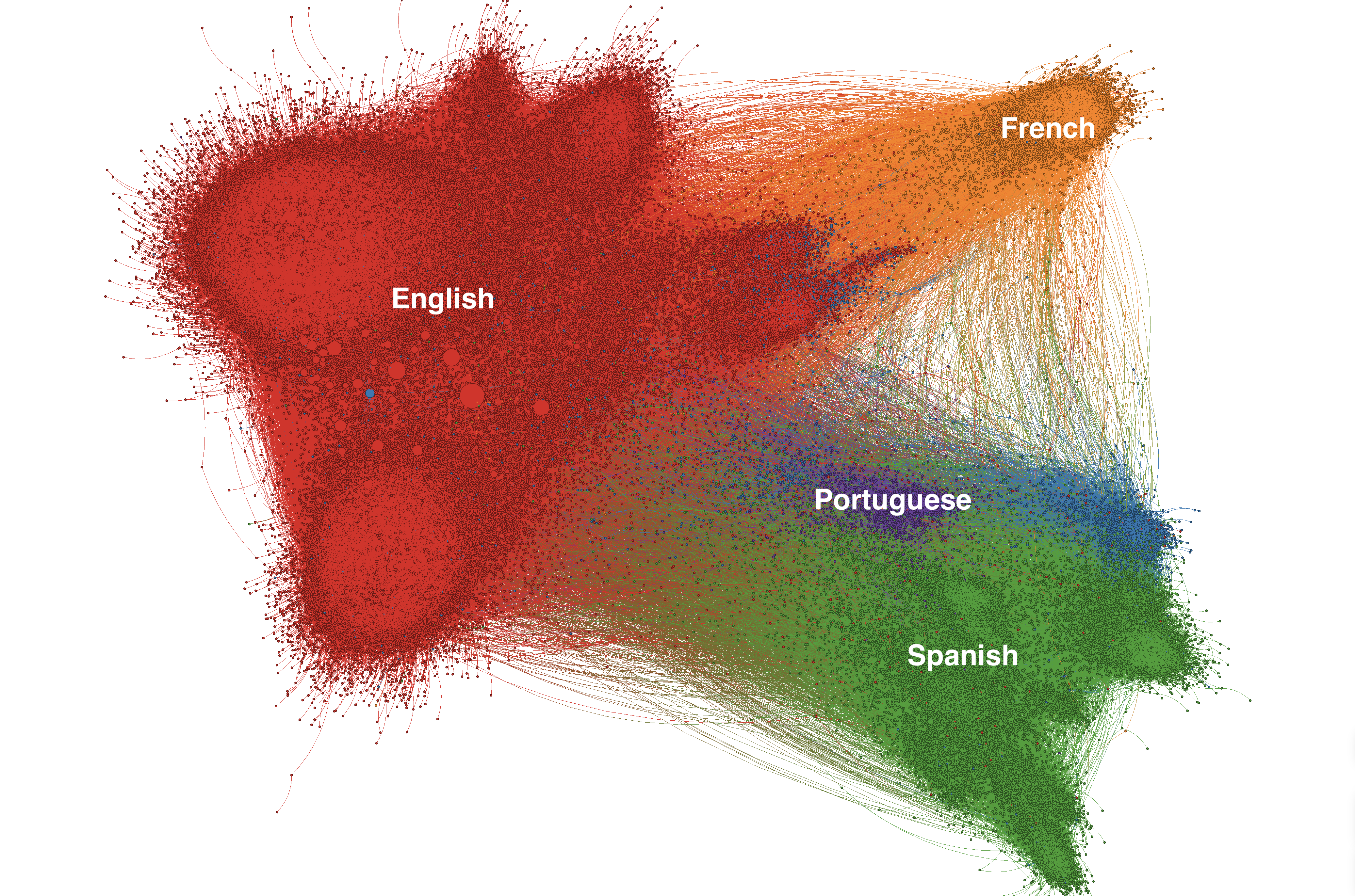}
  \caption{Visualizing the sparsified k-core ($k=100$) of the mention network, colored by prominent language.}
  \label{fig:mention_network_lang2_annotated}
\end{figure}

In Figure \ref{fig:mention_network_lang2_annotated} we visualize the core of the mention network colored by language using the \texttt{Graphistry}\footnote{\url{https://www.graphistry.com/}} software.  We could also visualize larger networks with \texttt{ORA-PRO} or the  \texttt{sigmaNet}\footnote{\url{https://github.com/iankloo/sigmaNet}} package in R. In Figure \ref{fig:mention_network_lang2_annotated} we see the inter-connection between the English, French, Spanish, and Portuguese conversations.  We also can see the community groups that are clearly evident in the larger conversations, particularly the English and Spanish conversation.

\subsection{Identification of Influencers}

Given that we now have networks parsed from our COVID-19 stream, there are a number of network science techniques that we can use to help understand this network.  For now we will focus on the measuring which accounts (or nodes in the network) are the most influential (or central) to the network.  Measuring \emph{centrality} is an important step in network science, and there are many different methods that have been published on how to do this, with each technique measuring a slightly different definition of influence and importance.  For example, \textit{degree} centrality measures influence by number of connections whereas \textit{betweenness} centrality measures influence by those accounts that bridge communities.

For our analysis, we chose to measure centrality and influence by \textit{eigenvector} centrality  \cite{katz1953new}. Eigenvector centrality measures influence by finding accounts that have the most connections to influential accounts. In other words, not all links are created equal, and an account is more influential if it is connected to many nodes who themselves have high scores.  We believe that this mirrors the way that many people view influence in the physical world, and therefore use eigenvector centrality for our analysis.  We also find that eigenvector centrality, unlike measures like \textit{betweenness}, is computationally practical on large networks.  

The top influencers as measured by eigenvector centrality for both the mention and the retweet networks are shown below in Table \ref{tab:influtential_bots}.  As we will discuss later, the central accounts in the mention network are politicians and celebrities that we expect.  The central accounts in the retweet network, however, includes many bots, which we will elaborate on later.  

\begin{table}[htbp]
  \centering
  \caption{Identifying Influential Accounts (\emph{red} indicates bot-like accounts)}
    \begin{tabular*}{\textwidth}{c@{\extracolsep{\fill}} lc @{\hskip 0.5in} lc}
     & \multicolumn{2}{c}{Mention Network} & \multicolumn{2}{c}{Retweet Network} \\
\cmidrule{2-3}  \cmidrule{4-5}
    Centrality Rank & Screen Name & Bot Score & Screen Name & Bot Score \\ \hline
    1     & realDonaldTrump & 0.08  & realDonaldTrump & 0.08 \\
    2     & POTUS & 0.10  & WhiteHouse & 0.18 \\
    3     & SpeakerPelosi & 0.01  & RealJamesWoods & 0.00 \\
    4     & CNN   & 0.43  & TomFitton & 0.02 \\
    5     & WhiteHouse & 0.18  & RealCandaceO & 0.04 \\
    6     & WHO   & 0.03  & TrumpWarRoom & 0.15 \\
    7     & NYGovCuomo & 0.00  & DonaldJTrumpJr & 0.01 \\
    8     & JoeBiden & 0.01  & \textcolor[rgb]{ 1,  0,  0}{AngelWarrior321} & \textcolor[rgb]{ 1,  0,  0}{0.59} \\
    9     & RudyGiuliani & 0.02  & \textcolor[rgb]{ 1,  0,  0}{GmanFan45} & \textcolor[rgb]{ 1,  0,  0}{0.78} \\
    10    & nytimes & 0.44  & SJPFISH & 0.49 \\
    11    & RealJamesWoods & 0.00  & jsolomonReports & 0.09 \\
    12    & VP    & 0.09  & \textcolor[rgb]{ 1,  0,  0}{alley167} & \textcolor[rgb]{ 1,  0,  0}{0.59} \\
    13    & DonaldJTrumpJr & 0.01  & michaeljohns & 0.06 \\
    14    & BreitbartNews & 0.16  & \textcolor[rgb]{ 1,  0,  0}{darhar981} & \textcolor[rgb]{ 1,  0,  0}{0.66} \\
    15    & \textcolor[rgb]{ 1,  0,  0}{FoxNews} & \textcolor[rgb]{ 1,  0,  0}{0.55} & \textcolor[rgb]{ 1,  0,  0}{superyayadize} & \textcolor[rgb]{ 1,  0,  0}{0.71} \\
    16    & CDCgov & 0.03  & \textcolor[rgb]{ 1,  0,  0}{gaye\_gallops} & \textcolor[rgb]{ 1,  0,  0}{0.62} \\
    17    & YouTube & 0.03  & \textcolor[rgb]{ 1,  0,  0}{RedPillMaC} & \textcolor[rgb]{ 1,  0,  0}{0.76} \\
    18    & IngrahamAngle & 0.02  & \textcolor[rgb]{ 1,  0,  0}{Julietknows1} & \textcolor[rgb]{ 1,  0,  0}{0.69} \\
    19    & RealCandaceO & 0.04  & trump\_noodle & 0.44 \\
    20    & marklevinshow & 0.13  & \textcolor[rgb]{ 1,  0,  0}{ConservaMomUSA} & \textcolor[rgb]{ 1,  0,  0}{0.70} \\
    21    & TomFitton & 0.02  & \textcolor[rgb]{ 1,  0,  0}{MissILmom} & \textcolor[rgb]{ 1,  0,  0}{0.76} \\
    22    & TuckerCarlson & 0.01  & \textcolor[rgb]{ 1,  0,  0}{zeusFanHouse} & \textcolor[rgb]{ 1,  0,  0}{0.64} \\
    23    & AOC   & 0.03  & \textcolor[rgb]{ 1,  0,  0}{JodyBelcher7} & \textcolor[rgb]{ 1,  0,  0}{0.85} \\
    24    & SenSchumer & 0.01  & \textcolor[rgb]{ 1,  0,  0}{Wyn1745} & \textcolor[rgb]{ 1,  0,  0}{0.70} \\
    25    & HillaryClinton & 0.00  & NevadaElJefe & 0.43 \\
    26    & seanhannity & 0.02  & \textcolor[rgb]{ 1,  0,  0}{KeishaJake} & \textcolor[rgb]{ 1,  0,  0}{0.83} \\
    27    & Acosta & 0.01  & dbongino & 0.02 \\
    28    & dbongino & 0.02  & \textcolor[rgb]{ 1,  0,  0}{ShawnG927} & \textcolor[rgb]{ 1,  0,  0}{0.72} \\
    29    & GavinNewsom & 0.01  & BlondieVex & 0.42 \\
    30    & TheDemocrats & 0.01  & \textcolor[rgb]{ 1,  0,  0}{Orcusa1} & \textcolor[rgb]{ 1,  0,  0}{0.57} \\
    31    & \textcolor[rgb]{ 1,  0,  0}{MSNBC} & \textcolor[rgb]{ 1,  0,  0}{0.62}  & marklevinshow & 0.13 \\
    32    & BillGates & 0.00  & NoodleSparklez & 0.44 \\
    33    & gatewaypundit & 0.40  & \textcolor[rgb]{ 1,  0,  0}{inthecopa} & \textcolor[rgb]{ 1,  0,  0}{0.51} \\
    34    & GovWhitmer & 0.01  & \textcolor[rgb]{ 1,  0,  0}{cajun4trump} & \textcolor[rgb]{ 1,  0,  0}{0.71} \\
    35    & BarackObama & 0.01  & \textcolor[rgb]{ 1,  0,  0}{BMcAdory9} & \textcolor[rgb]{ 1,  0,  0}{0.72} \\
    36    & TrumpWarRoom & 0.15  & \textcolor[rgb]{ 1,  0,  0}{TrumperSeaney} & \textcolor[rgb]{ 1,  0,  0}{0.70} \\
    37    & OANN  & 0.21  & \textcolor[rgb]{ 1,  0,  0}{TheRISEofROD} & \textcolor[rgb]{ 1,  0,  0}{0.71} \\
    38    & GOP   & 0.04  & \textcolor[rgb]{ 1,  0,  0}{Aliciastarr001} & \textcolor[rgb]{ 1,  0,  0}{0.78} \\
    39    & jsolomonReports & 0.09  & \textcolor[rgb]{ 1,  0,  0}{Jali\_Cat} & \textcolor[rgb]{ 1,  0,  0}{0.77} \\
    40    & charliekirk11 & 0.02  & \textcolor[rgb]{ 1,  0,  0}{phillyeaglesfa1} & \textcolor[rgb]{ 1,  0,  0}{0.80} \\
    41    & washingtonpost & 0.40  & \textcolor[rgb]{ 1,  0,  0}{go4itbas} & \textcolor[rgb]{ 1,  0,  0}{0.80} \\
    42    & tedcruz & 0.01  & \textcolor[rgb]{ 1,  0,  0}{up\_weekly} & \textcolor[rgb]{ 1,  0,  0}{0.83} \\
    43    & senatemajldr & 0.01  & \textcolor[rgb]{ 1,  0,  0}{HLAurora63} & \textcolor[rgb]{ 1,  0,  0}{0.77} \\
    44    & ABC   & 0.43  & \textcolor[rgb]{ 1,  0,  0}{AnthemRespect} & \textcolor[rgb]{ 1,  0,  0}{0.71} \\
    45    & \textcolor[rgb]{ 1,  0,  0}{CBSNews} & \textcolor[rgb]{ 1,  0,  0}{0.56} & \textcolor[rgb]{ 1,  0,  0}{MAGAPATRIOT\_TGM} & \textcolor[rgb]{ 1,  0,  0}{0.86} \\
    46    & NYCMayor & 0.01  & \textcolor[rgb]{ 1,  0,  0}{GA\_peach3102} & \textcolor[rgb]{ 1,  0,  0}{0.58} \\
    47    & FLOTUS & 0.04  & \textcolor[rgb]{ 1,  0,  0}{trumptrain1111} & \textcolor[rgb]{ 1,  0,  0}{0.76} \\
    48    & \textcolor[rgb]{ 1,  0,  0}{AngelWarrior321} & \textcolor[rgb]{ 1,  0,  0}{0.59} & \textcolor[rgb]{ 1,  0,  0}{LaylaAlisha11} & \textcolor[rgb]{ 1,  0,  0}{0.78} \\
    49    & RepAdamSchiff & 0.00  & mitchellvii & 0.13 \\
    50    & mitchellvii & 0.13  & \textcolor[rgb]{ 1,  0,  0}{MBOKSR\_MAGA} & \textcolor[rgb]{ 1,  0,  0}{0.89} \\ \hline
    \end{tabular*}%
  \label{tab:influtential_bots}%
\end{table}%

\subsection{Understanding Suspended Accounts}

It is often important to evaluate the quantity and nature of suspended accounts in any event oriented stream.  Twitter and most social media companies suspend accounts that frequently violate their terms of service.  These violations could include frequently posting violent, racist, or other unauthorized content.  It also could be that the accounts display unauthorized automated activity (i.e. they are a bot).  By identifying and evaluating suspended accounts, we get a sense of how the social media company has been ``cleaning'' this particular stream.  

To identify suspended accounts, we  ``re-hydrate'' account IDs in order to see if the account still exists.  If the account doesn't exist, we will then test to see if the account was suspended by Twitter or deleted by the user.  The workflow begins by identifying all unique account ID's in the stream.  We then ``rehydrate'' them in batch mode using the Twitter Rest API.  Using batch mode is a fast method to ``re-hydrate'', but does not provide any feedback for missing accounts.  Having rehydrated account IDs, we then identify those that are missing with $Missing = Total - Rehydrated$.  There are several reasons why an account could be missing.  The two most common are that the account was deleted by the user or was suspended by Twitter.  To determine which of these is the case, we individually attempt to rehydrate the missing IDs (not in batch mode).  This provides a detailed response that indicates whether the account was deleted or suspended.  

Given the size of the COVID-19 stream, we randomly sampled 1 million users from the available 27 million unique users in the stream.  After ``re-hydrating'' in batch mode, we determined that 70,842 were missing.  Using this number, we estimate that $7.0842 \pm 0.05$\% of the accounts in the stream have been deleted or suspended (estimated using 95\% confidence interval).  

Next we want to estimate the number that have been suspended.  To do this we now individually attempt to ``rehydrate'' the missing IDs.  When we do this, we find that 9,639 accounts had been suspended by Twitter.  Using this number, we estimate that Twitter has suspended $0.9639 \pm 0.019$\% (estimated using 95\% confidence interval).   Using these IDs, we made another pass through our stream, and determined that these 9,639 accounts produced 93,841 tweets that contain the terms we were filtering for the stream.  Running Bot-Hunter Tier 1 algorithm on this data we find that 73.4\% of the suspended users have strong bot-like characteristics. 

It is often helpful to sample several of the suspended accounts and view their tweets to determine if they were participating in information operations, and if so what was the message and who was the target audience.  For example, the tweets of suspended account @Scopatumanigga are provided in Table \ref{tab:Scopatumanigga}.  Our first observation is that nearly all statuses are retweets, which is highly indicative of bot behavior.  Given the account screen name and content, it appears that this likely bot account is attempting to infiltrate and influence African American virtual communities on Twitter.  The likely intent is to amplify racial divides in order to create instability in the United States.  

\begin{table}[htbp]
  \centering
  \caption{Tweets from suspended account @Scopatumanigga}
  \resizebox{\linewidth}{!}{%
    \begin{tabular}{l}
    \hline
    RT @tonyhawk: I’ve been sick lately (not “sick AF” just sick) with symptoms other than COVID-19. But I know two friends in the U.S. with co… \\
    RT @mvrlyns: so after the Coronavirus blows over, will y’all continue to practice good hygiene and sanitation? ... or will y’all go back to… \\
    RT @KenichiAL: Joe Biden: The tests for the coronavirus should be free Bernie Sanders: The vaccines and treatment for the coronavirus shou… \\
    RT @workerism: Haven't been able to stop thinking about this. A US pharma company with a potential COVID-19 vaccine is in court trying to P… \\
    RT @CrypticNoOne: 123movies would never do this \\
    RT @Carnage45\_\_: Black athletes give back during every crisis. I'm not saying it doesn't happen but I don't be seeing Tom Brady, Mike Trout… \\
    RT @shanalala\_: The elite getting tested without any symptoms and commoners with all the symptoms are denied tests. \\
    RT @elliecampbbell: I know it’s necessary to stop the spread of covid-19, but self isolation, no school etc and everything being shut/cance… \\
    RT @EricHaywood: There are 16 people in this photograph \\
    RT @eugenegu: @realDonaldTrump There it is. I’ve been deathly afraid of this exact moment where Trump turns to racism and xenophobia and ca… \\
   RT @baeonda: I’m 22 years old and I tested positive for COVID-19. I’ve been debating on posting, but I want to share my experience especi… \\
   RT @SocialistWitch: Coronavirus is not mother nature's 'cure' for 'evil humans'. The Earth doesn't suffer from humanity. It suffers from… \\
   RT @RaeOfLite: Hi. Yes, it originated in China, but the technical term is Covid-19 Your mom originated from the back of a Buick Skylark… \\
    @swevenpjm @ashleytwo @lydiakahill How's this covid19 treating you \\
    Bernie Sanders on his way to the hospital after he sees this tweet URL \\
    RT @ChapatiPapa: So they cashed out. Hoarded supplies. Moved money to companies they believe could make the vaccine, and testing kits all t… \\
    RT @Claireific: Hey remember that time I said that “tell me about Tuskegee” should be a required interview question for medical applicants… \\
    RT @Nigensei: Empty hotels all over the city of Las Vegas and they’re putting the homeless in a fucking parking lot. \\
    RT @DecaturDane: SECOND WAVE??? URL \\
    RT @hoodcuIture: I swear Christians and colonizers never stop!! How an ISOLATED GROUP get infected \\
    RT @BreeNewsome: whyyyyyyyyyyy do we accept such a lower quality of life in this country in exchange for nothing but slogans \&amp; confetti \\
    RT @\_iamtiredLord: this baby was shot 20 times in the head by a grown ass man because he wanted her girlfriend who rejected him. THEY ARE 1… \\
    RT @thespinsterymc: The family members of the Johnsons have made it clear that medical neglect killed them. Stop romanticizing this antibla… \\
    RT @LackingSaint: got to be honest, it's starting to feel like we're just doing things we feel like doing and saying it's in support of hea… \\ \hline
    \end{tabular}}%
  \label{tab:Scopatumanigga}%
\end{table}%

\section{Bot Detection}

Bot detection is a critical step for social cybersecurity workflows.  As discussed in Chapter 7, bot detection often helps to delineate an information warfare campaign, as well as illuminate lines of effort (topics) and target audiences.  It also sheds light on the scale, level of sophistication, and at times attribution for the operation.  

In this section we will discuss how to appropriately deploy bot detection algorithms for social cybersecurity.  We will start by estimating the accuracy of our algorithms on the given data stream as well as selecting an appropriate threshold for the data at hand (in our case the COVID-19 stream).  We will discuss where each of the Bot-Hunter tiers should be used in the workflow.  Having run Bot-Hunter on all 27 million accounts, we will use it to find influential bots, lines of effort, target audiences, and foreign influence.

\subsection{Determine appropriate thresholds}

Before using any bot detection tool on a given event or topic oriented data stream, the analysts should verify it's accuracy on the stream as well as determine an appropriate threshold.  As we discussed in Chapter 7, training data matters for bot detection.  We need to verify that the model that we are trying to use, and its respective training data, are appropriate and predictive for our event stream.  In our case that means verifying that the bot detection model works on our COVID-19 stream.  To make this evaluation, we need a small labeled dataset from our event stream.  For the COVID-19 data, we created a list of all unique user ID's that were found in the data, and then randomly sampled 200 accounts from this list.  We then manually labeled the accounts using a custom workflow that we've developed.  After manually labeling the 200 accounts, we evaluated the proposed bot-detection algorithms on the data.

With labels and bot-detection scores in hand, we can measure performance with various metrics (accuracy, precision, recall, F1 score, ROC-AUC, etc).  These scores will tell us how well our models are generalizing to COVID-19 data.  The scores are shown in Table \ref{tab:virus_model_results} for default settings of $threshold = 0.5$ for all models.  From Table \ref{tab:virus_model_results} we first and foremost determine that Bot-Hunter Tier 1 should be our primary bot detection model, with a higher F1 score and good balance of precision and recall.  From Table \ref{tab:virus_model_results} we can also determine that the Botometer model as well as the Bot-Hunter Tier 2 model don't seem to generalize as well to the COVID-19 Twitter stream.  The Bot-Hunter Tier 0 model, which appearing to perform well, will only be used on specific tasks because it is only able to predict English speaking accounts and because it tends to have a higher false positive rate (in this test the false-positive rate is three times larger than the Bot-Hunter Tier 1).  

\begin{table}[htbp]
  \centering
  \caption{Model Performance on the COVID-19 Twitter Stream}
    \begin{tabular*}{\textwidth}{l@{\extracolsep{\fill}}cccccc}
          & Threshold & F1 & Accuracy & Precision & Recall & ROC AUC \\ \hline
    Bot-Hunter Tier 0 & 0.5 & 0.849 & 0.750 & 0.812 & 0.890 & 0.556 \\
    Bot-Hunter Tier 1 & 0.5 & 0.823 & 0.736 & 0.938 & 0.734 & 0.742 \\
    Bot-Hunter Tier 2 & 0.5 & 0.559 & 0.486 & 0.986 & 0.390 & 0.681 \\
    Botometer & 0.5 & 0.112  & 0.197 & 0.733 & 0.060 & 0.475 \\ \hline
    \end{tabular*}%
  \label{tab:virus_model_results}%
\end{table}%

\begin{figure}[htbp]
\centering
\begin{subfigure}{0.49\linewidth}
  \centering
  \includegraphics[width=\linewidth]{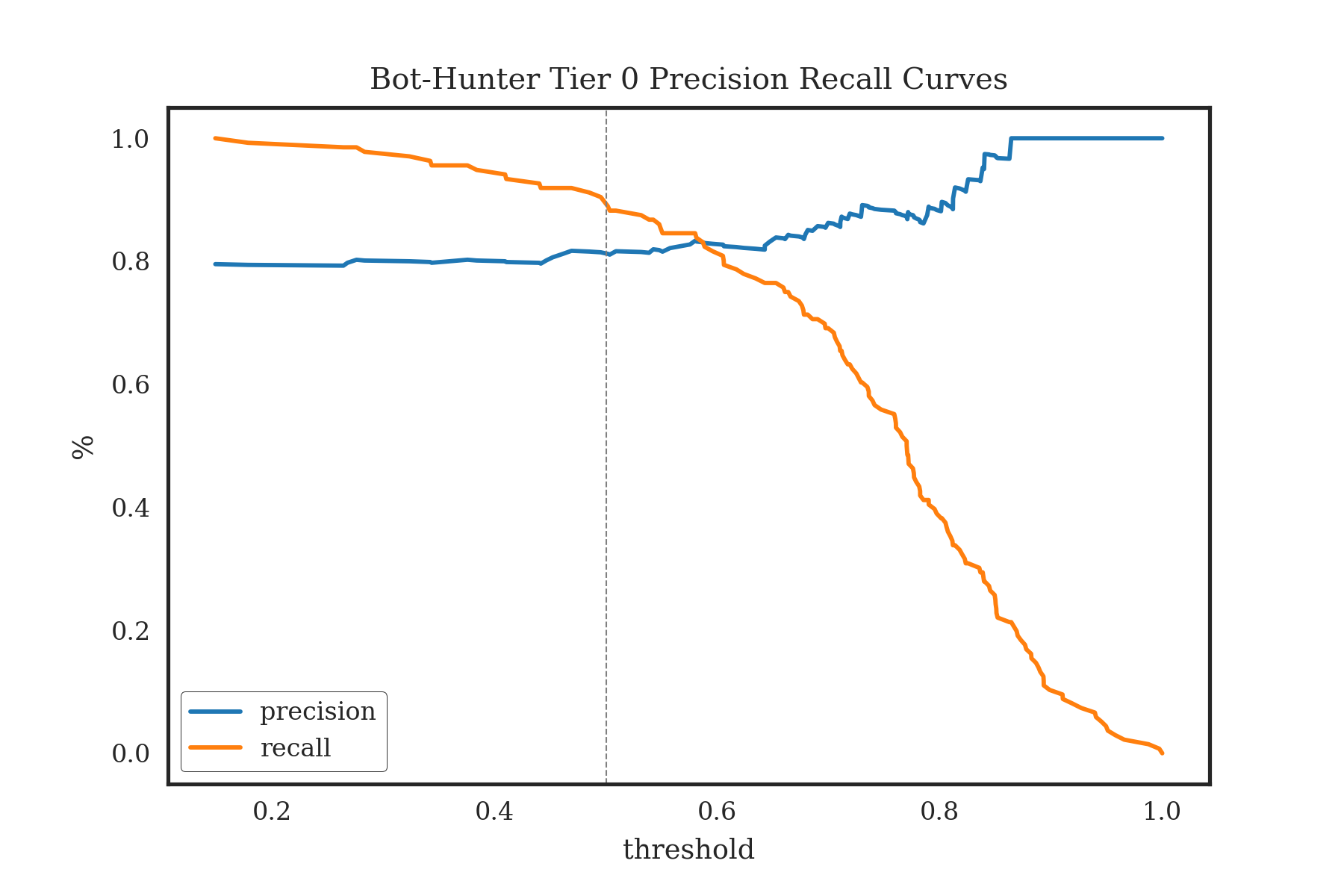}
  \caption{Tier 0 Precision-Recall Curves}
  \label{fig:P-R_tier0}
\end{subfigure}%
\begin{subfigure}{0.49\linewidth}
  \centering
\includegraphics[width=\linewidth]{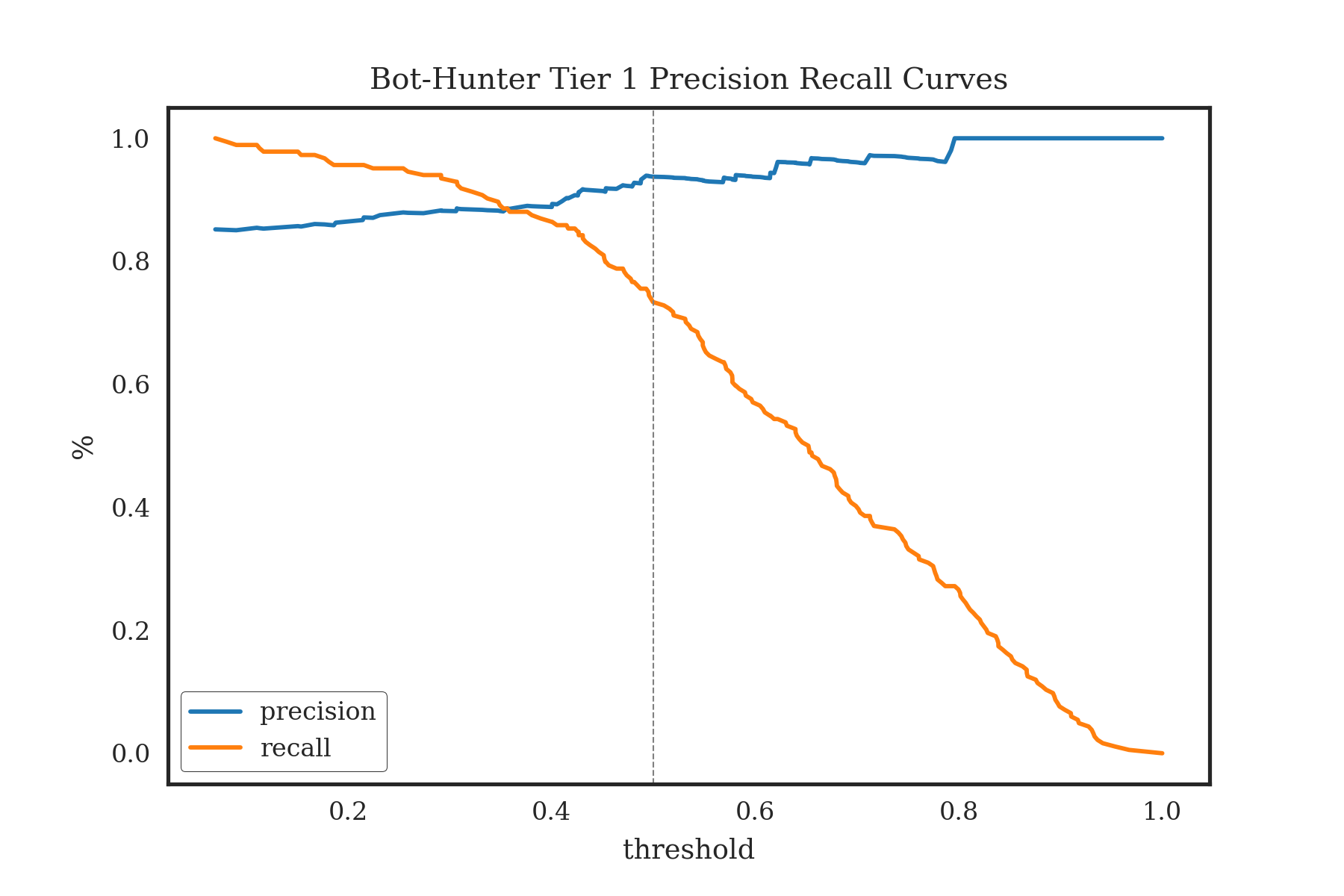}
  \caption{Tier 1 Precision-Recall Curves}
  \label{fig:P-R_tier1}
\end{subfigure}
\begin{subfigure}{0.49\linewidth}
  \centering
  \includegraphics[width=\linewidth, ]{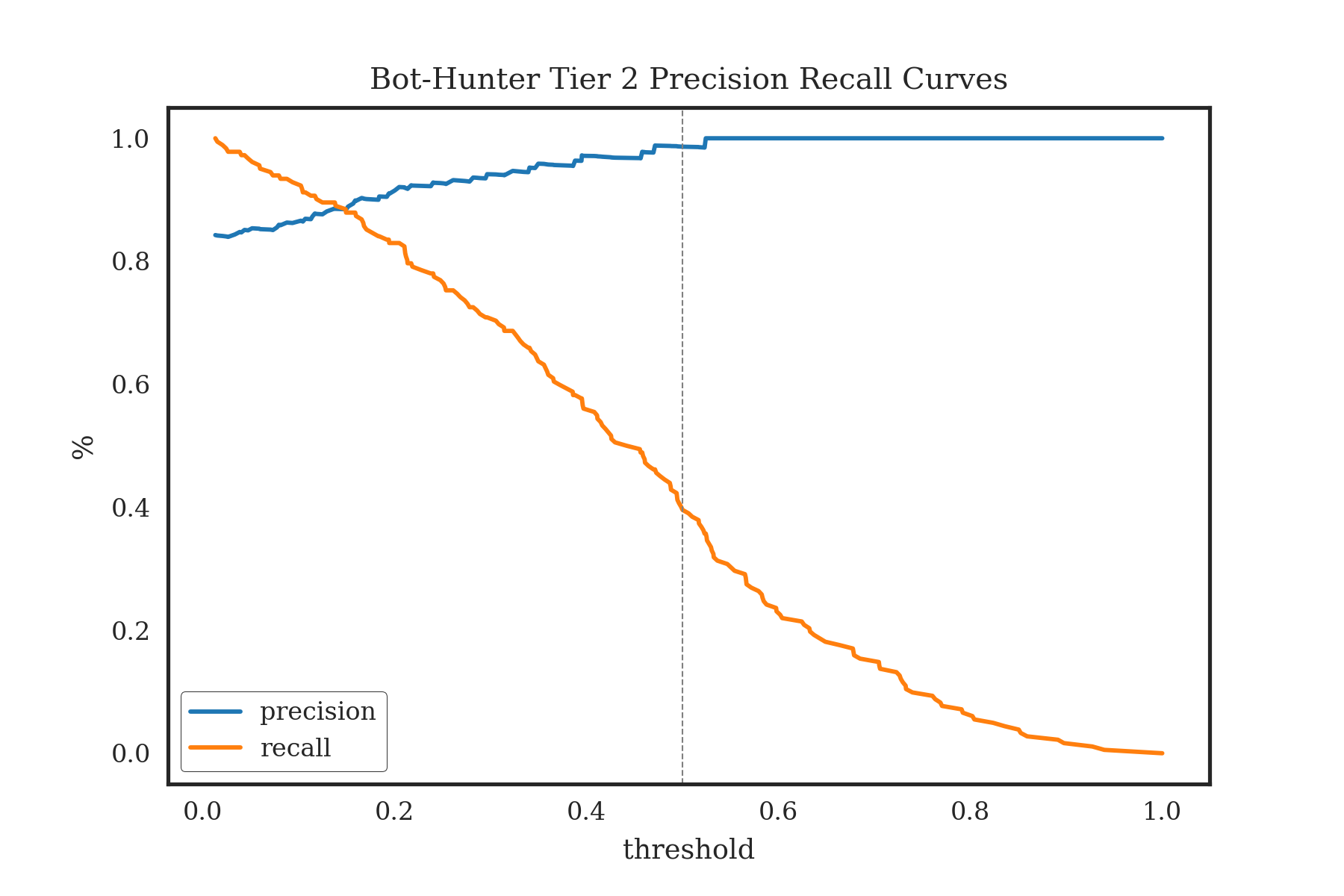}
  \caption{Tier 2 Precision-Recall Curves}
  \label{fig:P-R_tier2}
\end{subfigure}
\begin{subfigure}{0.49\linewidth}
  \centering
  \includegraphics[width=\linewidth]{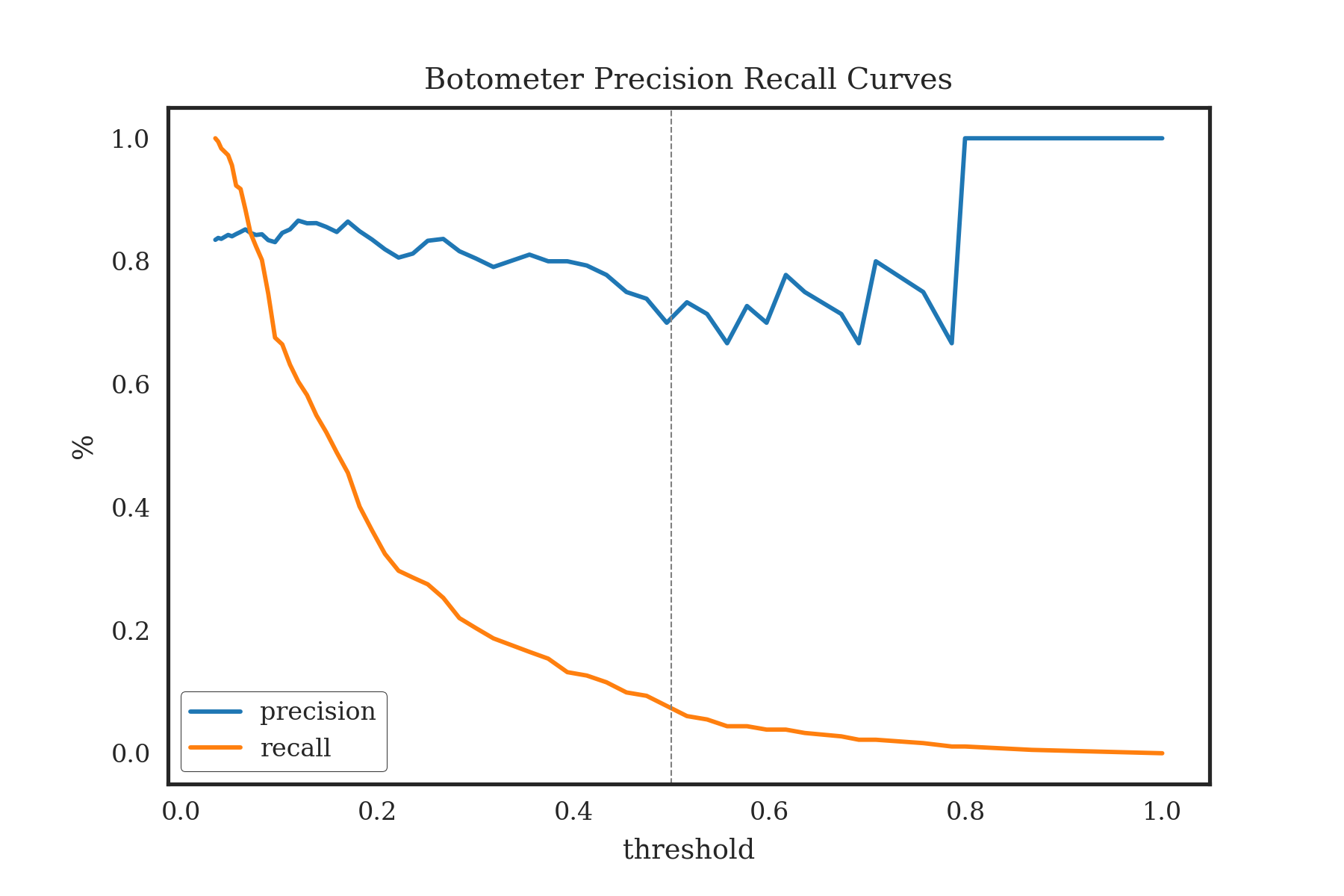}
  \caption{Botometer \cite{ferrara2016rise} Precision-Recall Curves}
  \label{fig:P-R_botometer}
\end{subfigure}
\caption{Precision Recall Curves on 200 Manually Labeled Accounts Randomly Sampled from COVID-19 Stream}
\label{fig:P_R Curves}
\end{figure}

These scores, however, are sensitive to the threshold that we choose.  In order to choose an appropriate threshold, we use our labels and bot detection scores and plot precision-recall curves as shown in Figure \ref{fig:P_R Curves}.  Remember that \textit{precision} and \textit{recall} often have an inverse relationship.  As \textit{precision} increases, \textit{recall} decreases, and vice versa. \textit{Recall} monotonically decreases, whereas \textit{precision} does not monotonically increase. The exact choice of the threshold will depend on the context and any related policy decisions.  If the policy decision requires a low false-positive rate, then choose a threshold with high \textit{precision}.  If the policy decision or analytical goal requires a low false-negative rate, then choose a threshold with higher \textit{recall}.  For most tasks it is best to have a balance of \textit{precision} and \textit{recall}, which is why we often use F1 score to measure the performance of bot detection algorithms.

In choosing a bot detection threshold, our goal for the COVID-19 stream is to characterize the entire conversation.  We want to have robust characterization of the entire forest, not necessarily precise analysis of individual trees. This goal requires a good balance between precision and recall.  As indicated above, our Bot-Hunter Tier 0 Text model tends to produce a high false positive rate.  For this reason, we will use a threshold of 0.7 for this model in our COVID-19 Stream, which cuts false positive rate by a third.  For Bot-Hunter Tier 1 we will retain the 0.5 default threshold, since it provides a good balance with precision and recall.  For Botometer and Bot-Hunter Tier 2, we will use a threshold of 0.3 in order to increase recall.  The adjusted performance is provided in Table 

\begin{table}[htbp]
  \centering
  \caption{Model Performance on the COVID-19 Twitter Stream with Adjusted Thresholds}
    \begin{tabular*}{\textwidth}{l@{\extracolsep{\fill}}cccccc}
          & Threshold & F1 & Accuracy & Precision & Recall & ROC AUC \\ \hline
    Bot-Hunter Tier 0 & 0.7 & 0.762 & 0.663 & 0.861 & 0.684 & 0.634 \\
    Bot-Hunter Tier 1 & 0.5 & 0.823 & 0.736 & 0.938 & 0.734 & 0.742 \\
    Bot-Hunter Tier 2 & 0.3 & 0.805 & 0.716 & 0.941 & 0.703 & 0.741 \\
    Botometer & 0.3 & 0.325  & 0.294 & 0.804 & 0.203 & 0.477 \\ \hline 
    \end{tabular*}%
  \label{tab:virus_model_results_adjusted}%
\end{table}%

\begin{figure}[htbp]
\centering
\begin{subfigure}{0.49\linewidth}
  \centering
  \includegraphics[width=\linewidth]{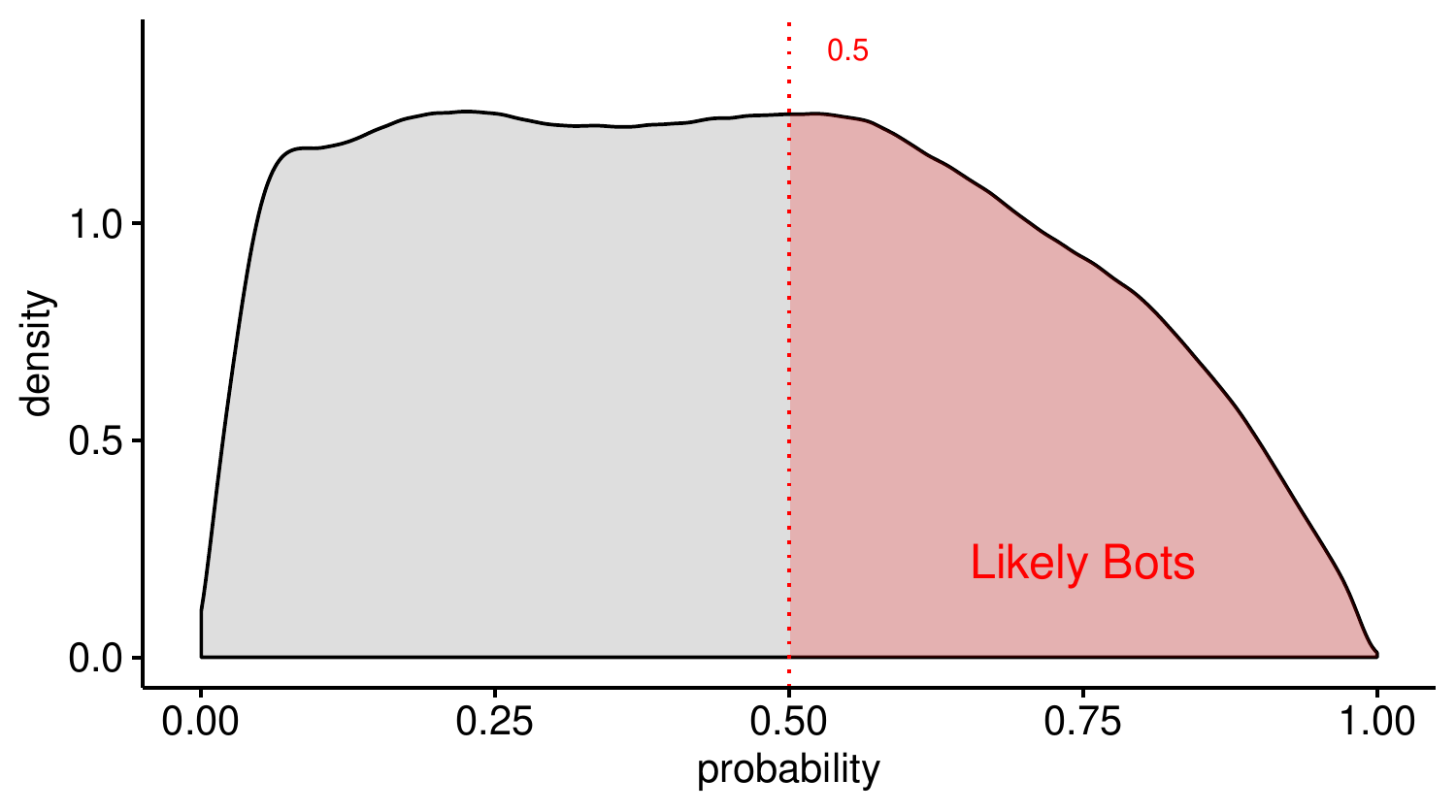}
  \caption{$threshold = 0.50$}
  \label{fig:bot_density_50}
\end{subfigure}%
\begin{subfigure}{0.49\linewidth}
  \centering
  \includegraphics[width=\linewidth]{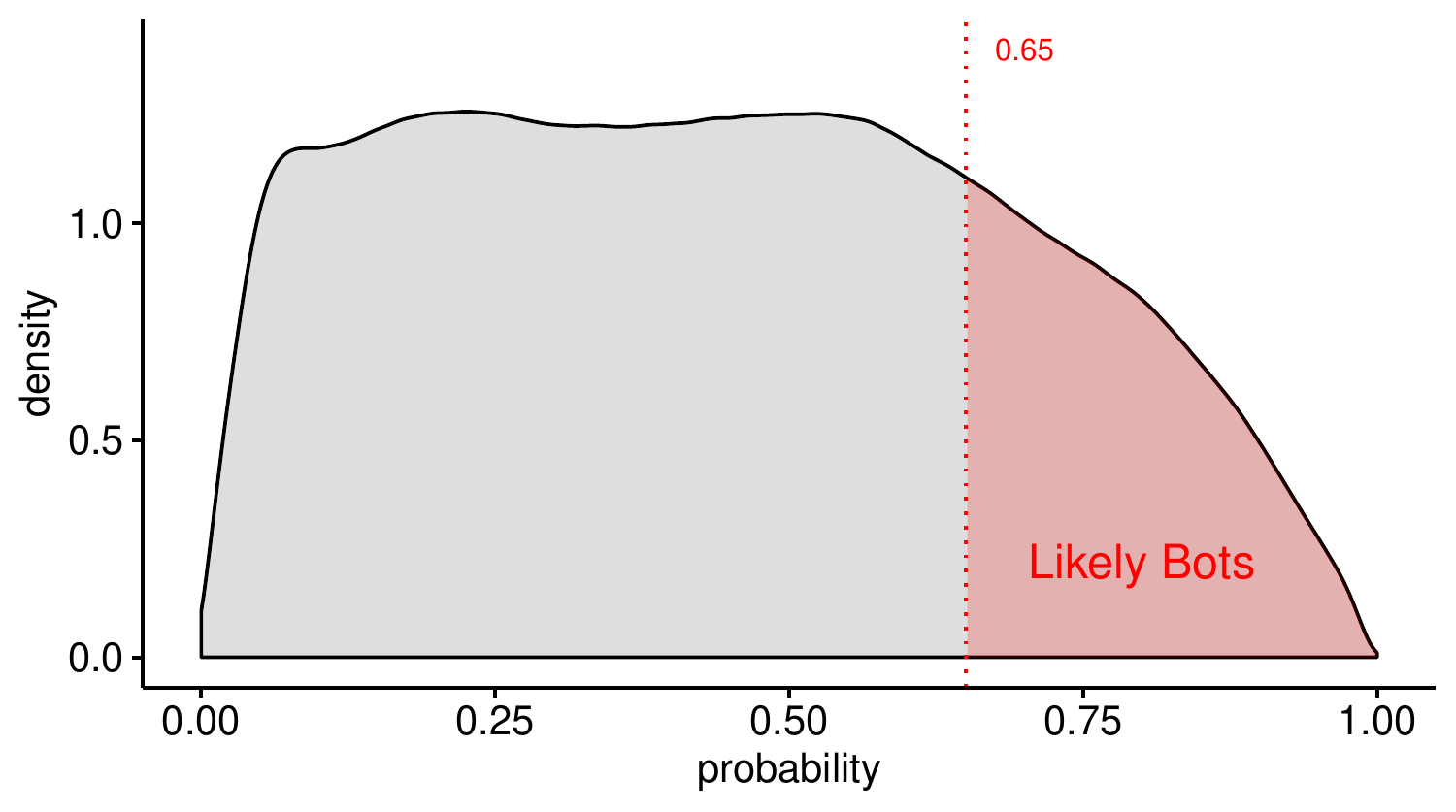}
  \caption{$threshold = 0.65$}
  \label{fig:bot_density_65}
\end{subfigure}
\caption{Bot-Hunter Tier 1 Probability Density }
\label{fig:bot_density}
\end{figure}

Since the Bot-Hunter Tier 1 algorithm is our primary algorithm, we've visualized the probability distribution for all COVID-19  Accounts in Figure \ref{fig:bot_density_50} with $threshold = 0.5$ and $threshold = 0.65$.  Here we see a large number of human accounts, as well as a large number of accounts that are in the middle, with decreasing numbers of high probability bots.  The $threshold = 0.5$ is our default model, while we can at times use $threshold = 0.65$ for increased precision.  

Note that for each actor they will have a botscore for each bot detection tool.  These algorithm identified bots can be more accurately described as actors with bot-like characteristics. Depending on which tool is used and which threshold is used the number of ``bots'' that are identified will vary.  For example, on any given day depending on what is used, the number of ``bots'' may vary from approximately 25\% to 48\%.

\subsection{Bots in the Network}

\begin{figure}[htbp]
\centering
\begin{subfigure}{\linewidth}
  \centering
  \includegraphics[width=\linewidth]{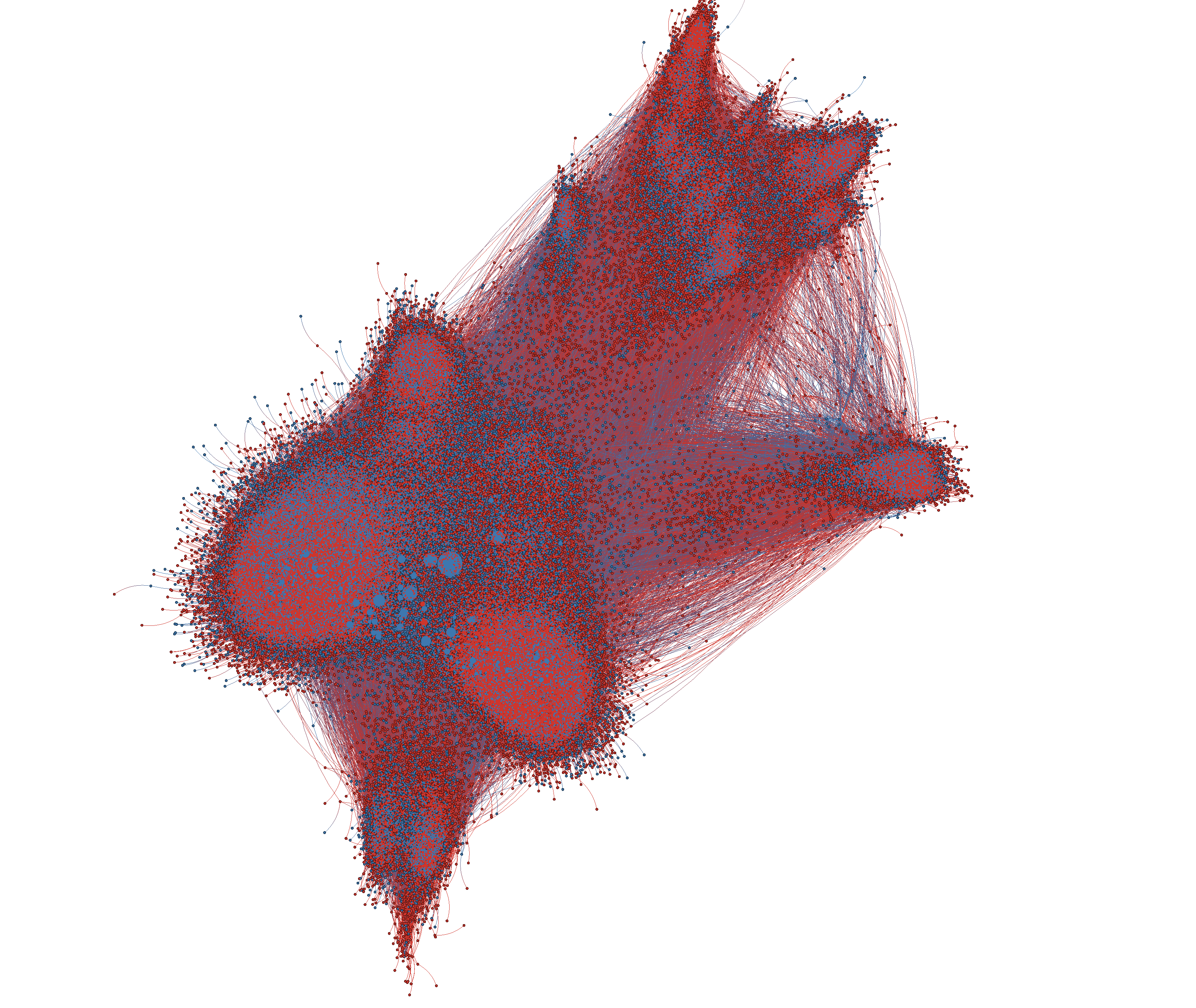}
  \caption{Core of Mention Network}
  \label{fig:mention_network_bots8}
\end{subfigure}%

\begin{subfigure}{\linewidth}
  \centering
\includegraphics[width=\linewidth]{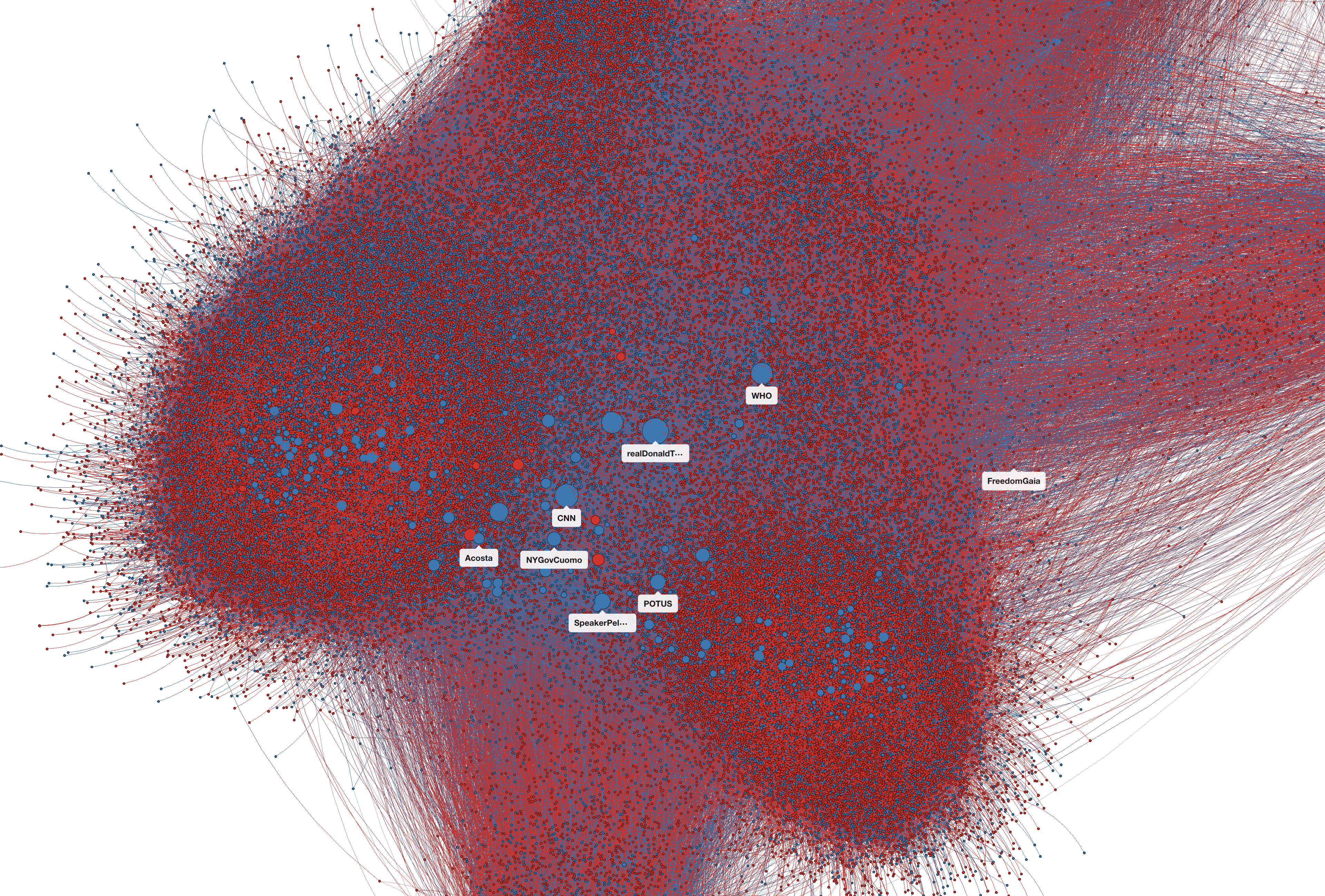}
  \caption{Zoom in of influential accounts}
  \label{fig:mention_network_bots6}
\end{subfigure}
\caption{Visualize a k-Core ($k=100$) of the Mention Network (\emph{red} indicates accounts with bot-like behavior)}
\label{fig:mention_network_bots}
\end{figure}

Now that we have bot prediction thresholds, we can begin to use the models to understand the COVID-19 stream and the conversation and actors involved in it.  In Figure \ref{fig:mention_network_bots} we visualize the sparsified core of the mention network ($k_{core} = 100$), colored by bot prediction with Bot-Hunter Tier 1 and $threshold = 0.5$.  We see that bot-like accounts are highly embedded in the core of the conversation, and connect to and mention influential accounts in an effort to manipulate these personalities and their followers.  In this visualization, we have also zoomed in to get a better feeling of the structure of the network and so that we can highlight the location of prominent English speaking accounts.  

\subsection{Bots by Account Creation}

The most surprising bot detection analysis that we found was in regards to account creation date.  Anytime we analyze a list of accounts, it is often enlightening to visualize a histogram of their account creation date.  The Twitter JSON contains a field in the user object that records the date, hour, minute, and second that the account was originally created.  This date will be some date between March 21, 2006 (the day Twitter started) and the current date.  We've found it best to bin this by day.  Each bar in the resulting histogram will contain all accounts that were created on that day.  Any large spike of accounts created around the same time should cause us to dig deeper to check for the presence of a bot ``army''.

\begin{figure}[htb]
\centering
  \includegraphics[width=\linewidth]{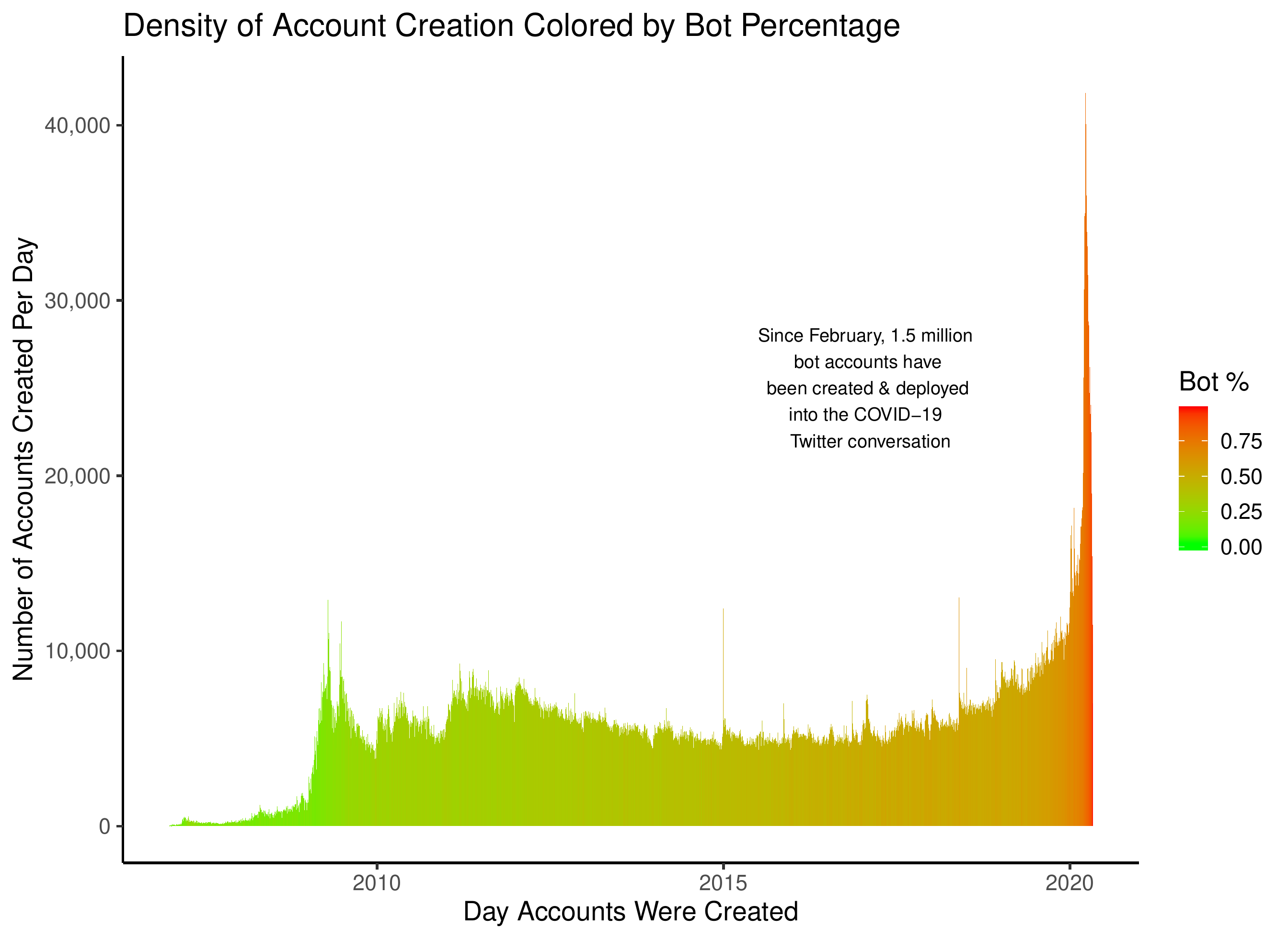}
  \caption{Density of Account Creation Date Colored by Bot Proportion}
  \label{fig:account_creation_timeline}
\end{figure}

In Figure \ref{fig:account_creation_timeline} we visualize the account density plot for COVID-19 colored by bot percentage.  The coloring indicates what portion of accounts in that bar have a Bot-Hunter Tier 1 score greater than 0.5.  Green indicates bars that have few bots, while red indicates bars that have higher proportion of bots.  

From Figure \ref{fig:account_creation_timeline} we see that a large number of bot-like accounts have been created since the pandemic began and then immediately deployed into the conversation.  In fact, of the 27 million accounts participating in the conversation, 1.5 million have been created since 1 February AND have a Bot-Hunter Tier 1 score greater than 0.5.  Undoubtedly part of this surge in accounts is created by individuals who are stuck at home and decided to create a Twitter account.  A significant portion, however, appears to be bot armies.  These bot armies are produced by a number of actors with a variety of agendas, but likely all involve manipulation of the marketplace of beliefs, ideas, and collective action.

\subsection{Intersect Influence with Bot Scores}

Next we want to intersect our measure of influence (eigenvector centrality) with a bot prediction score in order to identify influential bots.  In Table \ref{tab:influtential_bots} we list the top 50 most influential accounts in the \textit{mention} network and \textit{retweet} network as measured by eigenvector centrality.  Table \ref{tab:influtential_bots} also provides the Bot-Hunter Tier 1 bot score, with \emph{red} text indicating accounts that have a score greater than 0.5.  Looking at the macro-level comparison, we see more bots are involved with the \textit{retweet} network than the \textit{mention} network.  This makes sense given that bots are often used to scale amplification, and the easiest way to amplify is with retweets.  We also see many more verified politician, news, and government accounts in the mention network.  Many news accounts have ``bot-like'' behavior, with @FoxNews, @MSNBC, and @CBSNews accounts surpassing the 0.5 threshold.  It has been documented that many news and celebrities can have bot-like behavior \cite{gilani2017classification}, which is supported in our analysis.  We also see in the retweet network that several accounts that are not classified as bots nonetheless have bot probability that approaches the 0.5 threshold, and are still likely bots (examples include @SJPFISH  and @NoodleSparklez). We also note that it is possible for users to employ software or a bot on occasion from the same account.  This hybrid form that is human and bot we refer to as a cyborg.  Cyborgs will also tend to exhibit bot like characteristics; however, they are likely to be lower in their values than a totally automated account.



\subsection{Topic Modeling (Lines of Effort)}

\begin{figure}[htb]
\centering
\begin{subfigure}{0.33\linewidth}
  \centering
  \includegraphics[width=\linewidth]{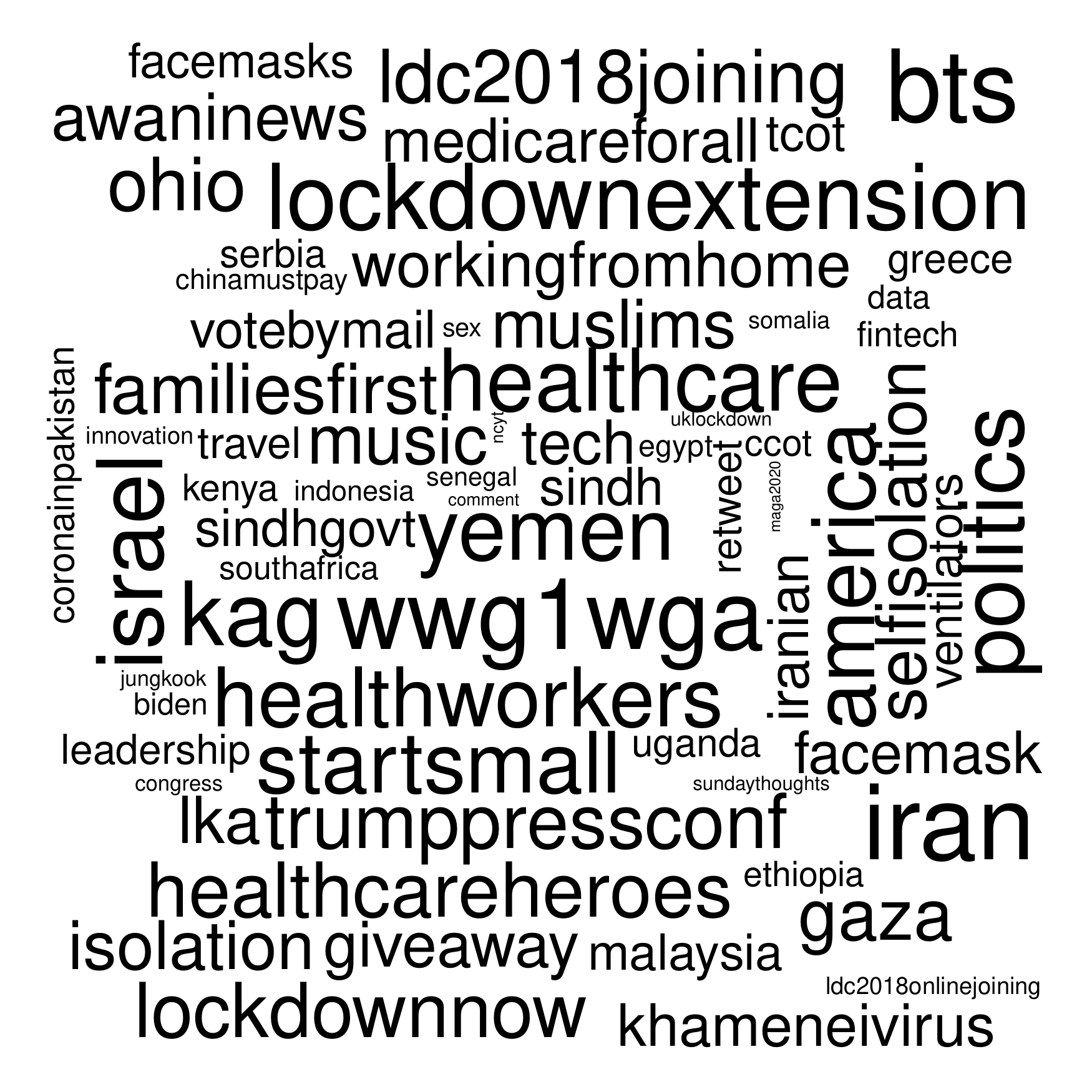}
  \caption{Topic One (40.8 \% bot)}
  \label{fig:lda_cloud1}
\end{subfigure}%
\begin{subfigure}{0.33\linewidth}
  \centering
  \includegraphics[width=\linewidth]{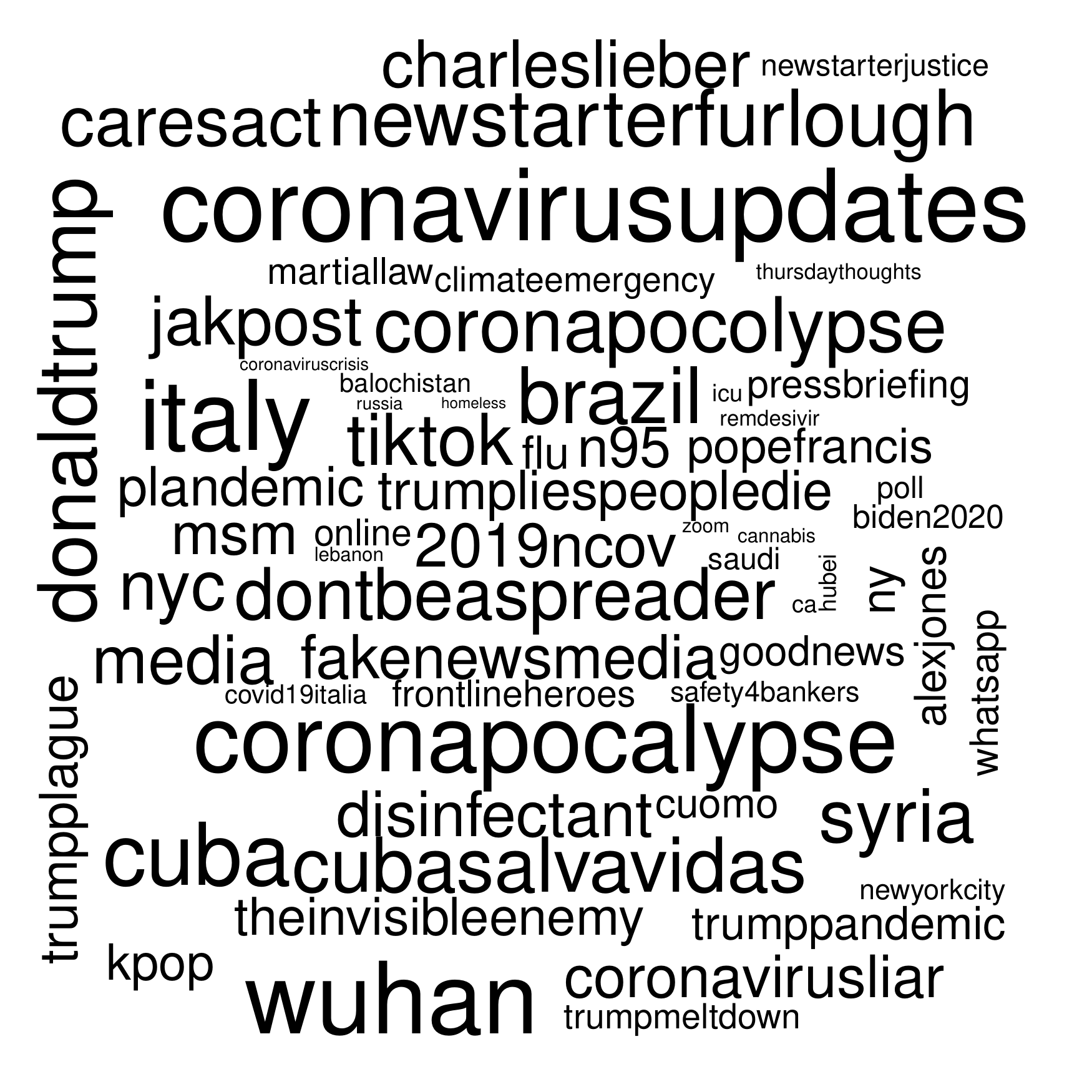}
  \caption{Topic Two (43.2\% bot)}
  \label{fig:lda_cloud2}
\end{subfigure}
\begin{subfigure}{0.33\linewidth}
  \centering
  \includegraphics[width=\linewidth, ]{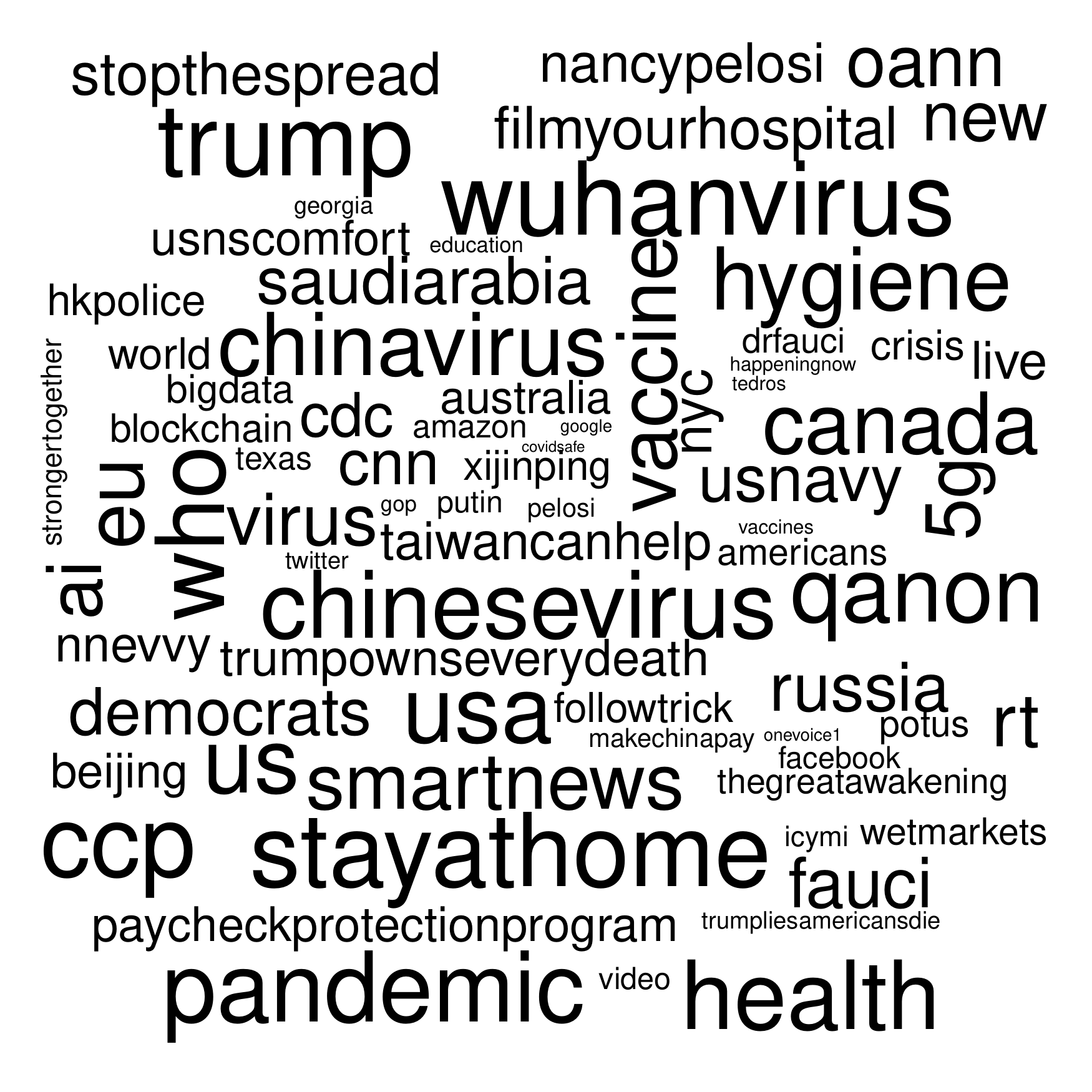}
  \caption{Topic Three (47.9\% bot)}
  \label{fig:lda_cloud3}
\end{subfigure}
\begin{subfigure}{0.33\linewidth}
  \centering
  \includegraphics[width=\linewidth]{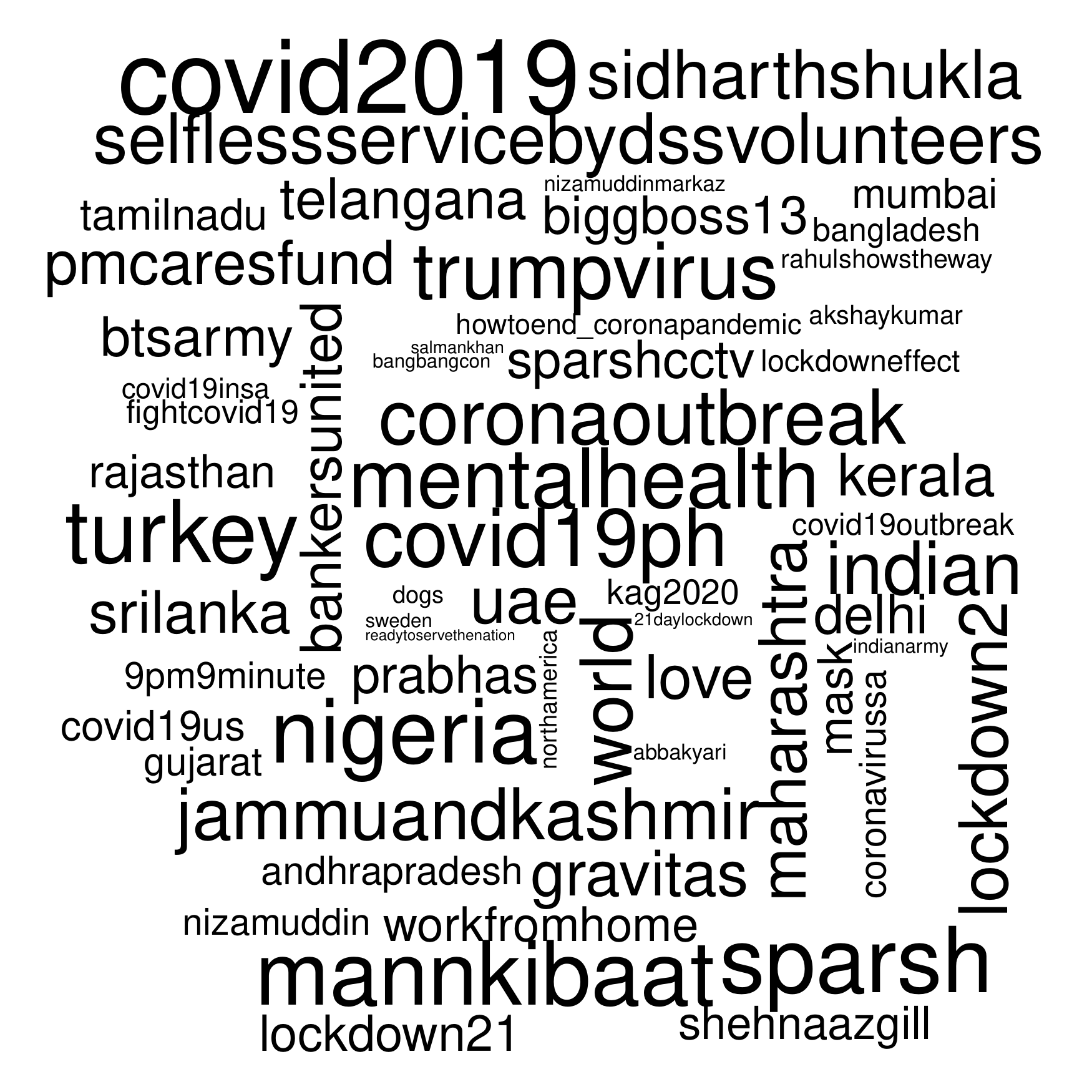}
  \caption{Topic Four (51.0\% bot)}
  \label{fig:lda_cloud4}
\end{subfigure}
\begin{subfigure}{0.33\linewidth}
  \centering
  \includegraphics[width=\linewidth]{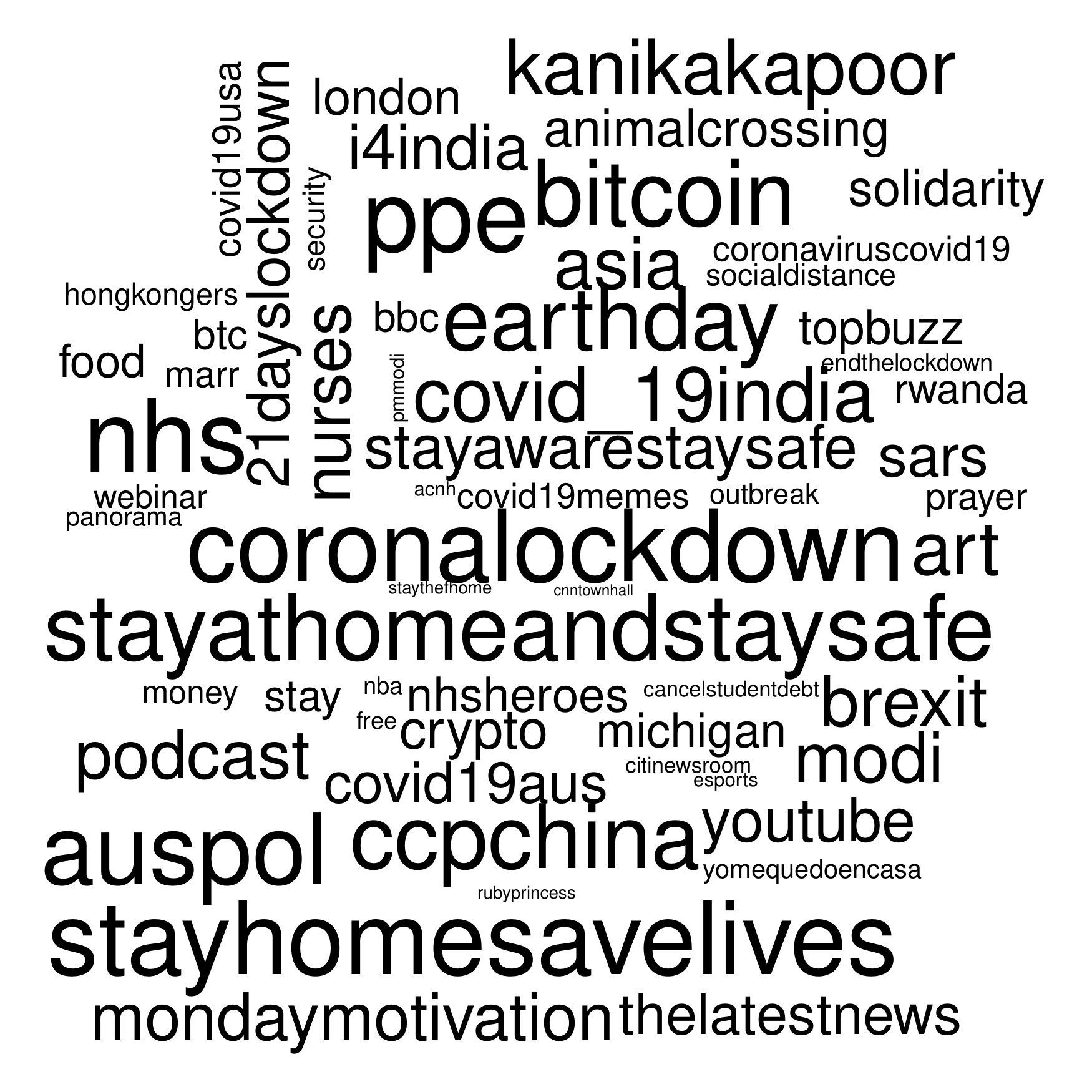}
  \caption{Topic Five (35.0\% bot)}
  \label{fig:lda_cloud5}
\end{subfigure}
\begin{subfigure}{0.32\linewidth}
  \centering
  \includegraphics[width=\linewidth]{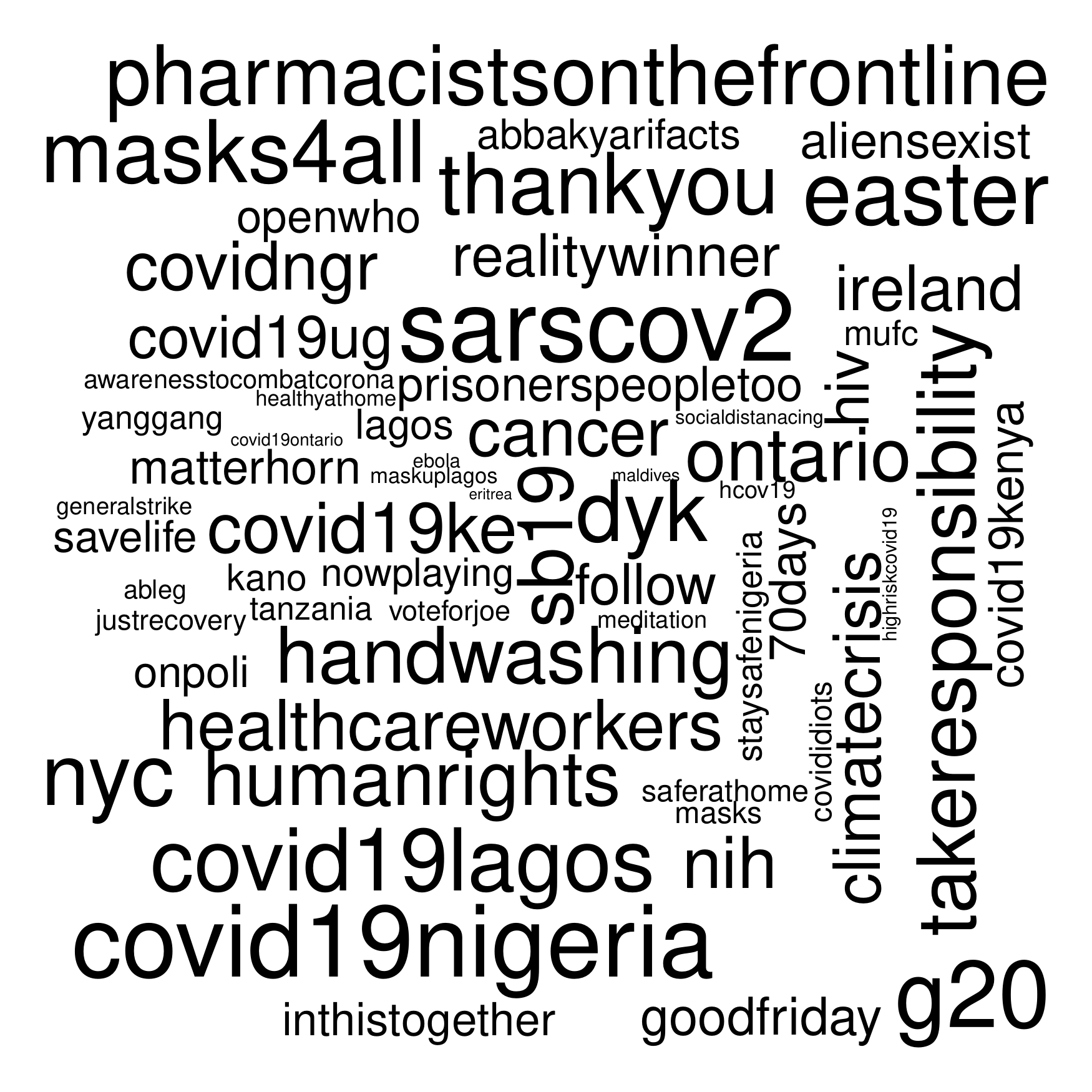}
  \caption{Topic Six (38.1\% bot)}
  \label{fig:lda_cloud6}
\end{subfigure}
\caption{Wordclouds of LDA Topic Groups (corpus consists of hashtags concatenated by user)}
\label{fig:lda_cloud}
\end{figure}

Next we want to try to separate the stream into topic groups.  We will do this with Latent Dirichlet Allocation (LDA) model \cite{blei2003latent}.  To do this, we concatenated all English hashtags by account and then performed LDA with $k=5$.  We chose to concatenate hashtags rather than use raw text for computational tractability and because hashtags provide tokens that capture the essence of topic and meaning. Word Clouds of the resulting five topic groups are shown in Figure \ref{fig:lda_cloud}.   We see that Topic Groups are differentiated in some ways by geography and in other ways by politics.  This topic groups allow us to segment the conversation and focus on a topic of interest, for example a certain geography (the Nigerian conversation) or a specific political affiliation (the conservative political conversation). 

The choice of $k$ is just as much an art as a science.  If you want to get a view of the macro topics, use a smaller $k$, like we did here with $k=5$.  If you want to extract a very specific conversation (i.e. the liberal conversation in Canada), you will have to increase $k$ in order to sufficiently isolate the topic of interest.  

\subsection{Community Detection (Target Audiences)}

Similar to topic analysis, we also want to look at community groups.  While topic analysis looked at semantic topics that seem to cluster together, community groups look at accounts that tend to cluster together regardless of topic.  We ran louvaine community detection \cite{blondel2008fast} on both the retweet and mention network, and then looked at influential accounts and top hashtags for each of the top 10 communities.  A summarization of the community groups in the retweet network are summarized in Table \ref{tab:louvaine_community}.  Here we see once again that some community groups are geographically and/or linguistically oriented, while others are politically affiliated.  

\begin{table}[htbp]
  \centering
  \caption{Top 10 Louvaine Community Groups}
   \resizebox{0.75\linewidth}{!}{%
    \begin{tabular}{m{1.5cm}m{3cm}m{4cm}m{2.5cm}}
  Size & Influential Accounts & Top Hashtags & Description \\ \hline
          3,700,533 \newline 36.7\% Bots & NCDCgov \newline ThatBrightDude\newline DrOlufunmilayo\newline akortainment\newline Mawunya\_ & covid19nigeria\newline nigeria\newline wuhan\newline coronavirusinsa\newline africa & African Discussion \\ \hline
          2,868,851 \newline 27.6\% Bots & JoeBiden\newline kylegriffin1\newline ryanstruyk\newline nytimes Yamiche & trump\newline china\newline maga\newline hydroxychloroquine\newline trumpvirus & Left Leaning \newline American \newline Conversation \\ \hline
              2,055,555 49.2\% Bots & DiazCanelB\newline BrunoRguezP\newline PresidenciaCuba\newline AleLRoss198\newline GuerreroCuba & cuba\newline quedateencasa\newline eeuu\newline ecuador\newline venezuela & Cuba Conversation \\ \hline
              1,685,303 \newline  23.0\% Bots  & SkyNews\newline JamesMelville\newline carolecadwalla\newline NursingNotesUK\newline piersmorgan & nhs\newline stayhomesavelives\newline ppe\newline coronavirusuk\newline china\newline borisjohnson & British Conversation \\ \hline
              1,510,246 \newline  59.8\% Bots  & ABSCBNNews\newline rapplerdotcom\newline inquirerdotnet\newline cnnphilippines\newline ANCALERTS & ja(new coronavirus)\newline china\newline ch(new virus)\newline ja(new corona)\newline ch(declaration of emergency) & Asian Conversation \\ \hline
              1,456,655 \newline  52.0\% Bots  & narendramodi\newline rajbirrohilla\newline Shrish\_1987\newline deepaggarwal1\newline im\_sumit\_verma & indiafightscorona\newline india\newline china\newline tablighijamaat\newline pakistan\newline coronavirusindia & Indian Conversation \\ \hline
               1,330,143 \newline  43.1\% Bots  & ldpsincomplejos\newline Alvisepf\newline matthewbennett\newline Santi\_ABASCAL\newline hermanntertsch & últimahora\newline venezuela\newline quedateencasa\newline españa\newline yomequedoencasa\newline china & Spanish \newline Conversation \\ \hline
              1,041,880 \newline  46.0\% Bots  & Conflits\_FR\newline Drmartyufml\newline le\_Parisien\newline AiphanMarcel\newline Brevesdepresse & confinement\newline macron\newline masques\newline france\newline chloroquine\newline chine\newline italie & French conversation \\ \hline
              1,001,444 \newline  39.1\% Bots  & realDonaldTrump\newline WhiteHouse\newline RealJamesWoods\newline TomFitton\newline RealCandaceO\newline TrumpWarRoom &  wuhanvirus\newline qanon\newline fakenews\newline hydroxychloroquine\newline foxnews\newline  trump\newline ccpvirus & Right Leaning \newline American \newline Conversation \\ \hline
                872,213 \newline  65.4\% Bots  & RizmaWidiono\newline AriestaRiico\newline Zul\_\_88\newline dwiyanaDKM\newline jokowi & ko(corana19\newline coronaoutbreak\newline ko(let's stay home)\newline coronavirusitalia & Korean \newline Conversation \\ \hline
    \end{tabular}}%
  \label{tab:louvaine_community}%
\end{table}%

\section{Account Characterization}

Now that we've explored bots in the data, we will begin to look at other ways to characterize accounts.  These include analysis of biased or questionable sources, abusive languages, and use of national flags.

\subsection{Questionable Sources}

\begin{figure}[htbp]
\centering
\begin{subfigure}{\linewidth}
  \centering
  \includegraphics[width=\linewidth]{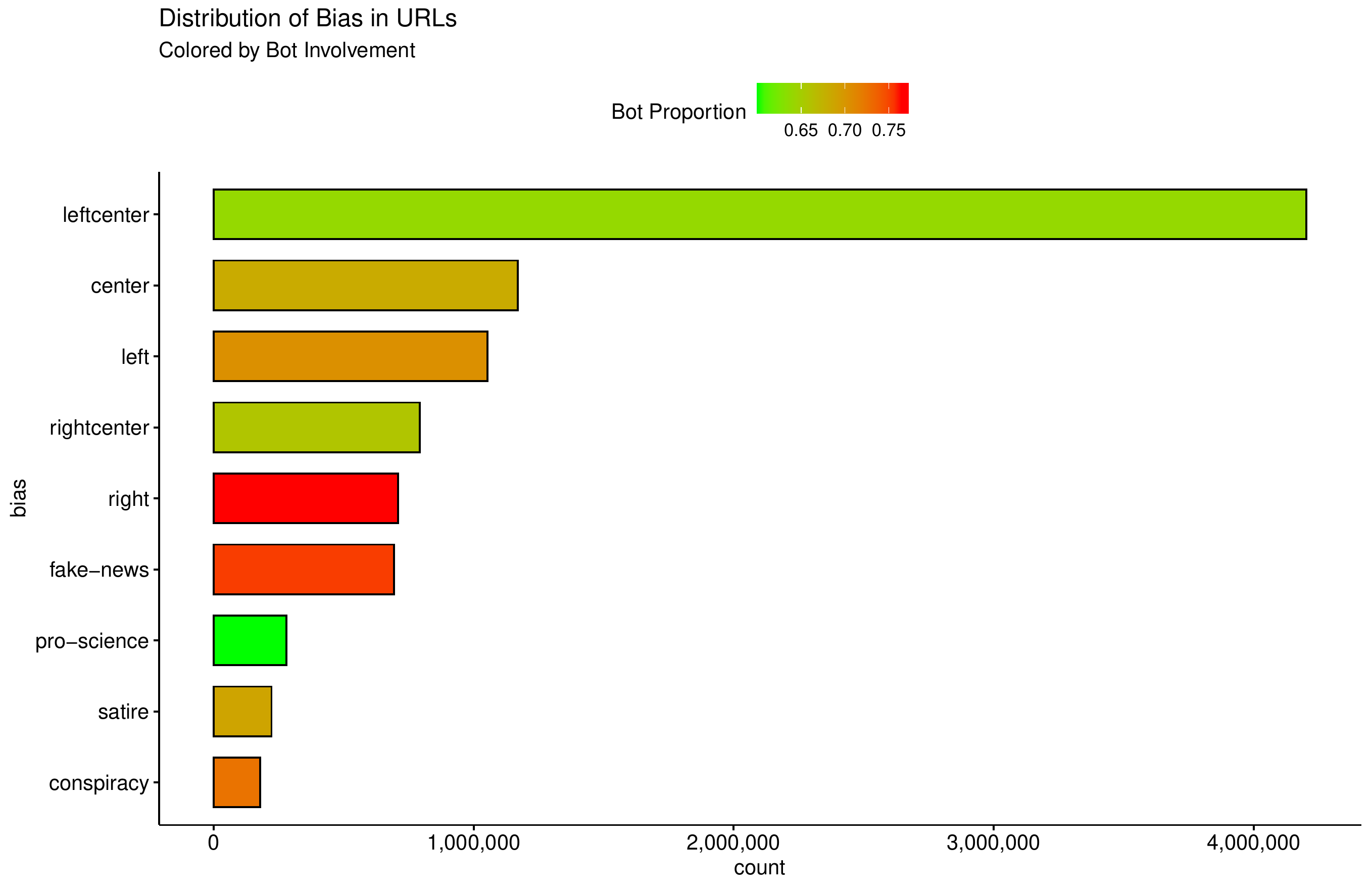}
  \caption{Distribution of Bias in URLs, colored by bot Bot-Hunter Tier 1 bot proportion)}
  \label{fig:slant_bot}
\end{subfigure}%
\vspace{0.5cm}
\begin{subfigure}{\linewidth}
  \centering
  \includegraphics[width=\linewidth]{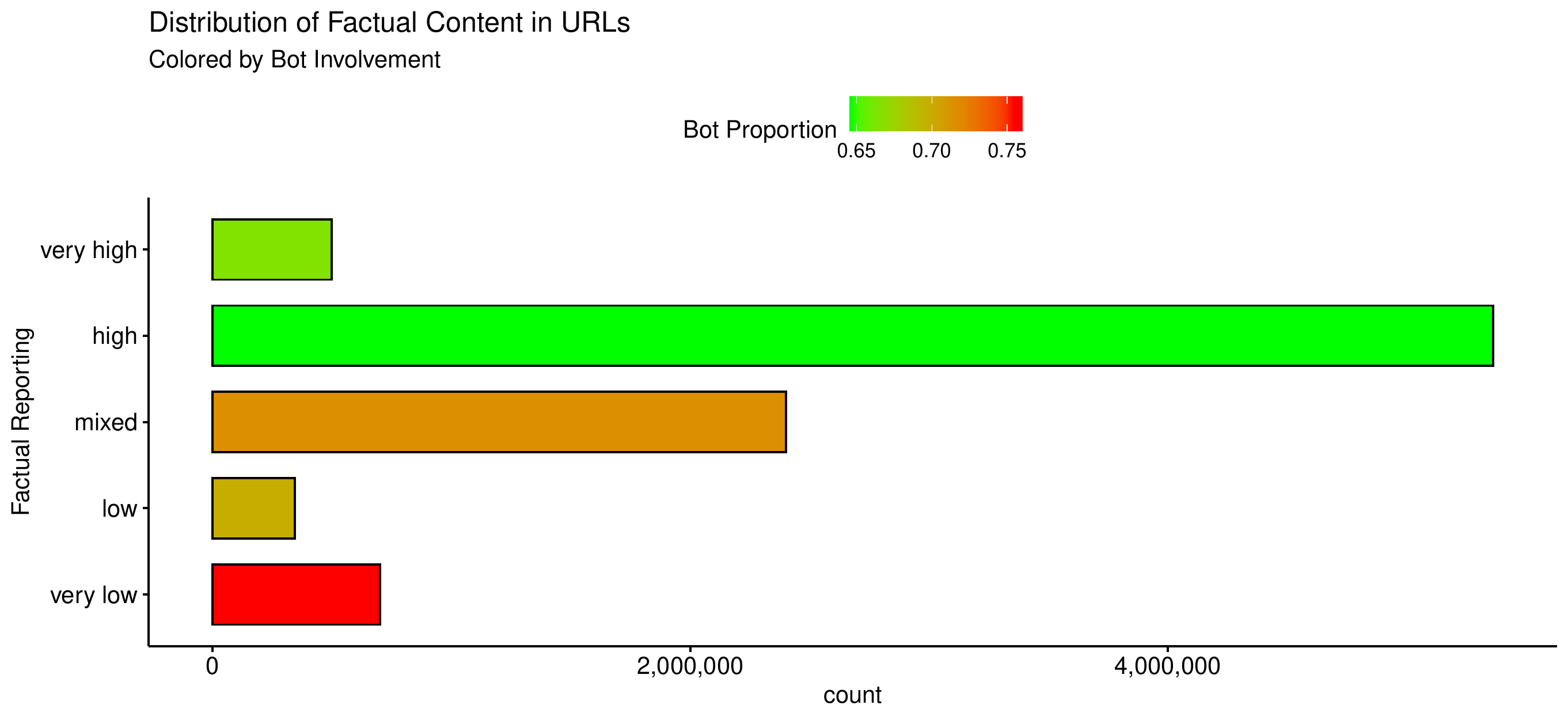}
  \caption{Distribution of Factual Content in URLs, colored by bot Bot-Hunter Tier 1 bot proportion)}
  \label{fig:slant_bot2}
\end{subfigure}
\caption{Analysis of Bias and Factual Content Found in URLs. }
\label{fig:slant}
\end{figure}

Next we will look at the political bias found in the URLs.  To do this, we will use the dictionary approach presented in Chapter 11.  Using this approach, 18.2\% of the total URL domains were found in the dictionary lookup that we built in Chapter 11.  Having estimated the bias and factual content in the URLs, we visualize this distribution in Figure \ref{fig:slant} where the bar plots are colored by the proportion of bot involvement.

From this we first see that the highest number of URLs are coming from the Center and Center Left political bias, and generally have high factual content.  We do see the presence of fake, satire, and conspiracy theory sources in this stream, which are often correlated with the low factual content seen in Figure \ref{fig:slant_bot2}.  We also see that bots have a higher degree of correlation with URLs from the far-right and fake-news biases, and to a lesser extent from the far-left.  We also see high bot correlation with URLs containing low factual content.  These discoveries largely confirm our assumptions going into the analysis.

\subsection{Abusive Language}

\begin{figure}[htbp]
\centering
  \includegraphics[width=\linewidth]{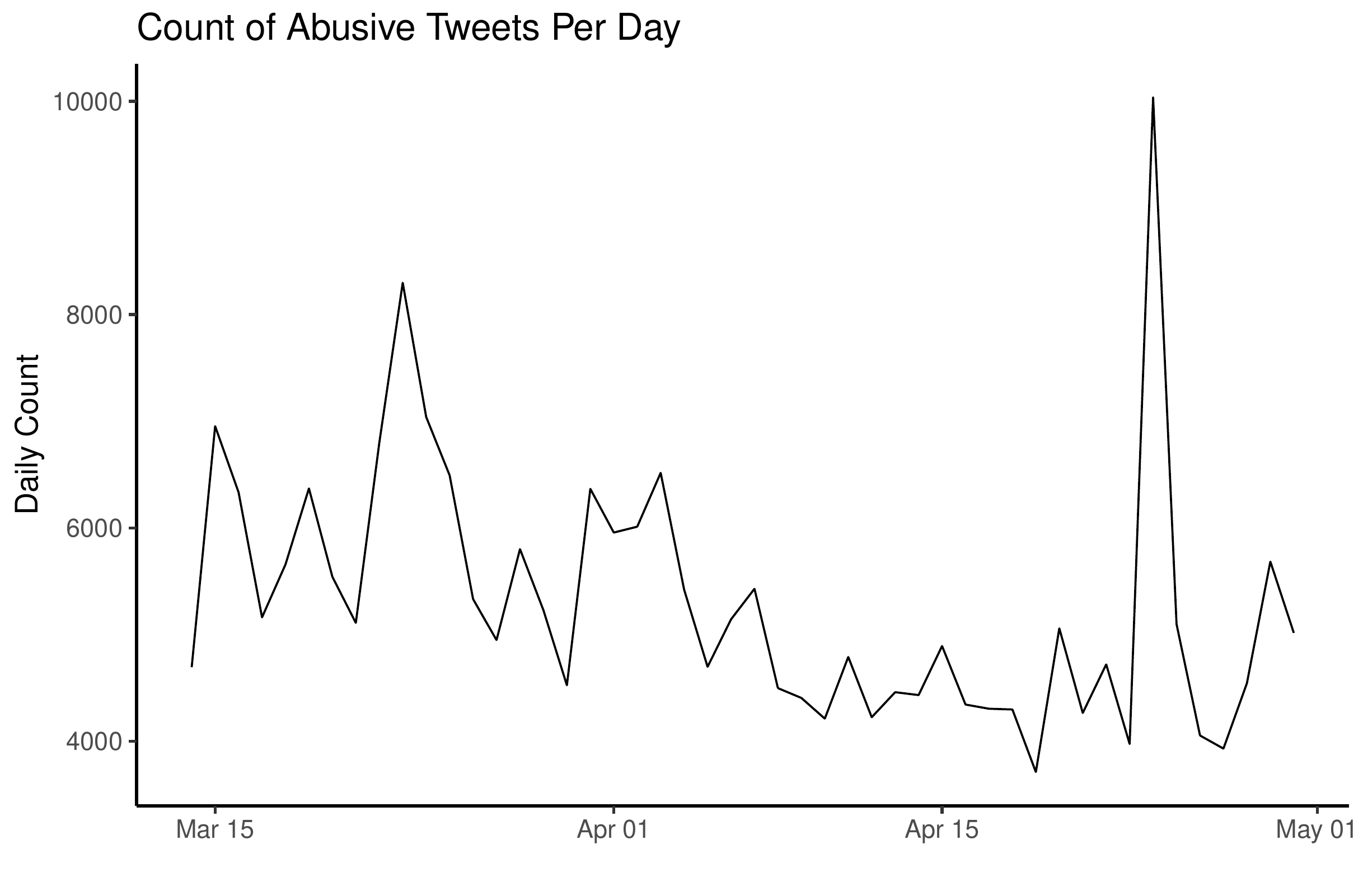}
  \caption{Daily count of abusive Tweets (Tweets that contain abusive terms) }
  \label{fig:abusive_viz}
\end{figure}

We used the multi-lingual dictionary based algorithm presented in Chapter 1111 to identify tweets that contain abusive language.  The daily volume of abusive is presented in Figure \ref{fig:abusive_viz}.  We did not normalize this visualization since our total daily count of tweets was held constant at 4.3 million tweets.  If this was not true, it would be appropriate to normalize this (plot proportion instead of raw count).  This allows us to identify events that seem to aggravate the population of active Twitter users.  The two prominent spikes are tied to political events and voices in the United States, with the first spike tied to actions by US Congress and the second spike tied to comments by the US Executive Branch.  

We found that 57.7\% of the accounts that share abusive content had bot like characteristics.  This is significantly higher than the 41.8\% of non-abusive accounts that have bot-like characteristics.  This means that within the COVID-19 stream bot-like accounts are used to produce or promote abusive content.

\subsection{High Flags}

\begin{figure}[htbp]
\centering
\begin{subfigure}{0.49\linewidth}
  \centering
  \includegraphics[width=\linewidth]{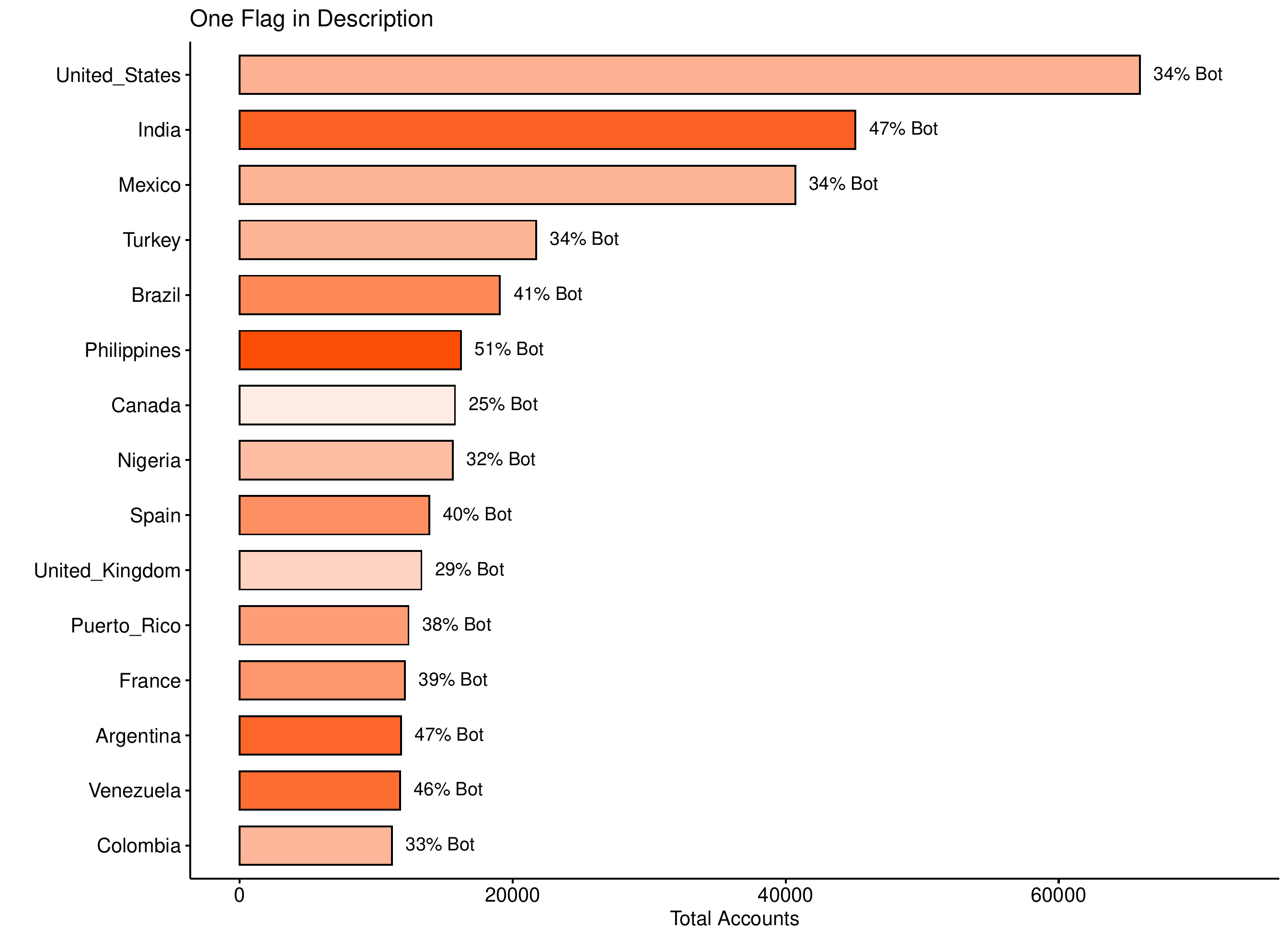}
  \caption{One Flag}
  \label{fig:flag1}
\end{subfigure}%
\begin{subfigure}{0.49\linewidth}
  \centering
  \raisebox{0.18in}{\includegraphics[width=\linewidth]{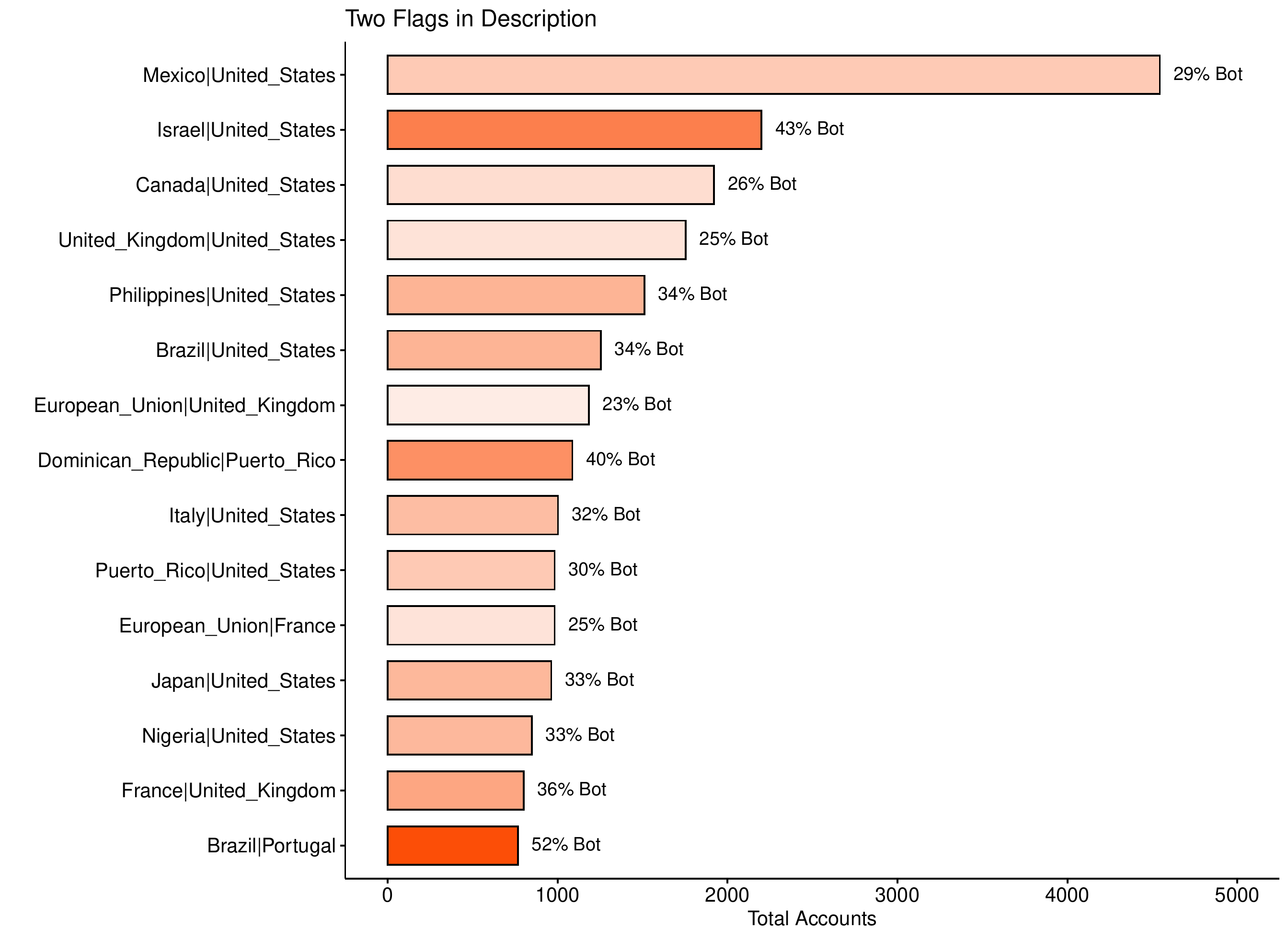}}
  \caption{Two Flags}
  \label{fig:flag2}
\end{subfigure}
\begin{subfigure}{0.49\linewidth}
  \centering
  \includegraphics[width=\linewidth, ]{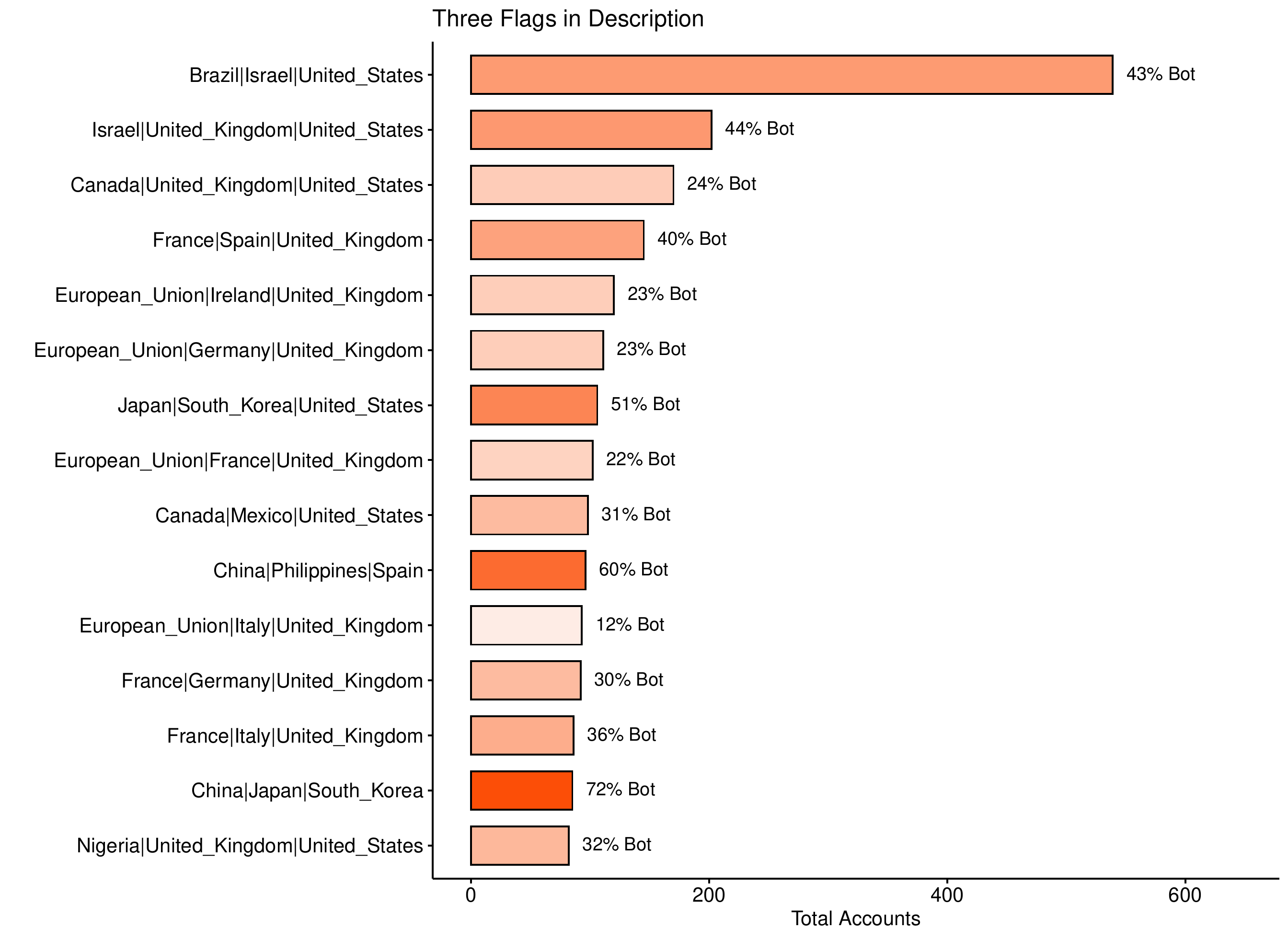}
  \caption{Three Flags}
  \label{fig:flag3}
\end{subfigure}
\begin{subfigure}{0.49\linewidth}
  \centering
  \includegraphics[width=\linewidth]{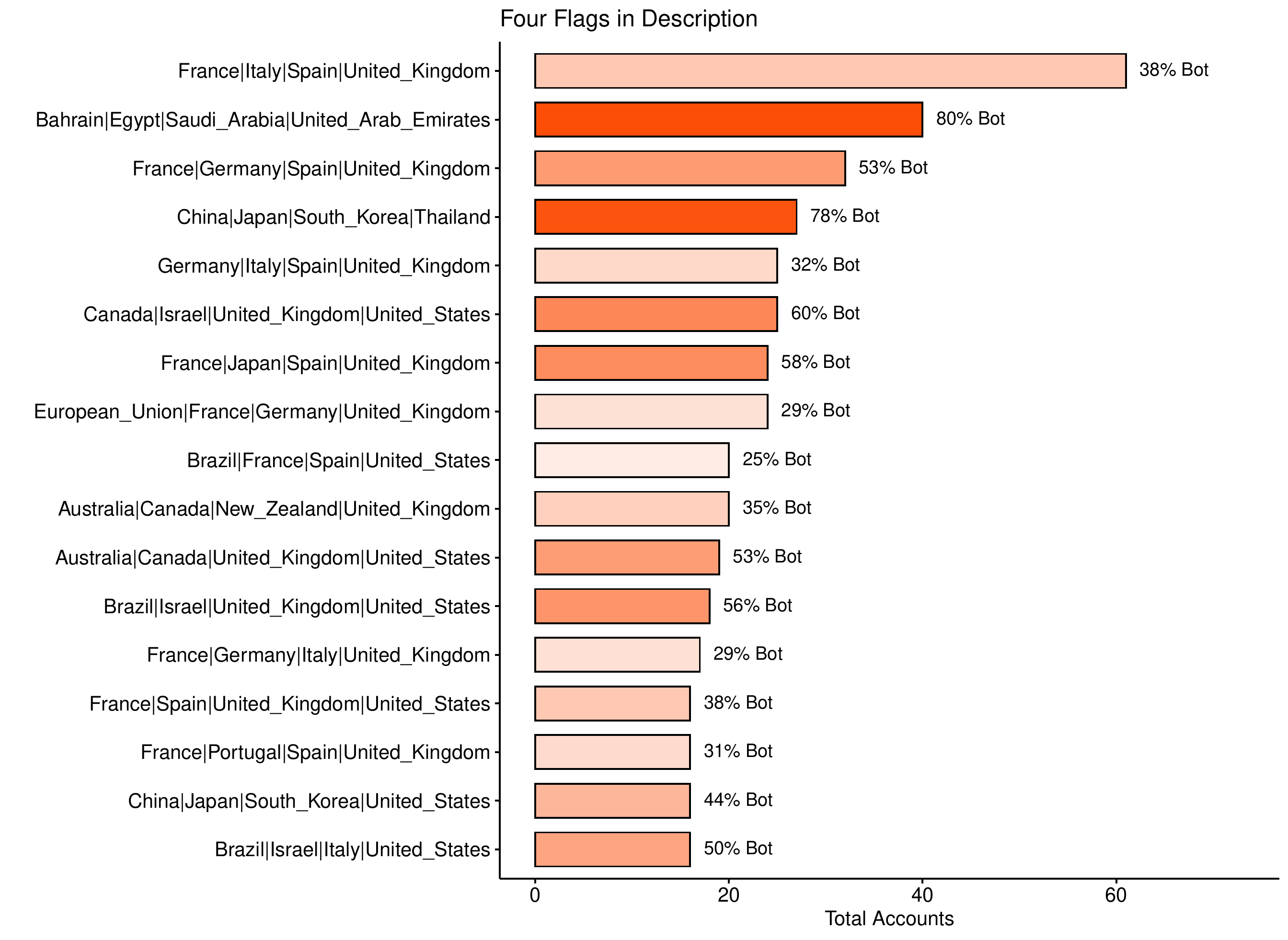}
  \caption{Four Flags}
  \label{fig:flag4}
\end{subfigure}
\begin{subfigure}{0.49\linewidth}
  \centering
  \includegraphics[width=\linewidth]{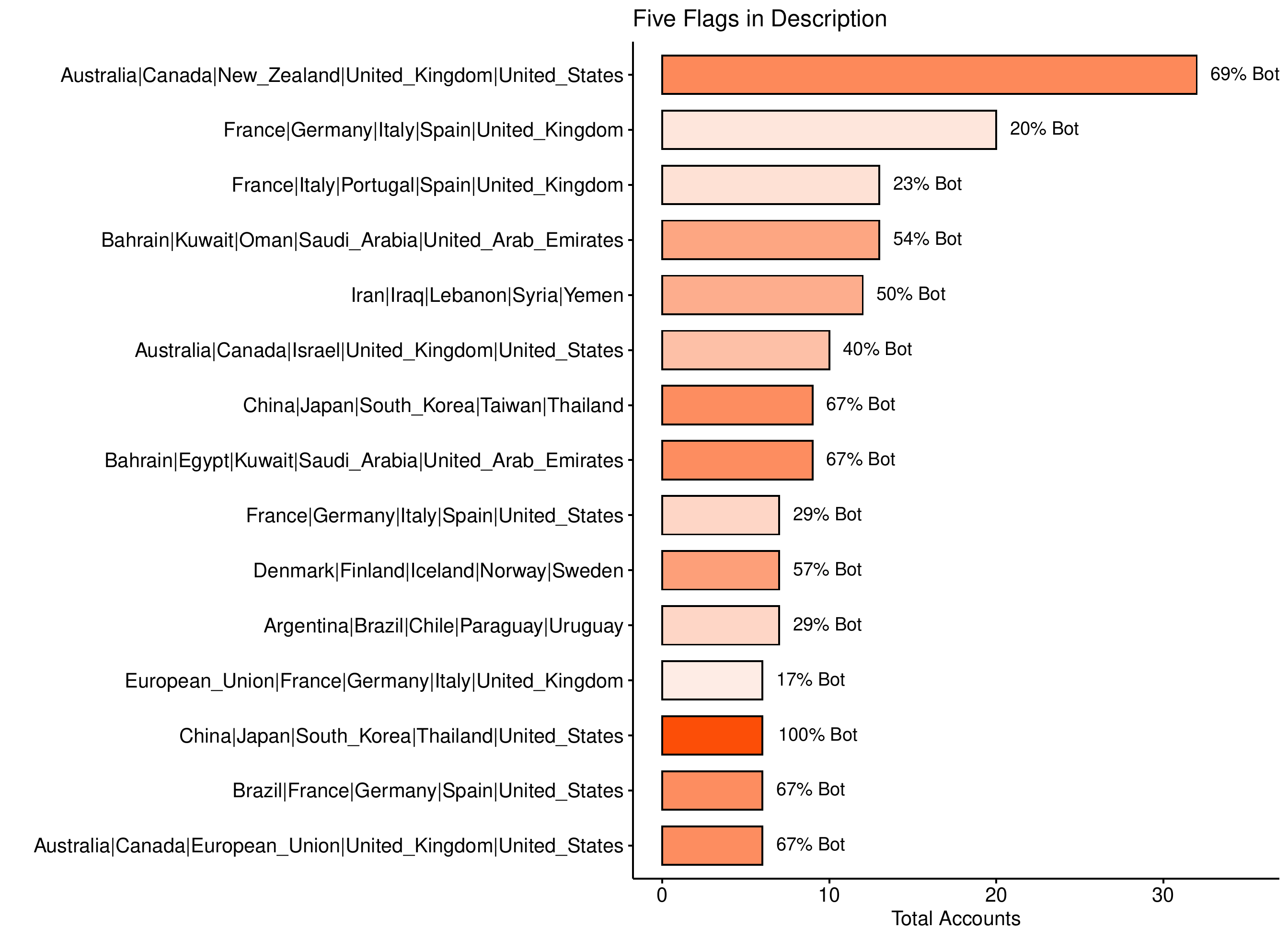}
  \caption{Five Flags}
  \label{fig:flag5}
\end{subfigure}
\begin{subfigure}{0.49\linewidth}
  \centering
  \includegraphics[width=\linewidth]{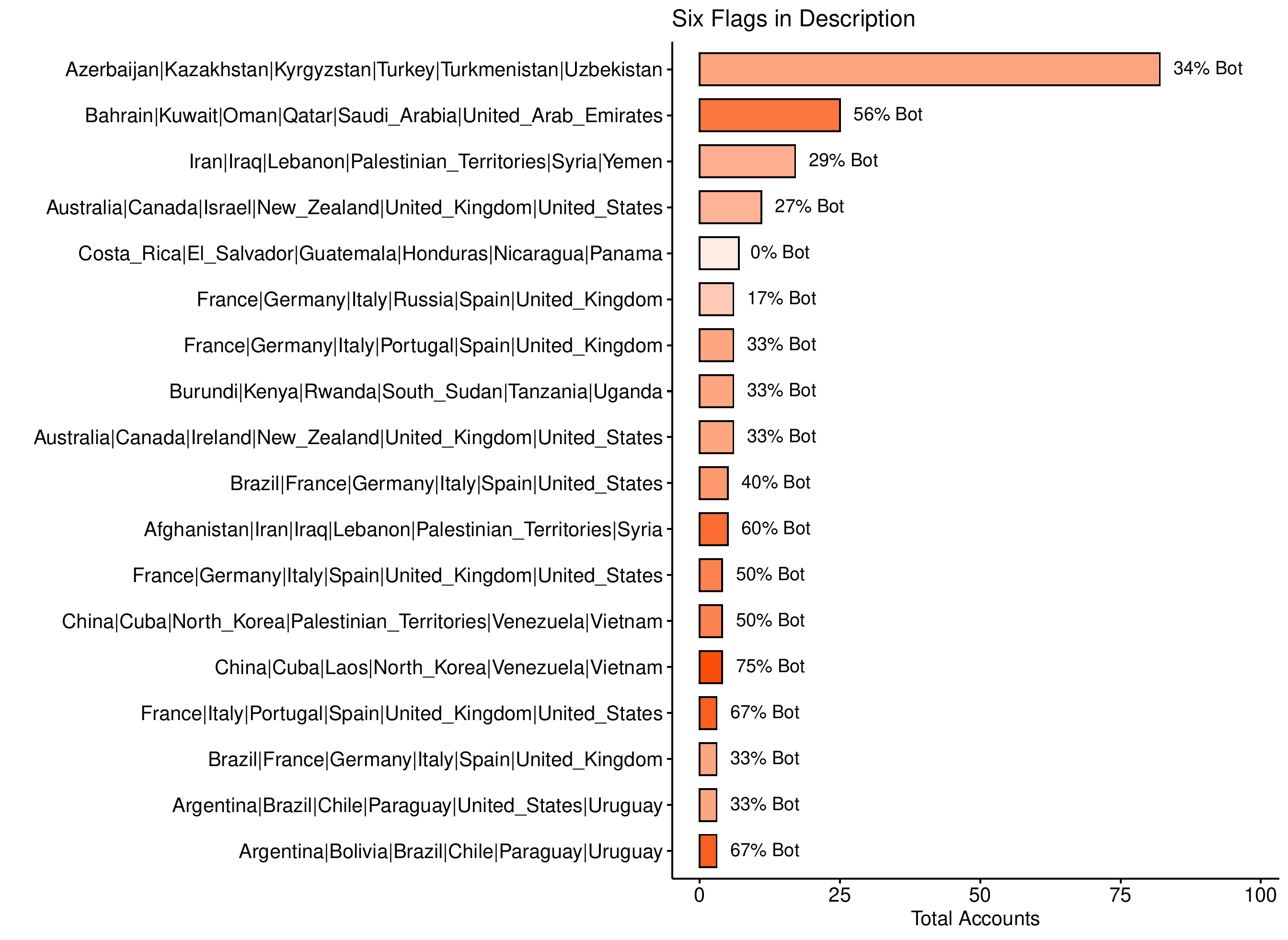}
  \caption{Six Flags}
  \label{fig:flag6}
\end{subfigure}
\caption{Distribution of flag combinations in user descriptions}
\label{fig:flag}
\end{figure}

As discussed in Chapter 11, at times flags in the user description can indicate suspicious accounts.  This is especially true with multiple flags.  To explore this in the COVID-19 stream, we extracted all flags in the account descriptions for all 27 million accounts.  We've plotted the distribution of these in Figure \ref{fig:flag}, with bars colored by proportion of bots-like accounts.  

Reviewing the distributions in Figure \ref{fig:flag}, nothing in the one and two flag distribution is unexpected or necessarily cause for further exploration.  Once we get to the three, four, five, and six flag distributions, however, these are likely suspicious accounts.  In particular many of these accounts have multiple Western nations (US, Canada, and European Nations), and may be used to manipulate multiple Western nations while appearing to be an expatriate.

\section{Image/Meme Analysis}

\subsection{Meme Detection}

\begin{figure}[htbp]
\centering
  \includegraphics[width=\linewidth]{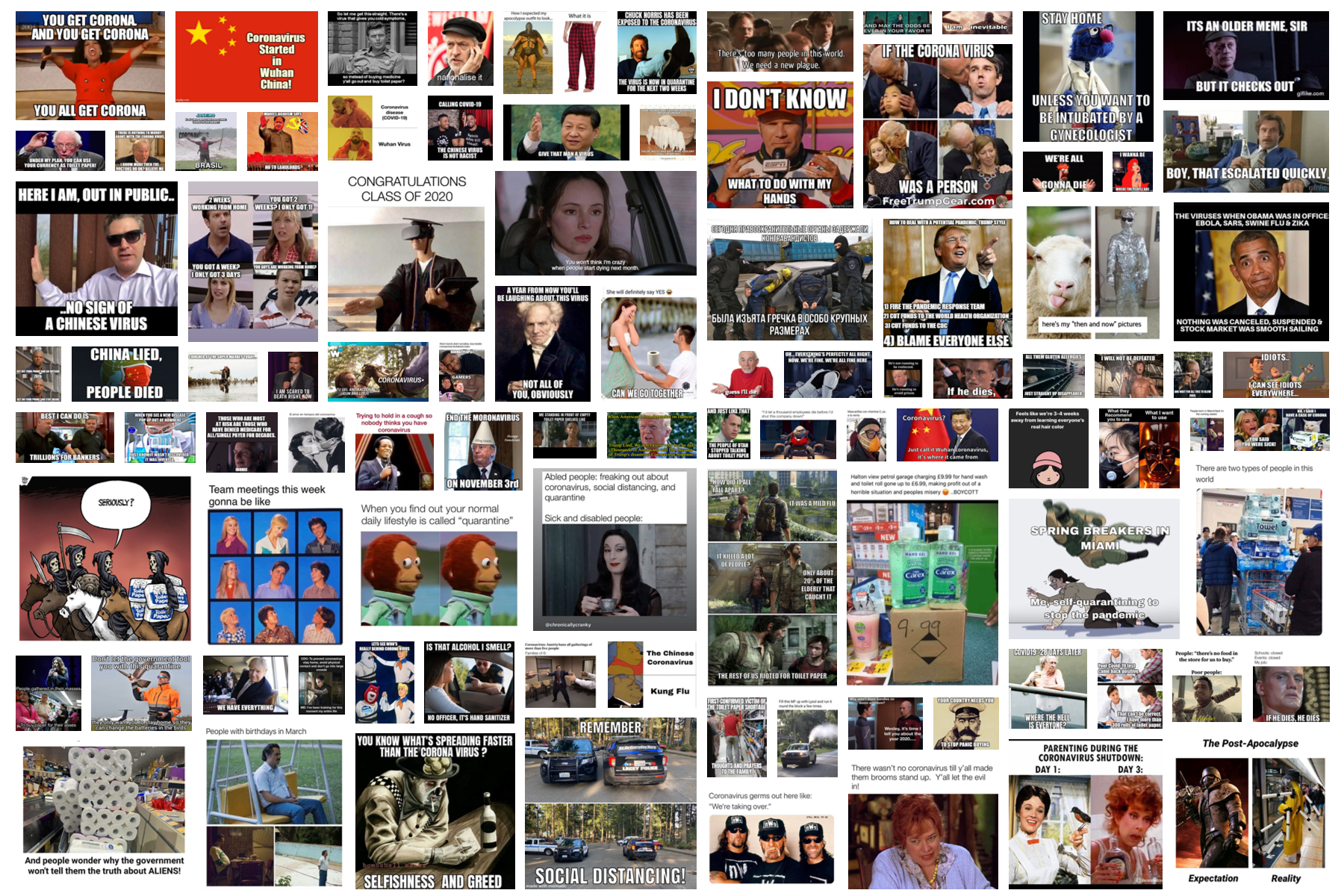}
  \caption{Memes extracted from a sample of 1 million COVID-19 related images.}
  \label{fig:virus_meme_collage}
\end{figure}

As discussed in Chapter 10, memes are a powerful way to connect a message to a target audience.  Memes evolve as they propagate through a society.  Given the size of the COVID-19 Stream, we sampled 1 million images (approximately 10\% of the total) and conducted meme classification on these.  The Meme-Hunter model classified 37,473 images as memes.  A collage of these memes is found in Figure \ref{fig:virus_meme_collage}.

Given the massive impact of COVID-19 on society and daily life, many memes were innocent humor designed to help folks get through some very tough circumstances.  We did find a number of political memes, however, some targeting domestic pandemic policy discussions and others targeting geo-political competition.  The domestic policy memes were trying to use image and text to argue for one of the competing priorities: namely the safety of society or the economic foundation of that society.  The geo-political memes were likely created by nation-states or nation-state proxies, with many memes created by Russia, China, and Iran, which will be discussed in more detail below.  

\subsection{Meme OCR and Topic Analysis}

\begin{figure}[htbp]
\centering
  \includegraphics[width=0.7\linewidth, trim = 4cm 4cm 4cm 4cm, clip]{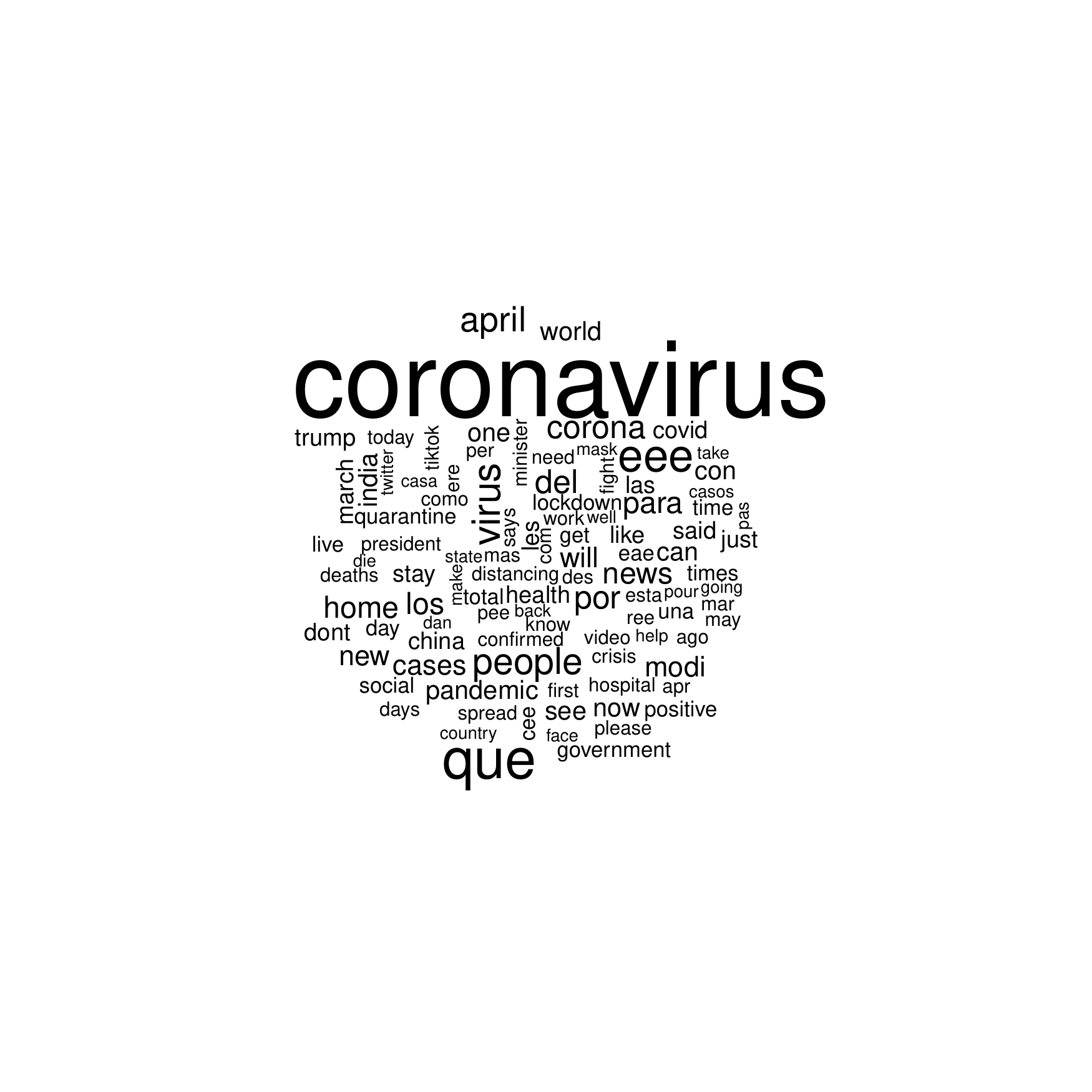}
  \caption{Wordcloud visualization of top words found in COVID-19 memes}
  \label{fig:meme_wordcloud}
\end{figure}

As discussed in Chapter 10, the \texttt{meme-hunter} suite of tools includes a special meme Optical Character Recognition (OCR) pipeline for meme images.  This was also presented in detail in \cite{BESKOW2020102170}.  We ran meme OCR on the extracted memes from 1 million images, and conducted wordcloud visual analysis of the results.  These results are found in Figure \ref{fig:meme_wordcloud}.

From these results we see that a number of general coronavirus memes.  We also see that a number of memes are targeting government leaders, ministers, and agencies.  

\subsection{Personalities}

\begin{figure}[htbp]
\centering
  \includegraphics[width=\linewidth]{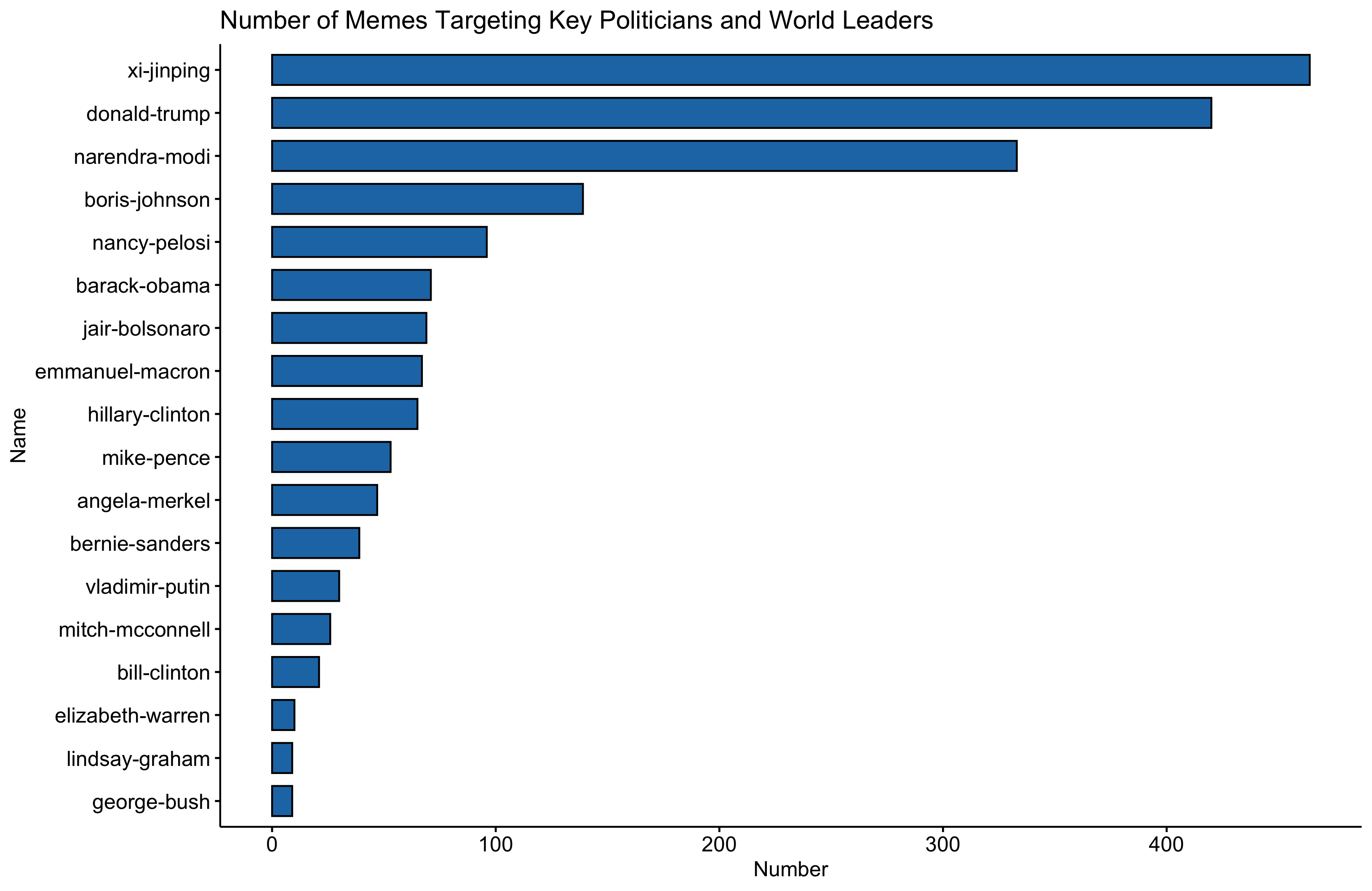}
  \caption{Distribution of memes on prominent US politicians and other world leaders (identified with open source facial recognition software)}
  \label{fig:face_detection_memes_virus}
\end{figure}

Next we used open source facial recognition software\footnote{\url{https://github.com/ageitgey/face_recognition}} to identify prominent politicians and world leaders in the memes.  Facial recognition software simply identifies personalities, but does not indicate whether the meme is supporting or attacking the specific personality found.  A distribution of memes about prominent US politicians and other world leaders is given in Figure \ref{fig:face_detection_memes_virus}.  

Here we see that prominent world leaders are the target of most of the memes in our sample, with Xi Jinping and Donald Trump in the first and second place positions, respectively.  This also highlights that much of the COVID-19 discussion and information conflict is between the US, China, and Europe. 

\subsection{Meme Evolution}

\begin{figure}[htbp]
\centering
  \includegraphics[width=\linewidth]{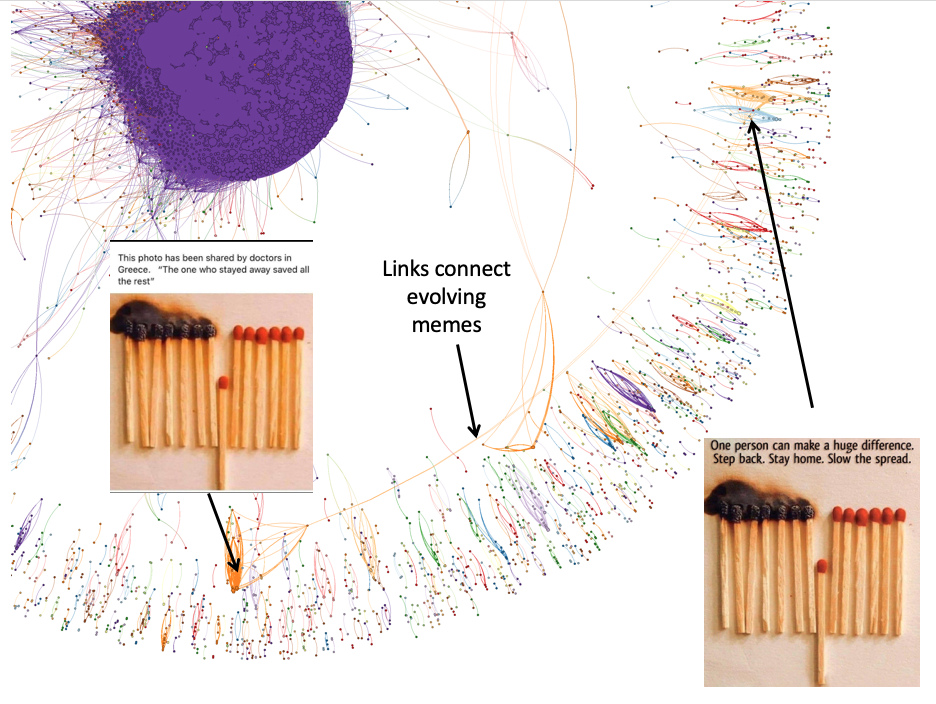}
  \caption{Evolution of Memes in the COVID-19 Stream}
  \label{fig:meme_evolution_virus}
\end{figure}

We next calculated the evolution of the 37,000 memes that Meme-Hunter classified in our sample.  The network was created by using a VGG deep learning model \cite{simonyan2014very} and extracting the last layer before softmax.  Using this 25,088 dimension vector to represent the image, we then conducted radius nearest neighbor graph learning with $distance = 1200$.  In Figure \ref{fig:meme_evolution_virus} we see that many of the darker image memes are clustered together in the center, with other prominent memes evolving in clusters that are separate components.   We've highlighted the evolution and links between two of the memes.  

\section{Sketch-IO}

We were able to use the COVID-19 data to test the Sketch-IO prototype application for sketching and analyzing information operation campaigns.  Given that our prototype application was not able to ingest the entire stream due to its size, we instead ingested all data produced by or propagating state sponsored media.  

\begin{figure}[htbp]
\centering
  \includegraphics[width=\linewidth]{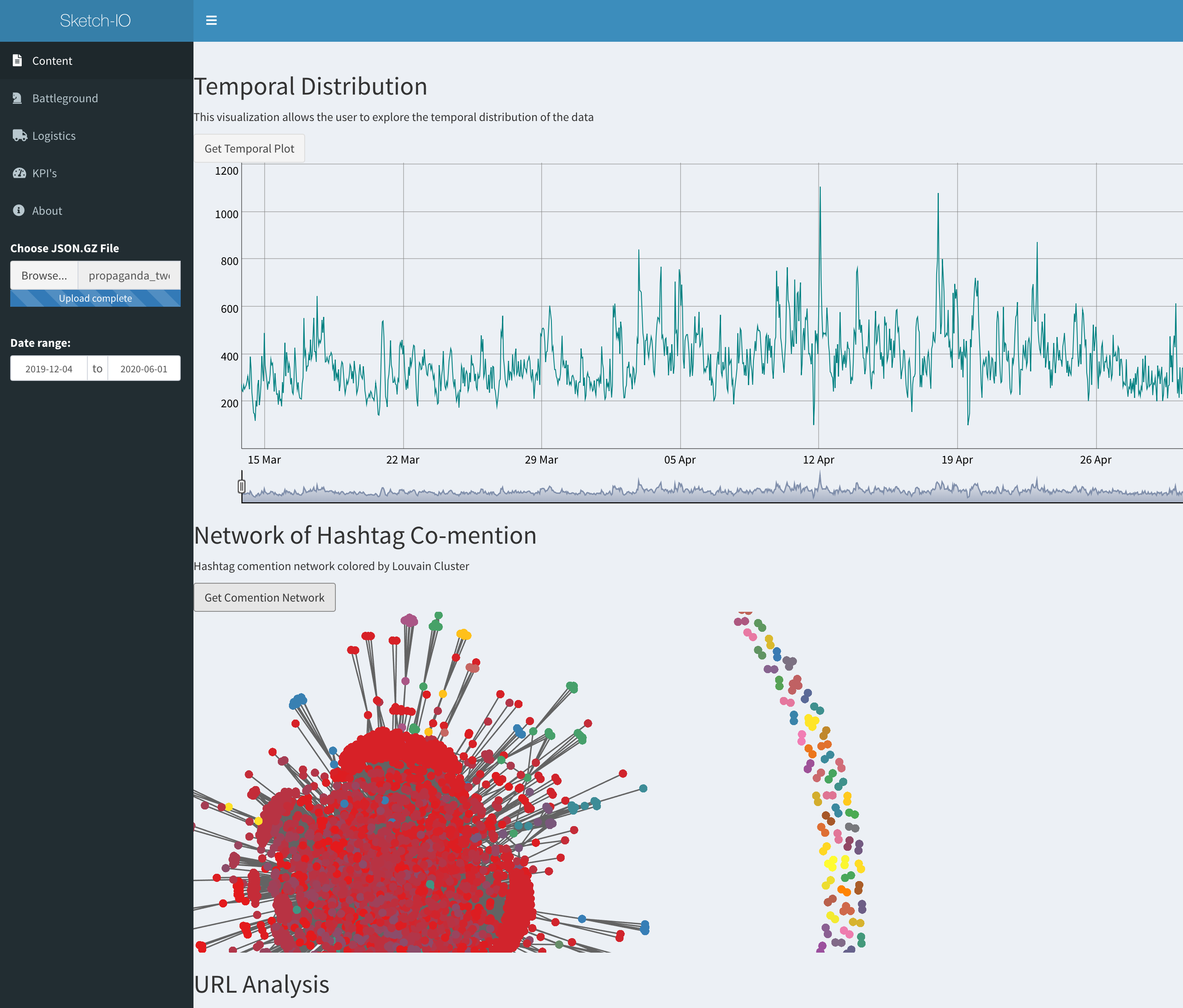}
  \caption{Demonstration of the Sketch-IO prototype dashboard on the COVID-19 stream.}
  \label{fig:sketch_IO}
\end{figure}

The Sketch-IO application proved effective and responsive at quickly performing a number of battle drills to analyze this data.  Example usage and screenshots are provided in Figure \ref{fig:sketch_IO}.


\section{Foreign Influence Operations}

In this section we will identify overt and seemingly overt propaganda operations by nation state actors.  Covert and black propaganda is much harder to detect and attribute to a nation state actor (black propaganda is designed to make the victim appear to be the perpetrator). 

\begin{figure}[htb]
\centering
  \includegraphics[width=\linewidth]{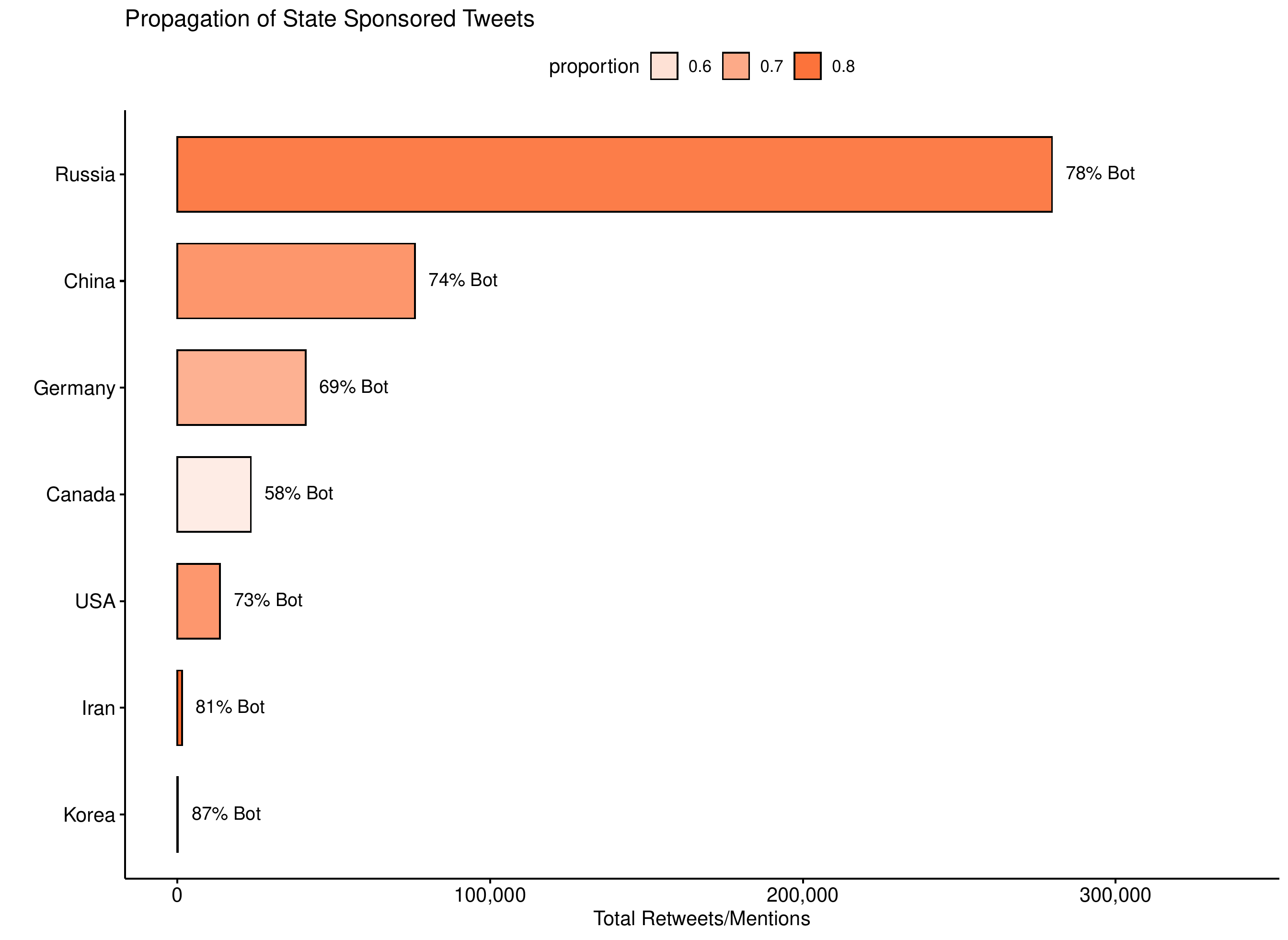}
  \caption{Propagation of State Sponsored Media by Country (colored by proportion of bot involvement)}
  \label{fig:propaganda_bar}
\end{figure}

We will use the \texttt{bot\_labeler} function for finding state sponsored media and state sponsored media amplification as a proxy for finding propaganda.  In Figure \ref{fig:propaganda_bar} we show the number of retweets of state sponsored media accounts as measured by \texttt{bot\_labler}.  This image is colored by percentage of bots that are retweeting these accounts.  As indicated in Chapter 11, these state sponsored media vary drastically in purpose and independence (the purpose of Russian RT is different than the US Voice of America).  That being said, we clearly see that Russia and China are investing heavily in producing and promoting state sponsored media and messages.  These messages are amplified by both bots and legitimate accounts.

\subsection{Chinese Propaganda}

To some extent, the COVID19 pandemic seems to be a turning point for the Chinese in regards to information operations.  Traditionally, Chinese information operations focused on positive narrative largely pushed by the ``50 cent Army'', a low-paid or otherwise coopted group of online netizens.  Traditionally, they did not conduct ``higher risk higher reward'' operations that relied on negative, antagonizing and controversial statements.  This slowly started to change with the Hong Kong protests in 2019, and fully changed with COVID-19 pandemic, with China seeming to adopt more aggressive IO practices and operations.  

\begin{figure}[htb]
\centering
  \includegraphics[width=\linewidth]{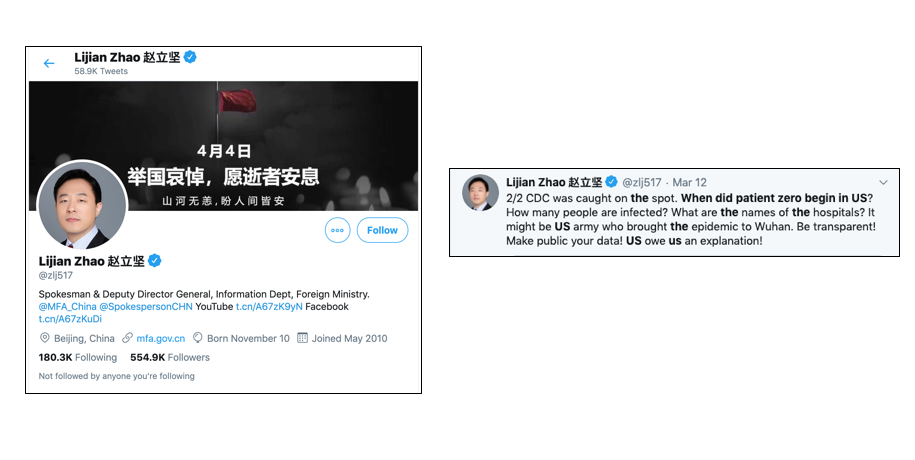}
  \caption{Foreign Minister Lijian Zhao's rise to director of Chinese information operations seems to correlate with a more aggressive approach than historical Chinese information operations (Tweet ID: 1238111898828066823).}
  \label{fig:propaganda_china2}
\end{figure}

In February 2020, Lijian Zhao was promoted to the Deputy Director General, Information Department, Foreign Ministry.  Foreign Minister Zhao has a history of aggressive information operations, and seems to be implementing this in the Chinese information operations as well as personally on social media.  On March 12, Zhao posted a tweet that suspected the US Army of bringing the coronavirus to Wuhan province in China (see Figure \ref{fig:propaganda_china2}).  Zhao seems to be a key personality leading and directing China's more aggressive approach to information operations.

Chinese propaganda is amplified by Chinese government representatives.  Chinese government officials around the world are able to amplify state sponsored media without questioning it, knowing that only approved messages are published.  In the COVID-19 stream, the Chinese Ambassador to Venezuela has the 2nd highest number of state sponsored mentions/retweets.  He retweeted or mentioned Chinese state sponsored Spanish and English COVID-19 content 769 times.  The increasing penetration of Chinese state sponsored media around the world is pushed by these legitimate accounts.  

It also appears that China often uses trolls instead of bots.  With easy access to human capital, the Chinese seem to prefer the control and nuance that trolls allow compared to a bot Army.  Many of the suspicious Chinese accounts contain enough nuance and temporal patterns to consider them a troll rather than a bot.    

\begin{figure}[htbp]
\centering
\begin{subfigure}{\linewidth}
  \centering
  \fbox{\includegraphics[width=\linewidth]{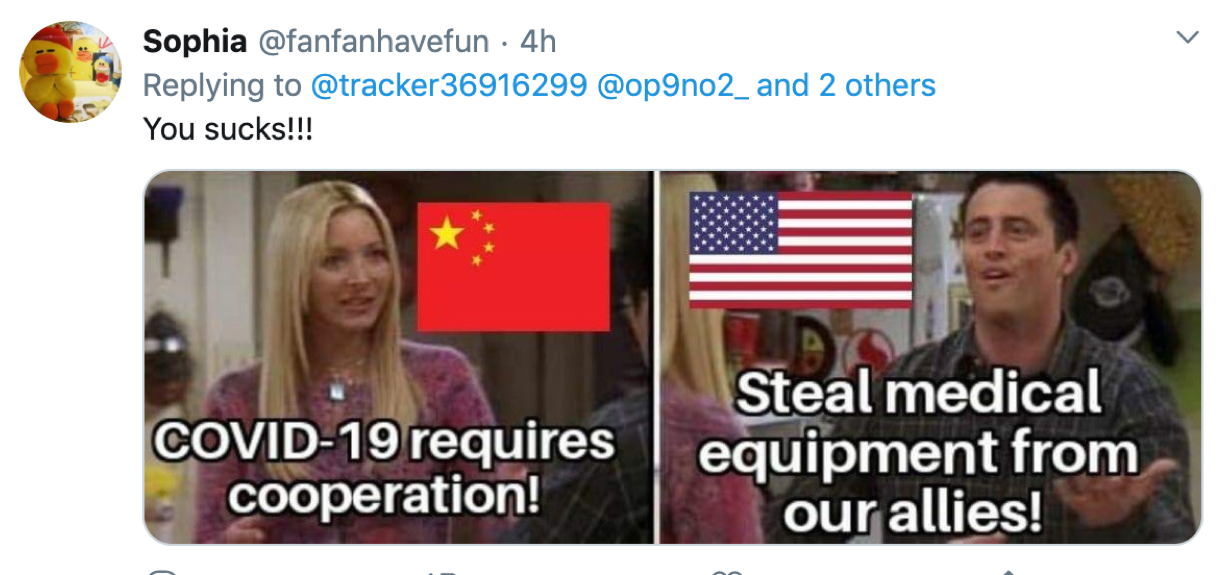}}
  \caption{Chinese meme connects propaganda message with American cultural artifact (Tweet ID: 1247081317784682496)}
  \label{fig:china_propaganda1}
\end{subfigure}%
\vspace{0.5cm}
\begin{subfigure}{\linewidth}
  \centering
  \fbox{\includegraphics[width=\linewidth]{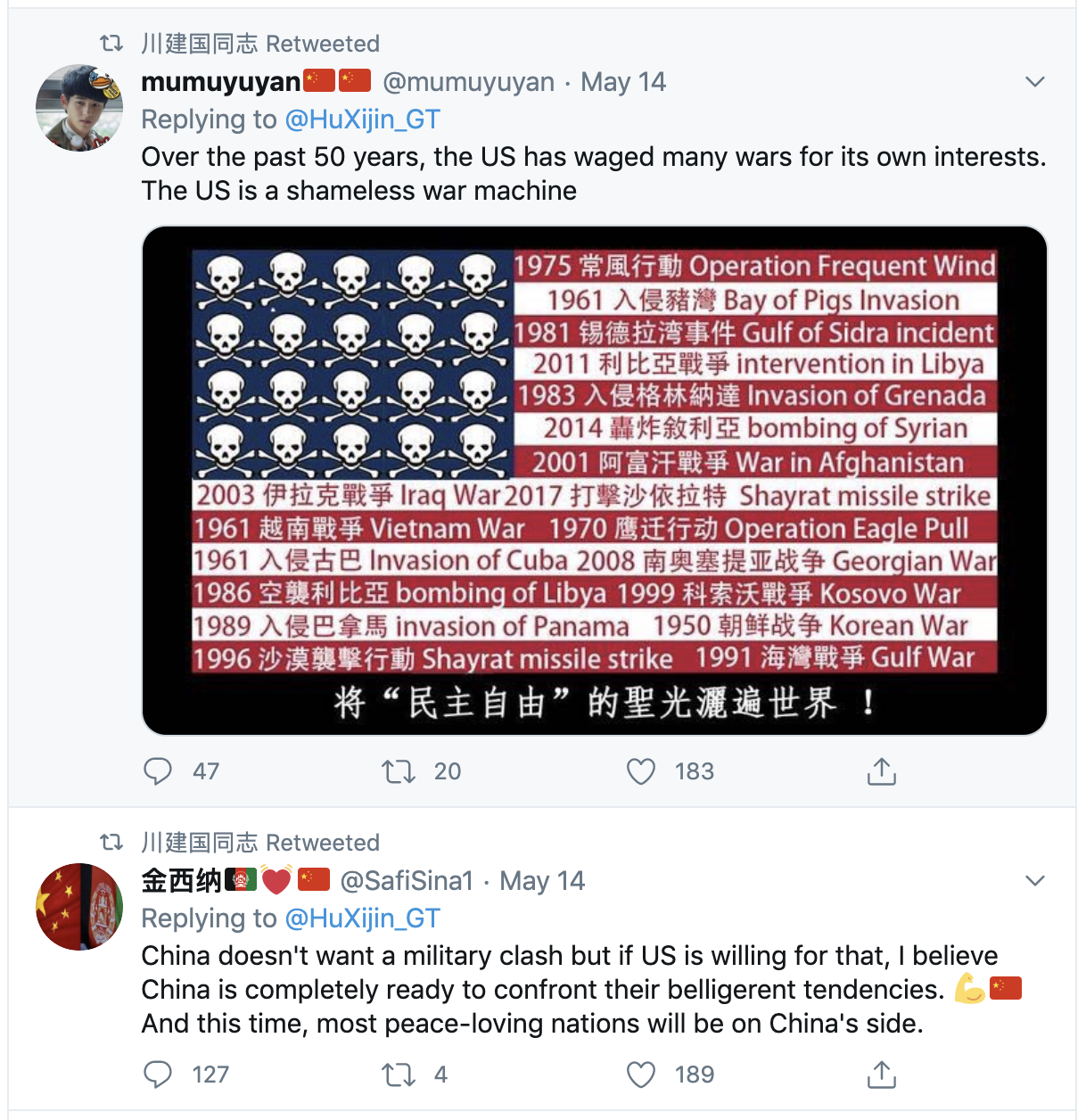}}
  \caption{Chinese information operation combines visual image with geo-political message (Tweet ID: 1260871418209660928) }
  \label{fig:china_propaganda13}
\end{subfigure}
\caption{Examples of Chinese Propaganda}
\label{fig:china_propaganda_example}
\end{figure}

We see increased use of meme warfare by China in the COVID-19 stream.  Historically, China has seemed hesitant to use memes, potentially because memes propagate through evolution, and this evolution is outside of the control of the state \cite{beskowIRA2020}.  China has especially been concerned with the evolution of memes within their own population, and has banned some memes \cite{BESKOW2020102170}.  In COVID-19, however, we see them developing and deploying memes in a way that is more akin to Russian information operations.  An example of a Chinese meme is provided in Figure \ref{fig:china_propaganda1}. Here we see a message tied to an image that has clear cultural relevance and traction within the target audience.   Additional examples of Chinese Memes that were found using the Bot-Match methodology are provided in Figure \ref{fig:china_bot_match_collage_clean} below.

\begin{figure}[htbp]
\centering
\begin{subfigure}{\linewidth}
  \centering
  \includegraphics[width=\linewidth]{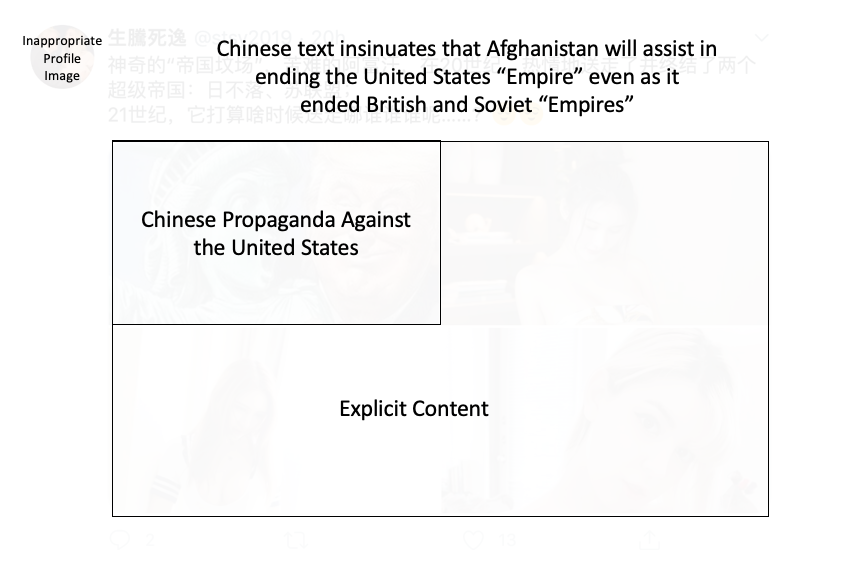}
  \caption{Chinese Propaganda Combined Image and Text Propaganda with Adult Content (Tweet ID: 1262737238451986432)}
  \label{fig:china_propaganda15}
\end{subfigure}%
\vspace{0.5cm}
\begin{subfigure}{\linewidth}
  \centering
  \includegraphics[width=\linewidth]{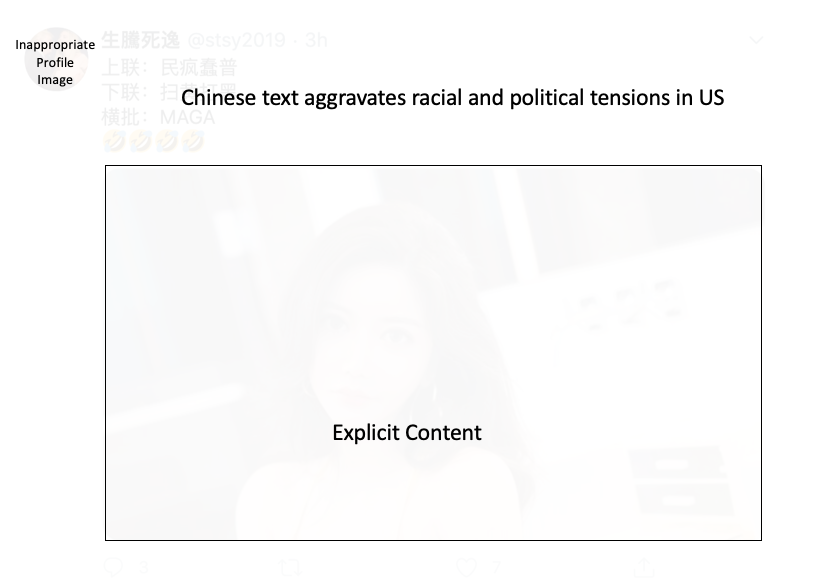}
  \caption{Chinese Propaganda Combined Anti-US Propaganda with Adult Content}
  \label{fig:china_propagand16.png}
\end{subfigure}
\caption{China trolls following Russian example by Mixing Propaganda with Explicit Content.  Given the Chinese text, this is arguably targeting domestic audiences in China in an attempt to strengthen nationalism and support for the CCP (Tweet ID: 1266697198177095687).}
\label{fig:china_explicit}
\end{figure}

We also found evidence that China is starting to interlace adult oriented and humorous content with their information in order to increase traffic, particularly from certain demographics.  Examples of this are found in Figure \ref{fig:china_explicit}).  This has historically been a key part of Russian IO, and it seems that China is increasingly adopting similar practices.  Explicit adult content is often designed to attract and manipulate the minds of young impressionable men.  

In the Chinese propaganda, we see some propaganda uses Chinese language while others uses English language content.  The English text is often accompanied by memes that connect the message with American audiences.  The English language propaganda is arguably targeting American audience, attacking leaders and institutions in America.  It is also designed to strengthen American's view of China, Chinese leadership and the Chinese Communist Party (CCP).  The propaganda that uses Chinese language text is arguably targeting Chinese audiences within China's borders as well as in other Asian countries.  This propaganda is designed to increase nationalistic fervor within China's borders.  

Within the Chinese propaganda that we viewed, the vast majority was singularly focused on the United States.  This differs from Russian information operations, which focuses more broadly on the West, adding European actors to their list of targets.  

\subsection{Russian Propaganda}

Within the COVID-19 stream, Russia appears to be staying with their historical information playbook.  With extensive experience in manipulating world opinion with their ``active measures'' throughout the Cold War, Russia has long had one of the most aggressive and well-resourced information operations capabilities among nation states.  Russia's information operations are tightly coupled with their other cyber operations.  

\begin{figure}[htbp]
\centering
\begin{subfigure}{\linewidth}
  \centering
  \includegraphics[width=\linewidth]{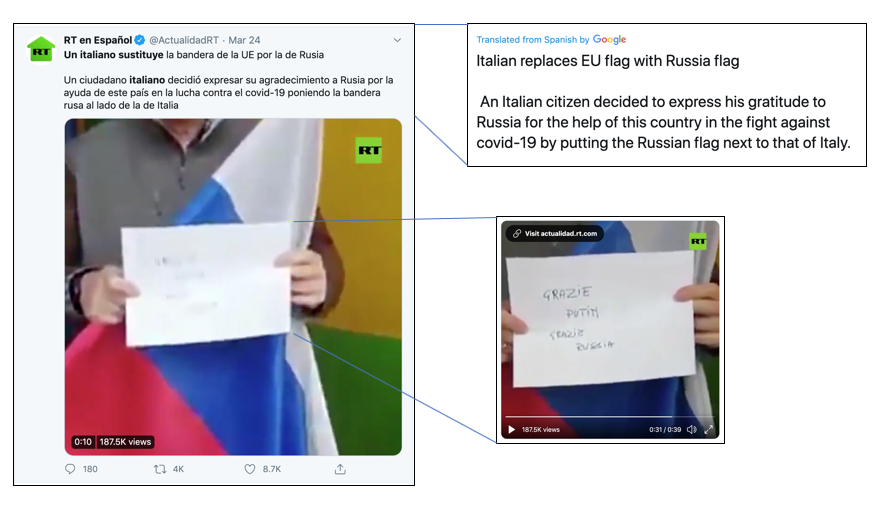}
  \caption{Russian Propaganda promotes story that Italians favor Russia and Russian leadership over European Union leadership (Tweet ID: 1243453201891725312)}
  \label{fig:propaganda_russia1}
\end{subfigure}%
\vspace{0.5cm}
\begin{subfigure}{\linewidth}
  \centering
  \includegraphics[width=\linewidth]{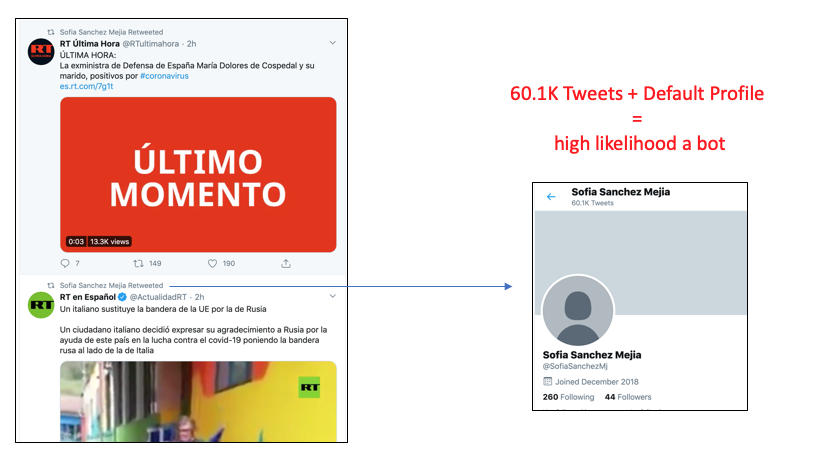}
  \caption{Russian Propaganda is pushed by large bot ``armies''}
  \label{fig:propaganda_russia2}
\end{subfigure}
\caption{Russian Information Operations rely on state sponsored media messages pushed by large bot armies}
\label{fig:propaganda_russia}
\end{figure}

Russian information operations rely on increasing penetration of their state sponsored media around the world.  RT, Sputnik, and other state sponsored media outlets offer news in many languages around the world.  These media outlets offer news stories that support Russian information narratives, as seen in Figure \ref{fig:propaganda_russia}.  In this Figure we see state state sponsored media promote stories that Russia provided more help to the Italian population than did the European Union.

As seen in Figure \ref{fig:propaganda_russia2}, Russian operations still use large and sophisticated bot ``armies'' to push their content.  This observation is supported by the quantitative analysis seen in Figure \ref{fig:propaganda_bar}, which shows the amplification of Russian state sponsored messages is 78\% bots.  

\begin{figure}[htbp]
\centering
\begin{subfigure}{0.8\linewidth}
  \centering
  \includegraphics[width=\linewidth]{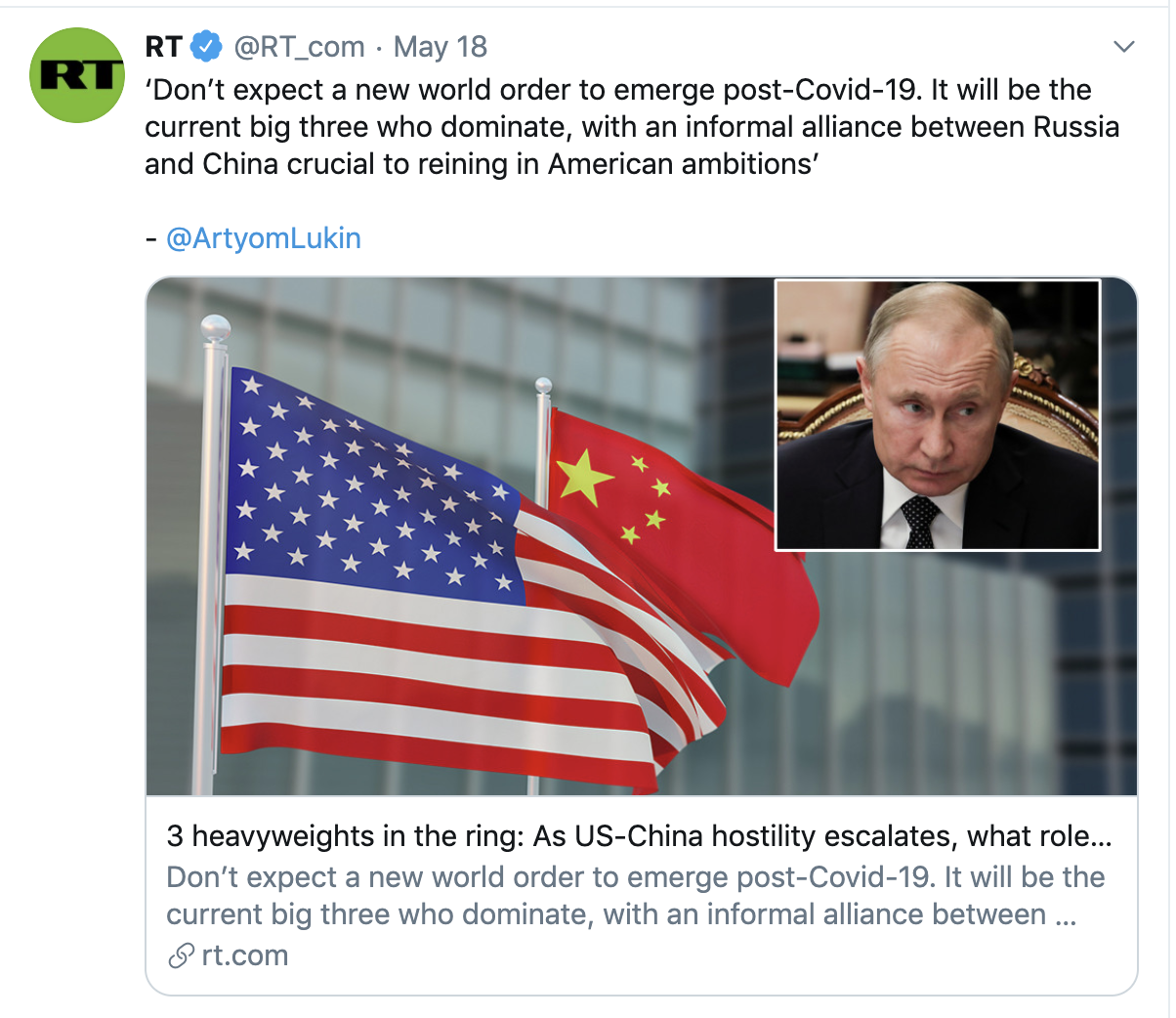}
  \caption{Russian State Sponsored Media promote geo-political messages that extend beyond reporting news (Tweet ID: 1262513293714849793) }
  \label{fig:russia_nationa_security}
\end{subfigure}%
\vspace{0.5cm}
\begin{subfigure}{0.8\linewidth}
  \centering
  \includegraphics[width=\linewidth]{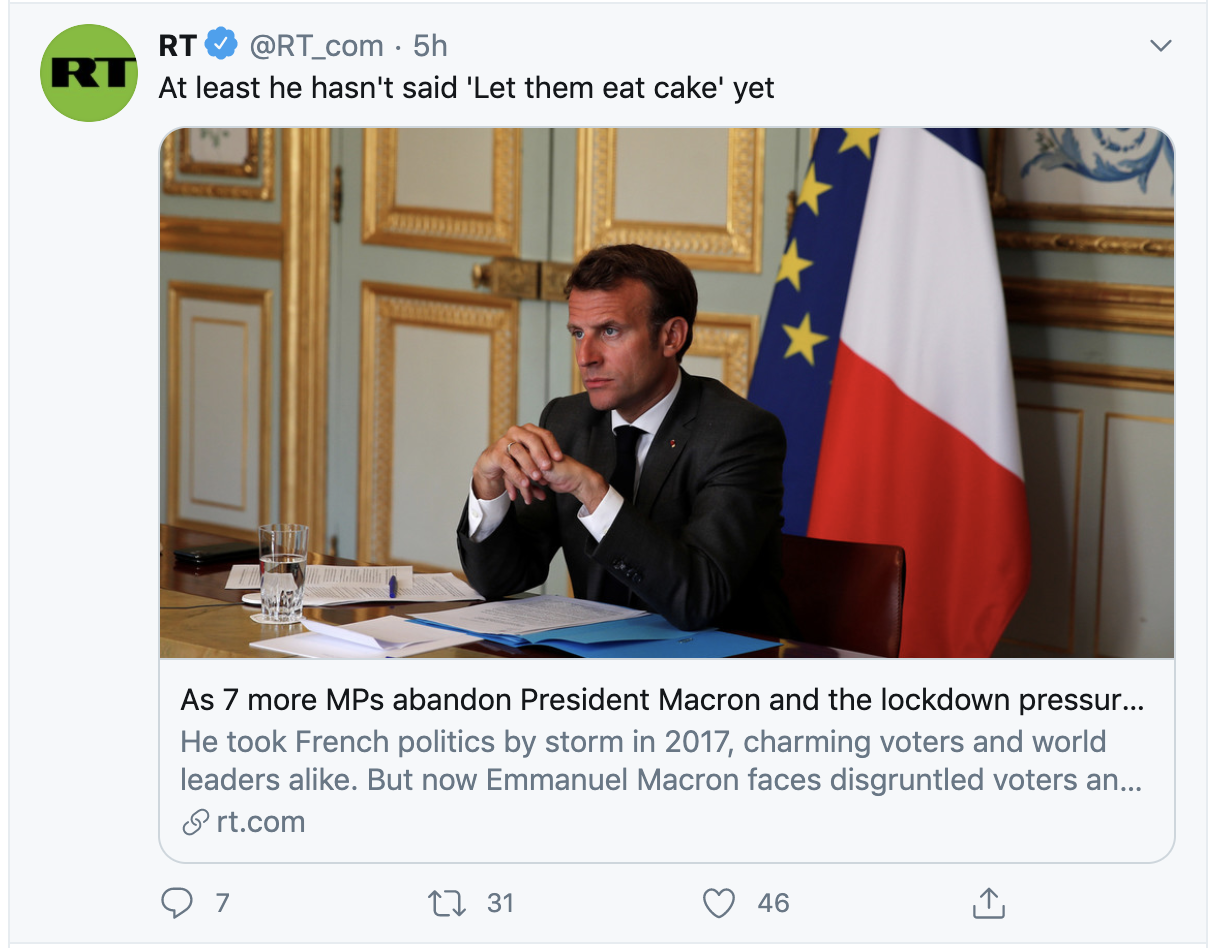}
  \caption{Russian State Sponsored media trolls European leaders (Tweet ID: 1262966278555176960)}
  \label{fig:russia_propaganda1}
\end{subfigure}
\caption{Examples of Russian Propaganda attacking the West}
\label{fig:propaganda_russia_state}
\end{figure}

As indicated above, Russian operations target the West in general, with European targets receiving almost as much emphasis as the United States.  This differs from Chinese operations, which seem to primarily target the United States, with smaller efforts directed at Europe.  In Figure \ref{fig:propaganda_russia_state} we see examples of Russian state media trolling the West.  While Russian state sponsored media advertise themselves as news organizations, this content is arguably well beyond reporting unbiased news.

\subsection{Iranian Propaganda}

\begin{figure}[htb]
\centering
  \includegraphics[width=\linewidth]{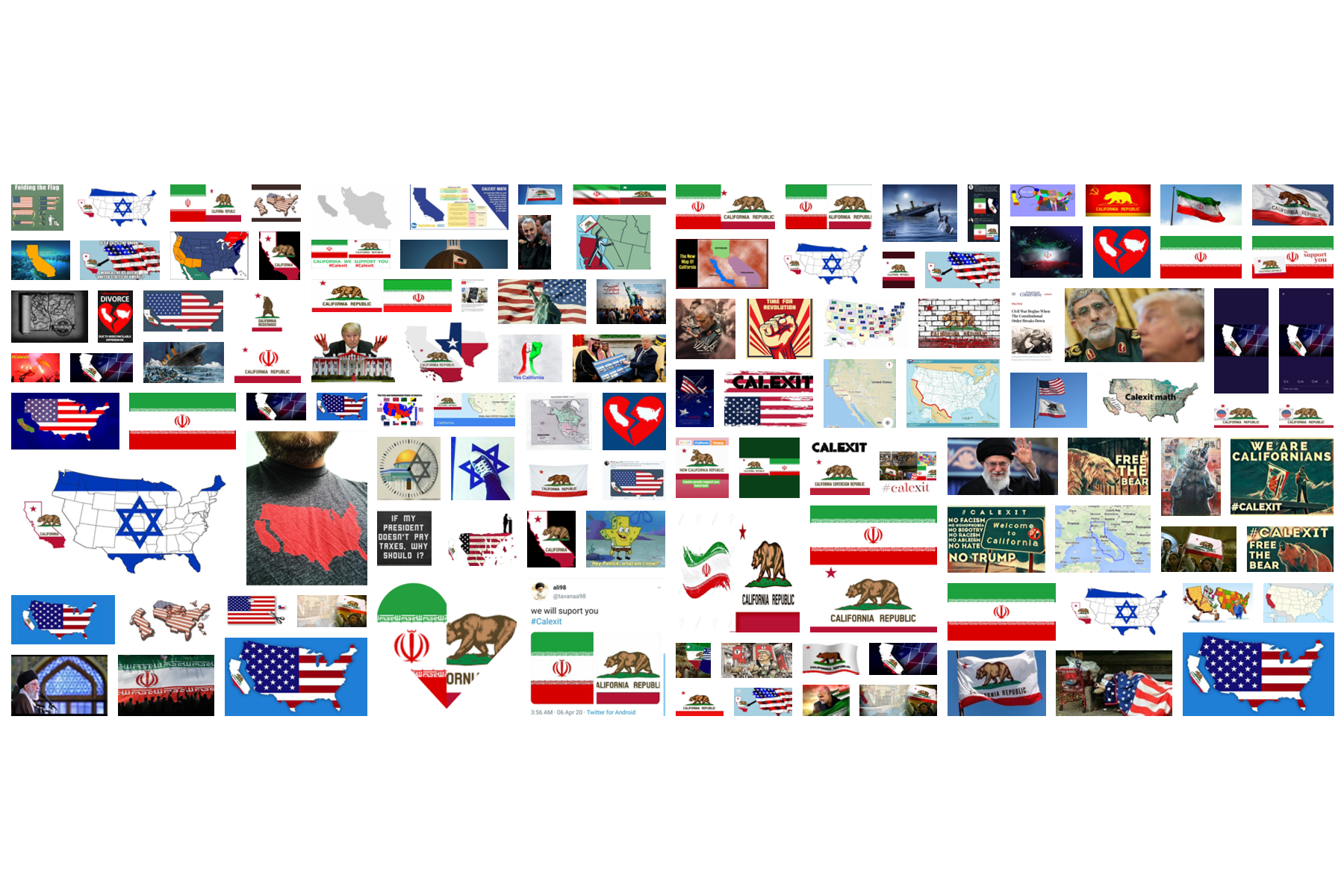}
  \caption{Memes used in an Iranian information operation initiated on 6 April 2020.  The attack encouraged California citizens and leadership to exit the United States.}
  \label{fig:CalExit_Collage}
\end{figure}

On 6 April 2020, Iran initiated a concerted attack on the United States trying to encourage California to exit the union. Most of this effort was tagged with \#CalExit. By the time the dust settled approximately 30K tweets had been launched by a several thousand strong bot/troll Army. This is not the first time that \#CalExit (or other similar messages like \#Texit), have become trending hashtags on Twitter largely due to foreign information operations.  The 6 April surge appears to be largely Iranian influence operation that was triggered by domestic political tension in the United States.  In the days and weeks preceding 6 April, tensions between US national leadership and California leadership intensified, and the governor of California referred to California as a ``nation-state'' \cite{Coronavi55:online}.  The Iranian government and/or proxies appeared to be monitoring these political tensions in the United States, and timed their \#CalExit campaign to capitalize on them.  This 24 hour information operation had some creative content with many human/troll accounts and limited automation.  The creative content can be seen in the meme collage in Figure \ref{fig:CalExit_Collage}.

\section{Bot Match to Find Foreign Influencers}

As discussed in Chapter 9, \texttt{bot-match} can be a very powerful tool for finding similar accounts given a seed account.  Bot-Match allows you to find similar accounts where similarity can be defined by network proximity, semantic proximity, or a combination of both.  The size of our network and tweet corpus limit the number of models available for measuring similarity.  We first concatenate all user text, thereby aggregating content to the account level.  Because of the size of our corpus, we then used cosine similarity on a document-term matrix (also known as a bag-of-words) with 4000 top words.  This was created for English tweets.  

\begin{figure}[htbp]
\centering
  \includegraphics[width=\linewidth]{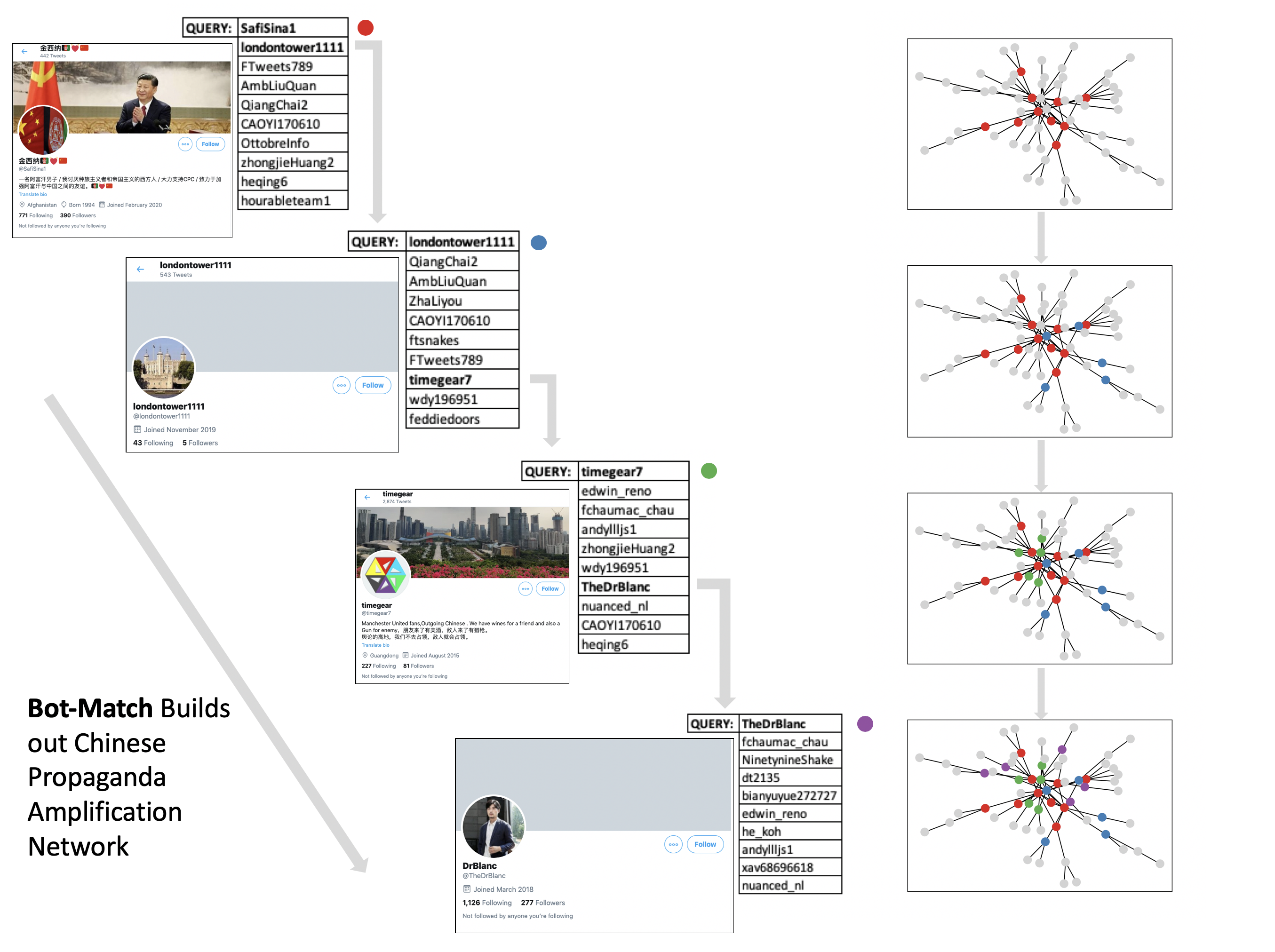}
  \caption{Bot-Match was recursively used to find accounts that are propagating and amplifying Chinese state propaganda.  In this image, @SafiSina1 (see Figure \ref{fig:china_propaganda13}) was used as the seed account.}
  \label{fig:bot_match_china}
\end{figure}

We used this to identify accounts that were propagating Chinese propaganda in English.  We used @SafiSina1 as our seed account (this account was discovered above in Figure \ref{fig:china_propaganda13}).  We illustrate in Figure \ref{fig:bot_match_china} how we recursively build out the Chinese propaganda network with Bot-Match.  Notice that we did not have to conduct any elaborate labeling or training process, we just needed to start with our seed node and the document-term matrix.  All accounts seen in Figure \ref{fig:bot_match_china} are amplifying Chinese state sponsored media, and are each embedded in slightly different networks.  

\begin{figure}[htbp]
\centering
  \includegraphics[width=\linewidth]{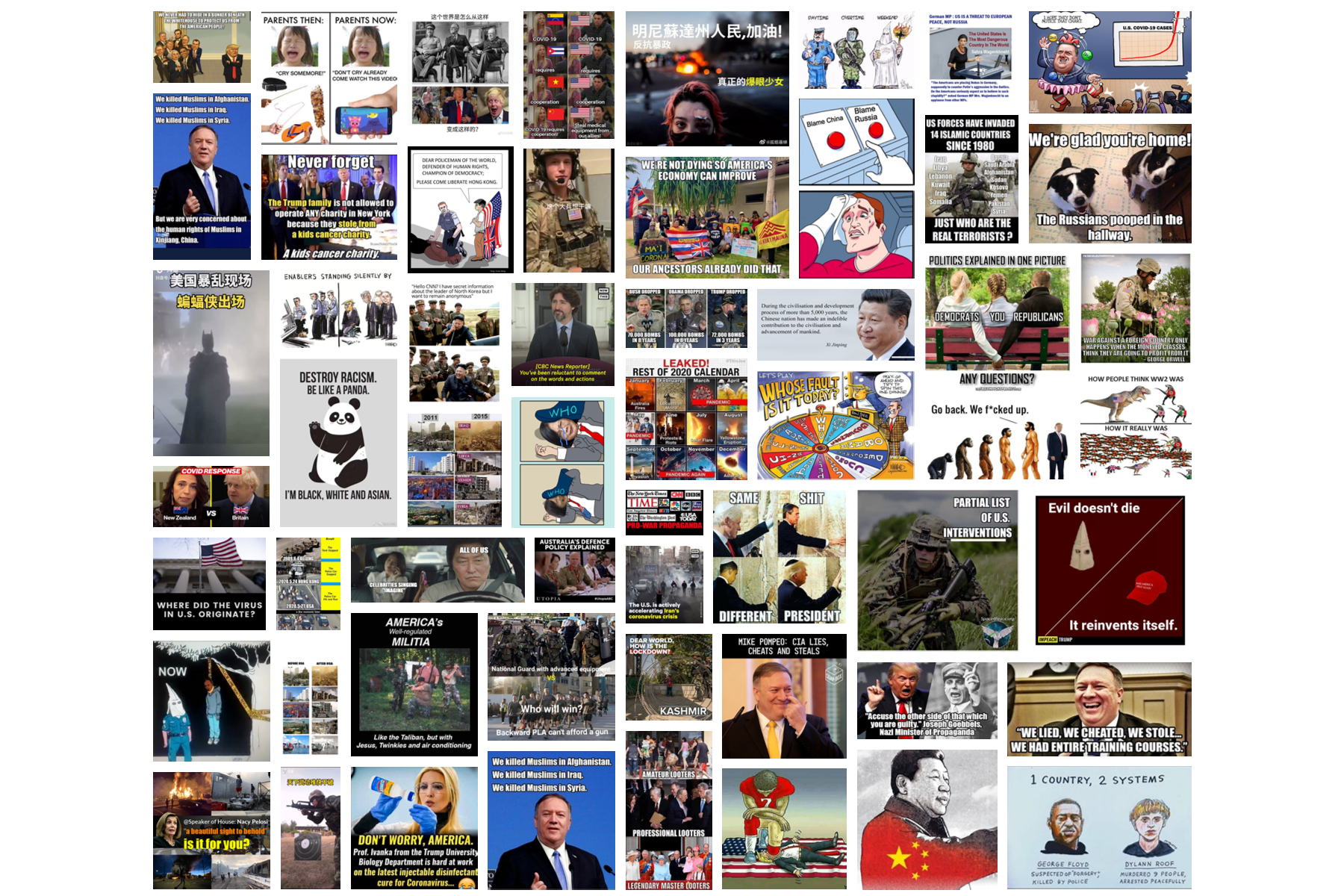}
  \caption{Sample of expansive use of Chinese memes propagated by 20 accounts discovered with Bot-Match}
  \label{fig:china_bot_match_collage_clean}
\end{figure}

Using the Bot-Match methods illustrated in Figure \ref{fig:bot_match_china}, we were able to identify approximately 20 additional accounts that appeared to be propagating Chinese propaganda targeting America.  We then used the Twitter REST API to scrape the timeline history of these accounts, extract image links, and run Meme-Hunter on the images shared by these accounts.  We found that indeed all of these accounts were conducting very targeted information operations against the United States around the COVID-19 pandemic as well as the American protest that followed the death of George Floyd at the hands of Minneapolis police officer.  A collage of a sample of these targeted memes is provided in Figure \ref{fig:china_bot_match_collage_clean}.

\section{Discussion of BEND Forms of Maneuver}

Throughout our analysis of foreign influence, we observed their use of the BEND forms of maneuver.  We continue to observe that Russia closely intertwines \textit{narrative} and \textit{network} maneuvers.  They conduct long and protracted efforts to infiltrate target audiences before interjecting narrative.  China, while working hard to control and manipulate the narrative, does not appear to infiltrate targeted subcultures.  They do appear to tie their narrative to the target audience culture as seen above with memes based on the Friends sitcom as well as memes that use the George Floyd protests to support pro-CCP policies.  With the limited network maneuver, it remains to be seen whether their information operations gain traction or simply become a ``shot in the dark.''  Iran also appeared to launch large attributed information campaigns focused on a specific narrative such as \#CalExit, without first preparing target networks.  Once again, these operations may become information warfare ``chaff'' with limited effects.

\section{Conclusion}

The primary goal of this case study was to illustrate social cybersecurity workflows in a relevant event.  Using the apt COVID-19 Twitter stream from 15 March to 30 April 2020, we demonstrated how to collect the data, conduct initial exploratory data analysis, conduct and use bot and meme classification and exploration, conduct account characterization, and demonstrate the role of Sketch-IO and the BEND Framework.  This chapter therefore serves as an example for social cybersecurity researchers on how to leverage these tools to identify and characterize offensive information operations targeting their society, institutions and culture.

In regards to the COVID-19 pandemic, we found large information operations trying to manipulate domestic and international perceptions, beliefs, and actions.  At the domestic level the information conflict was largely over pandemic policy, particularly whether public safety or the economy were more important.  At the international level we identified attempts to manipulate international perception of the origins of the disease as well as perceptions of each countries handling of the disease.  We also see nation-state efforts to amplify tensions and drive wedges in existing fissures in rival nations.  Throughout the data we see bots and trolls used to scale and spread narrative, therefore acting like a ``force-multiplier'' in information operations.  We see Russia continue to use memes, and China massively increase their use of memes in information warfare.  Both nations study their target audience and choose relevant cultural artifacts to connect their memes and messages to the target audience.  

Even as the COVID-19 coronavirus moved much of business and society to use virtual platforms for social interaction as well as business and collaboration, it also allowed many nation-states to increasingly use virtual platforms for competition in the information space with ramifications for geo-politics.  While the effects of these campaigns are hard to measure, their scale and persistence require social cybersecurity policy and process.

\section*{ACKNOWLEDGMENT}

This work was supported in part by the Office of Naval Research (ONR) Multidisciplinary University Research Initiative Award  N000140811186, Award N000141812108, ONR Award N00014182106 and the Center for Computational Analysis of Social and Organization Systems (CASOS). The views and conclusions contained in this document are those of the authors and should not be interpreted as representing the official policies, either expressed or implied, of the ONR or the U.S. Government.


\bibliographystyle{plain}
\bibliography{references.bib}

\end{document}